%% file: main.tex
\pdfoutput=1
\documentclass[10pt]{article}

\usepackage[preprint]{tmlr}

\usepackage{amsmath,amssymb,amsthm,mathtools}
\usepackage{booktabs,graphicx,xcolor,enumitem,microtype}
\usepackage[colorlinks=true,linkcolor=blue!60!black,citecolor=blue!60!black,urlcolor=blue!60!black]{hyperref}
\usepackage{url}
\usepackage[capitalise,noabbrev]{cleveref}
\crefname{assumption}{Assumption}{Assumptions}
\usepackage{caption}
\usepackage[section]{placeins}
\usepackage{needspace}
\newtheorem{theorem}{Theorem}
\newtheorem{proposition}[theorem]{Proposition}
\newtheorem{corollary}[theorem]{Corollary}
\newtheorem{lemma}[theorem]{Lemma}
\theoremstyle{definition}
\newtheorem{assumption}{Assumption}

\theoremstyle{remark}

\newcommand{\E}{\mathbb{E}}
\renewcommand{\P}{\mathbb{P}}
\newcommand{\R}{\mathbb{R}}

\title{Drift or Immaturity? Correcting for Maturity Matters;\\ Learning a Young Cohort's Own Speed Rarely Helps}

\hypersetup{pdftitle={Drift or Immaturity? Correcting for Maturity Matters; Learning a Young Cohort's Own Speed Rarely Helps},pdfauthor={Aaron Ontoyin Yin, Kelvin Amoaba, Zakari Salifu, Yigit Ihlamur, Fuat Alican, Ben Griffin}}
\author{\name Aaron~Ontoyin~Yin$^{1}$, Kelvin~Amoaba$^{1}$, Zakari~Salifu$^{1}$,\\ Yigit~Ihlamur$^{1}$, Fuat~Alican$^{1}$, Ben~Griffin$^{1,2}$\\
      \addr $^{1}$Vela Research, San Francisco, USA\qquad $^{2}$University of Oxford, Oxford, UK\\
      \addr \texttt{\{aaron,kelvin,zak,yigit,fuat,ben\}@vela.partners}\\
      \addr Code: {\upshape\url{https://github.com/Vela-Research/drift-or-immaturity}}}

\date{}

\begin{document}
\maketitle

\input{sections/abstract}
\input{sections/intro}
\input{sections/setup}
\input{sections/windows}
\input{sections/identification}
\input{sections/limits}
\input{sections/experiments}
\input{sections/discussion}
\input{sections/ai_use}

\bibliographystyle{tmlr}
\bibliography{refs}

\clearpage
\appendix
\input{sections/app_map}
\input{sections/app_scales}
\input{sections/app_proofs}
\input{sections/app_theory}
\input{sections/app_estimator}
\input{sections/app_experiments}
\input{sections/app_data}
\input{sections/app_related}
\input{sections/app_impact}

\end{document}

%% file: sections/abstract.tex
\begin{abstract}
Many outcomes, such as loan defaults or startup exits, take years to arrive, so recent cohorts' labels are provisional. Because outcomes not yet observed count as negatives, recent rates look low and are easily mistaken for drift, yet drift detectors and adaptive windows assume final labels. We show that adaptive training windows (Lepski's method) and the ADWIN drift detector, fed provisional labels, get worse with more data, settling on the youngest and most biased cohort even when nothing changes. Weighting cohorts by their expected maturity, the share of outcomes already observed, removes this failure and, with known maturities, is minimax-optimal up to a log factor. When cohorts differ in how fast outcomes arrive (their speed), corrections that assume one timing curve, such as the chain ladder of insurance reserving, can report change in the wrong direction. A cohort's own speed is identified from one snapshot in principle, but not knowing it multiplies the asymptotic variance of a young cohort's estimated rate by a factor set by its maturity, not sample size: with Weibull timing, two years into a seven-year horizon, matching the precision of a known speed needs 58 times more data. We test these ideas on 4.2 million mortgages and 1.9 million patent applications, and in a pre-registered replication on 1.4 million small-business loans and 3.9 million federal civil cases. In the replication, maturity-corrected labels cut the error of classifiers' predicted cohort rates by 73\% and 91\%. Cohort speeds gave no significant gain in any data set (a 15\% gain was ruled out only on small-business loans). A standard competing-risks hazard model beat our step-by-step estimator on patents and civil cases and was about as accurate on mortgages and, as registered, small-business loans. In our experiments, correcting for maturity matters most; estimating a young cohort's own speed rarely helps.
\end{abstract}

%% file: sections/intro.tex
\section{Introduction}\label{sec:intro}

Many outcomes that machine learning systems predict arrive late: a startup may take a decade to exit, and a mortgage that defaults may do so years after origination. Such data come in \emph{cohorts} of units that entered in the same period, and at any snapshot (the date at which the data are extracted) old cohorts have revealed most of their outcomes and young cohorts few. A label that counts outcomes not yet observed as negatives is therefore \emph{provisional}, and a model that treats provisional labels as final underestimates recent cohorts' rates.

\begin{figure}[t]
\centering
\includegraphics[width=1.0\linewidth]{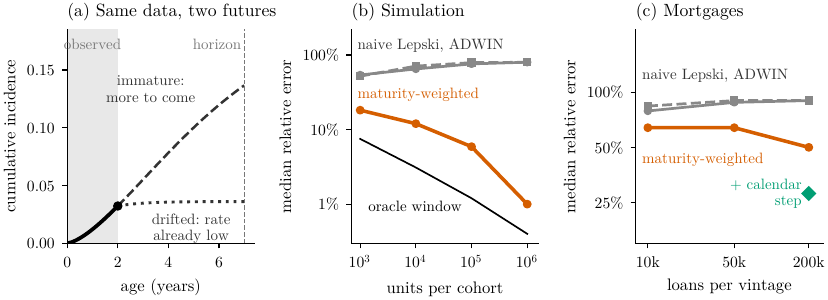}
\caption{\textbf{More data make adaptive windows worse when labels are provisional.} (a) A young cohort's events observed so far fit both an immature cohort and one whose rate has dropped. (b) Median relative error for the youngest cohort's rate over 972 simulated settings: fed provisional labels, Lepski's rule (solid grey) and ADWIN (dashed grey) pick ever shorter windows as data grow and their error rises, while the maturity-weighted rule (orange, \cref{sec:windows}) improves; black: oracle fixed window. (c) Mortgages: error for the 7-year default rate of the preceding year's vintage (origination year), median over the snapshot--subsample runs; the diamond adds calendar covariates (the calendar step; \cref{sec:calendar}).}
\label{fig:teaser}
\end{figure}

The effect can be large: default classifiers trained on a mortgage snapshot's provisional labels predict the default rate of the six youngest vintages (origination years) with a median error of 59--66\%, too low (\cref{tab:classifiers}). Have recent loans become safer, or have they not had time to default? A young cohort's data alone cannot say (\cref{fig:teaser}a), and both can be at work: during the 2008 housing crisis, default rates moved tenfold while the most exposed vintages were still young. The most recent cohorts always have the shortest follow-up, and real change would appear there first, so immaturity and drift affect the same cohorts whenever a model is retrained on recent data, a drift detector monitors incoming labels, or cohorts are compared.

Existing tools each handle part of this problem. Drift detection under label delay assumes labels that are late but final \citep{zliobaite2010,krempl2014,grzenda2020}. Delayed-feedback and positive--unlabelled methods correct provisional labels, but with one delay distribution for all cohorts \citep{chapelle2014,yang2021esdfm,elkan2008,bekker2020}. Reweighting units by their chance of being followed up (censoring weights) fails when no unit of a young cohort has yet been followed to the horizon \citep{robins2000,kubota2026}. Fixed outcome windows that include young units count their pending outcomes as failures \citep{chen2025vcbench}, and training only on mature cohorts \citep{xu2026pitfalls} ignores the cohorts where drift appears first. Insurance reserving completes each young cohort with the growth that older cohorts showed from the same age to the horizon (the chain ladder; \citealp{mack1993}), or pools cohorts weighted by their expected maturity over a window set by judgment or heuristics (Cape Cod; \citealp{stanard1985,gluck1997}). Epidemic nowcasting lets reporting delays change over time \citep{hohle2014,mcgough2020}, cure models separate how many units will have the event from when but are fragile under short follow-up \citep{meeker1987,maller1996,yu2004,rubio2023}, and demographers completing young fertility cohorts find that complex methods rarely beat simple ones \citep{bohkewald2018}. None of this work says what provisional labels do to adaptive windows and drift detectors, or how much one snapshot can reveal about a young cohort whose speed may have changed.

To test each idea on real data we build one estimator step by step and switch the steps on one at a time (\cref{sec:method}): the chain ladder; then each unit's exact age; then calendar covariates; then a speed for each cohort, smoothed towards those of its older neighbours. With all steps switched on it is the \emph{full method}, which we use as an ablation tool and do not propose as a method.

\paragraph{Contributions.} We analyse cohorts that differ in the level of their outcomes (how many units eventually have the event) and in their speed (how fast those events arrive).
\begin{itemize}[leftmargin=*,itemsep=1pt,topsep=2pt]
\item \textbf{Provisional labels break adaptive windows} (\cref{sec:windows}). Adaptive window rules of Lepski's type and the ADWIN detector, fed provisional labels, select the youngest cohort once the sample is large, even without drift (\cref{thm:A}). With known maturities, weighting cohorts by maturity gives a rule that adapts to unknown drift at the minimax rate up to a logarithmic factor in the typical case (\cref{thm:B}), with an error floor when maturities are misstated (\cref{cor:misspec}).
\item \textbf{One snapshot identifies a cohort's speed in principle} (\cref{sec:identification}), extending a reserving model in which cohorts run through one timing curve at different speeds \citep{lee2017operational} to cumulative incidence with competing exits (outcomes, such as prepayment, that rule the event out) and a fixed horizon (\cref{thm:id}). Equal-speed corrections err in a predictable direction (\cref{prop:cl}), and calendar shocks that differ across groups are separable from cohort change (\cref{thm:calendar}).
\item \textbf{In practice, young cohorts barely reveal their speed} (\cref{sec:limits}). Not knowing it inflates the asymptotic variance of the horizon rate (the event rate by a fixed age, such as 7 years) by a factor set by maturity and the onset of events, not by sample size, and a lower bound shows that the loss also holds at finite sample sizes (\cref{thm:precision}).
\item \textbf{On real data, maturity correction helps and cohort speeds do not} (\cref{sec:experiments}). We use backtests (estimates made at past snapshots and scored against outcomes observed later) on 4.2 million mortgages and 1.9 million patent applications, with default classifiers and drift detectors, and a pre-registered replication on small-business loans and federal civil cases (\cref{sec:replication}).
\end{itemize}

Correcting for maturity matters most: it cuts the error of the rates that classifiers predict for young cohorts, on mortgages by more than half after 2012 and by 18--42\% in the 2006--2012 crisis, and by 73--91\% in the replication. Cohort-specific speeds gave no significant gain in any data set, consistent with \cref{thm:precision}, although the replication could not rule out a 15\% gain on civil cases. A standard competing-risks hazard model, with one age baseline shared by all cohorts, was more accurate than the full method on patents and civil cases and about as accurate on mortgages; on small-business loans the result depends on how placeholder payoff dates are treated. \Cref{sec:discussion} turns these findings into a guide for practice; \cref{app:related} reviews related work in more detail.

%% file: sections/setup.tex
\section{Setup}\label{sec:setup}

\begin{figure}[!htb]
\centering
\includegraphics[width=1.0\linewidth]{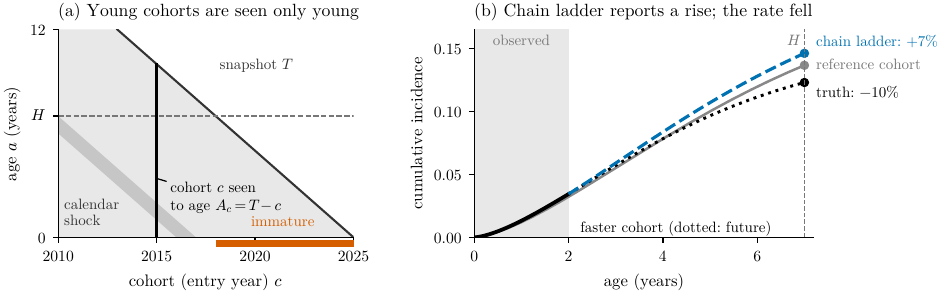}
\caption{\textbf{Young cohorts are seen only at young ages, and an equal-speed correction can reverse the sign of drift.} (a) Cohort $c$ is observed to age $A_c = T - c$ (shaded); cohorts that entered after $T - H$ are immature; a calendar shock runs along a diagonal. (b) A cohort 30\% faster than the reference, seen to age 2, has a 7-year rate 10\% below the reference's, but the chain ladder reports 7\% above it (\cref{prop:cl}; Weibull timing, shape 1.5, median 5 years).}
\label{fig:schematic}
\end{figure}

We define cohorts, the rate we want to estimate, maturity, and a model in which cohorts differ in level and speed (\cref{tab:notation} lists the symbols).

Units enter in cohorts $c$, such as founding or origination years. At the snapshot $T$, cohort $c$ has been followed for $A_c = T - c$ (\cref{fig:schematic}a). A unit may experience the \emph{event} of interest (exit, default, grant) or a \emph{competing exit} that rules it out (shutdown, prepayment, abandonment). On its observed window, each cohort reveals the cumulative incidence functions
\[
G_c(a) = \P(\text{event by age } a \mid c), \qquad G^{\mathrm{comp}}_c(a) = \P(\text{competing exit by age } a \mid c), \qquad a \in [0, A_c],
\]
for example through the Aalen--Johansen estimator \citep{aalen1978}, the competing-risks version of the Kaplan--Meier estimator, which handles censoring (follow-up that ends early for reasons unrelated to the outcome).

The target is a cohort's \emph{horizon rate} $p_H(c) = G_c(H)$, such as the 7-year default rate, and \emph{drift} means that it changes with $c$. A cohort is \emph{immature} if $A_c < H$, and its \emph{maturity} $m_c = G_c(A_c)/p_H(c)$ is the share of its horizon events already observed. A \emph{provisional label} marks a unit positive if its event has occurred by the snapshot. The share of positives, which we call \emph{events so far}, estimates $m_c\,p_H(c)$, not $p_H(c)$.

\paragraph{Level and speed.} We describe a cohort by a level and a timing curve that is shared up to a change of time scale,
\begin{equation}\label{eq:model}
K_c(a) := \psi\bigl(G_c(a)\bigr) = \theta(c)\,\Lambda_0\bigl(s(c)\,a\bigr),
\end{equation}
with $\Lambda_0$ increasing and $\Lambda_0(0) = 0$. A cohort with \emph{speed} $s(c) > 1$ accumulates events faster than the reference. The scale $\psi$ is either the identity, which we call the \emph{mixture-cure scale}, or $\psi(y) = -\log(1-y)$, the \emph{hazard scale}. On the mixture-cure scale, as in the mixture-cure models of biostatistics, $\theta(c)$ is the share of units that will eventually have the event and $\Lambda_0$ the distribution function of their event times; the precision results of \cref{sec:limits} are simplest there. On the hazard scale, $K_c$ is the cumulative \emph{sub-distribution hazard} \citep{fine1999}, the hazard of the event when units with a competing exit are kept among the units at risk; calendar shocks act multiplicatively on it (\cref{sec:calendar}), so the estimator of \cref{sec:experiments} works there. For rare events the two scales agree to first order (\cref{app:scales}). Calendar shocks act at calendar time $t = c + a$, so they hit different cohorts at different ages (\cref{fig:schematic}a).

%% file: sections/windows.tex
\section{Provisional labels break adaptive windows}\label{sec:windows}

Can an adaptive training window or a drift detector tell drift from immaturity? Lepski-type rules choose a window by comparing estimates over nested windows and stop once older data look different \citep{lepski1991,goldenshluger2011,mazzetto2023adaptive,han2024model}; change detectors such as ADWIN (adaptive windowing) cut where two sub-windows disagree \citep{bifet2007adwin}. Their guarantees assume final labels. With provisional labels, observed rates rise with age even without drift, so these comparisons detect age, and discarding older data removes the cohorts whose labels are most complete. We show that this failure grows with the data and that weighting cohorts by maturity removes it.

\paragraph{Model.} Index the cohorts from the youngest, $j = 1, \dots, N$, with $n$ units each, and assume a common timing curve, so that the maturities $0 < m_1 \le \dots \le m_N \le 1$ are a known function of age (in practice estimated from mature cohorts; \cref{app:proofB}). Unit $i$ of cohort $j$ has the independent provisional label $Y_{ij} \in \{0, 1\}$, with $\E Y_{ij} = p_j m_j$, where $p_j$ is the cohort's horizon rate. The target is $p_1$. The \emph{drift class} contains all sequences with $|p_j - p_{j+1}| \le \delta$ and $p_{\min} \le p_j \le p_{\max}$. Windows are $W_k = \{1, \dots, k\}$ and $S_j = \sum_i Y_{ij}$. \emph{Naive} rules compare the window means $\hat\mu_k = (nk)^{-1}\sum_{j \le k} S_j$.

\begin{theorem}[Provisional labels make adaptive windows collapse]\label{thm:A}
Let there be no drift, $p_j \equiv p$, and let $\Delta_m = m_2 - m_1 > 0$. Apply Lepski's rule, the Goldenshluger--Lepski rule, or ADWIN with cohorts as blocks, with comparison thresholds (the largest gap between two window means that the rule accepts) below $p\Delta_m/4$ (\cref{app:proofA}). With probability at least $1 - 2N\exp(-np\Delta_m^2/134)$:
\begin{enumerate}[label=(\roman*),itemsep=1pt,topsep=2pt]
\item the rule selects the youngest cohort alone, $\hat k = 1$;
\item its estimate is at most $p m_1 + p\Delta_m/8$, so its error is at least $p(1 - m_1) - p\Delta_m/8$;
\item every other window's estimate has a smaller error.
\end{enumerate}
\end{theorem}

Thresholds of order $\sqrt{p\log(N/\alpha)/(nk)}$, for error level $\alpha$, fall below $p\Delta_m/4$ once $n \gtrsim \log(N/\alpha)/(p\Delta_m^2)$. \emph{In words:} with enough data the rules correctly detect that the youngest cohort differs, but the difference is its age. The rules then keep only that cohort, so their error tends to $p(1 - m_1)$, above the bias $p(1 - \bar m_N)$ of pooling all cohorts ($\bar m_N$ is the mean maturity); more data make the failure more certain. The mechanism is elementary, since a consistent comparison eventually detects any difference in means, and ADWIN's own guarantee covers only a constant mean \citep{bifet2007adwin}; the theorem makes the failure precise. Work on label delay assumes labels that arrive late but final, or never \citep{zliobaite2010,grzenda2020,souza2015,alencar2026}, and delayed-feedback models that treat unconverted clicks as provisional negatives choose their waiting and training windows empirically \citep{ktena2019,yang2021esdfm}; we know of no analysis of adaptive windows or change detectors under provisional labels.

\paragraph{Maturity-weighted windows.} Pool the cohorts in a window and divide by their expected maturity. Over all cohorts this is the Cape Cod estimator of reserving \citep{stanard1985,gluck1997}, which fails when the level drifts (\cref{tab:baselines}); the window is what \cref{thm:B} chooses adaptively:
\[
\hat p_k = \frac{\sum_{j \le k} S_j}{n M_k}, \qquad M_k = \sum_{j \le k} m_j .
\]
It is unbiased for the maturity-weighted average of $p_1, \dots, p_k$ and has effective sample size $nM_k$. Let $L = \log(2N/\alpha)$ and let $\bar a_k = M_k^{-1}\sum_{j \le k} m_j (j-1)$ be the window's maturity-weighted age gap, so that its drift bias is at most $\delta \bar a_k$.

\begin{theorem}[Maturity-weighted Lepski rule]\label{thm:B}
\begin{enumerate}[label=(\roman*),itemsep=1pt,topsep=2pt]
\item Run Lepski's rule on $\hat p_k$ with data-driven Bernstein radii (\cref{app:proofB}). With probability at least $1 - \alpha$,
\[
|\hat p_{\hat k} - p_1| \le 20 \min_{k \le N} \Big\{ \delta \bar a_k + \sqrt{\tfrac{p_{\max} L}{n M_k}} + \tfrac{L}{n M_k} \Big\}.
\]
\item Every estimator $\tilde p$ has $\sup_{\text{drift class}} \P(|\tilde p - p_1| \ge \Delta^\ast/2) \ge 1/(4e)$, with $\Delta^\ast$ the largest $\Delta \le \min(p_{\max} - p_{\min}, p_{\max}/2)$ such that $2n\Delta^2 M_{\min(N, \lceil \Delta/\delta \rceil)} \le p_{\max}(1 - p_{\max})$.
\end{enumerate}
\end{theorem}

In the typical case, where $\Delta^\ast$ lies below both caps in its definition and the window that attains it holds at least $L$ expected events, the two bounds match up to an absolute constant and a factor $\sqrt{L/(1 - p_{\max})}$, about 2.5 for ten cohorts at $\alpha = 0.05$ (\cref{app:proofB}). \emph{In words:} once each cohort is divided by its expected maturity, the rule again trades drift bias against variance, and no estimator does better by more than a logarithmic factor; an immature cohort simply counts as a fraction $m_j$ of a mature one. The argument follows \citet{han2024model} and \citet{mazzetto2023adaptive}, who assume final labels; the same estimator corrects delayed conversions in a stationary bandit \citep{vernade2017}, and reserving sets the Cape Cod window by judgment or variance heuristics, without guarantees \citep{gluck1997,richman2026}.

\paragraph{In practice.} Lepski's rule accepts a longer window only if its estimate lies within a tolerance, or radius, of every shorter window's estimate. With the constants of part (i) the rule is conservative, so the experiments set each radius to $\kappa$ plug-in standard errors, with the Bonferroni-corrected Gaussian multiplier $\kappa = \sqrt{2\log(N/\alpha)}$ (in the replication of \cref{sec:replication}, the normal quantile $z_{\alpha/N}$); the rule is then consistent for any fixed $\kappa$ (\cref{app:proofB}).

\begin{corollary}[Misspecified maturities]\label{cor:misspec}
If the rule uses maturities $\tilde m_j$ with $|\tilde m_j/m_j - 1| \le \eta_j \le 1/2$, part (i) of \cref{thm:B} holds with $\tilde M_k = \sum_{j \le k}\tilde m_j$ in place of $M_k$, the constant 30 in place of 20, and $\delta\bar a_k + p_{\max}\max_{j \le k}\eta_j$ in place of $\delta\bar a_k$.
\end{corollary}

This is a standard misspecification bound. \emph{In words:} the fix is only as good as the maturities; an error $\eta_1$ in the youngest cohort's maturity leaves an error of order $p_{\max}\eta_1$ that no amount of data removes. Maturities come from older cohorts' timing, which is right only if young cohorts' outcomes arrive at the same speed. The next two sections ask what one snapshot reveals when they do not.

%% file: sections/identification.tex
\section{In principle, one snapshot identifies a cohort's speed}\label{sec:identification}

\Cref{sec:windows} needs each cohort's maturity, and the standard way to obtain it, the chain ladder, assumes that all cohorts share one timing curve: it scales each cohort's incidence so far by the development $\Lambda_0(H)/\Lambda_0(A_c)$ of the mature cohorts. This section asks what one snapshot reveals when cohorts also differ in speed, and what an equal-speed correction then gets wrong.

\begin{assumption}[Level and speed]\label{ass:R}
Model \eqref{eq:model} holds with $\theta(c), s(c) > 0$ and $s(c^\ast) = 1$ for a reference cohort $c^\ast$, a mature cohort that sets the time scale. On $(0, A_{c^\ast})$, $\lambda_0 = \Lambda_0' > 0$ and the elasticity $\epsilon(u) = u\lambda_0(u)/\Lambda_0(u)$ is strictly monotone.
\end{assumption}

The elasticity is the log--log slope of $\Lambda_0$. It is strictly decreasing for the Weibull, gamma, log-normal, log-logistic and generalised-gamma distribution functions (\cref{lem:elasticity}), and constant for a pure power law such as a Weibull cumulative hazard, which violates \cref{ass:R}. Our examples take $\Lambda_0$ to be a Weibull distribution function, which is identified; a Weibull cumulative hazard would not be. This exception is the classical fact that only the Weibull family is both a proportional-hazards and an accelerated-failure-time model \citep{kalbfleisch2002}, so that a change of level (which multiplies the hazard) and a change of speed (which rescales time) are indistinguishable.

\begin{theorem}[Identification with cohort speed]\label{thm:id}
Under \cref{ass:R}, suppose the reference cohort has been followed beyond each target cohort's horizon measured on the reference's time scale, $A_{c^\ast} \ge s(c)H$. Then one snapshot identifies the speed $s(c)$, the horizon rate $p_H(c)$, the level ratios $\theta(c)/\theta(c')$, and $\Lambda_0$ on $[0, A_{c^\ast}]$ up to a constant factor.
\end{theorem}

The reference's curve gives $\Lambda_0$ up to scale on its window. For any other cohort, the elasticity of the observed curve, $aK_c'(a)/K_c(a) = \epsilon(s(c)a)$, does not involve the level, so the log--log slope of the early curve determines the speed, and the horizon value follows from
\[
K_c(H) = K_c(A_c)\,\frac{\Lambda_0(s(c)H)}{\Lambda_0(s(c)A_c)} .
\]
\emph{In words:} on a log--log plot, a cohort's level shifts its curve up and its speed shifts it sideways. The two shifts look different unless the curve is a straight line, a power law. Every age in the window yields a value of $s(c)$, which makes the model testable (\cref{app:test}); we target a fixed horizon because an eventual rate depends on $\Lambda_0$ at its limit, which no finite window reveals.

On the mixture-cure scale, \eqref{eq:model} is the accelerated-failure-time cure model with a regression for the surviving fraction \citep{yamaguchi1992,li2002,lu2010}. It is also the operational-time model for reserving data (claims triangles) of \citet{lee2017operational}, whose identification condition, that $zf_2'(z)/f_2(z)$ not be constant near zero for their density $f_2$, is a density form of our elasticity condition. On the hazard scale, \eqref{eq:model} is the extended hazard model with cohort as a covariate \citep{etezadi1987,chen2001,rubio2019}. Identifying a curve's shift and stretch from its shape is the idea of shape-invariant regression \citep{lawton1972,kneip1988}. \Cref{thm:id} adds the step that a fixed-horizon target with competing exits needs and that these results leave implicit: one snapshot identifies $p_H(c)$ exactly when $s(c)H$ lies within the ages at which $\Lambda_0$ is identified (\cref{app:proofs-id}). Near zero, many timing curves behave like power laws, so the youngest cohorts are weakly identified (\cref{prop:powerlaw}); \cref{sec:limits} measures the cost.

\begin{proposition}[Equal-speed corrections err in a predictable direction]\label{prop:cl}
Let $R_c(\sigma) = \Lambda_0(\sigma H)/\Lambda_0(\sigma A_c)$ be the development from age $A_c$ to $H$ at speed $\sigma$. The chain ladder estimates $K_c(A_c)\,R_c(1)$, so the ratio of its estimate to the truth is $R_c(1)/R_c(s(c))$. If $\epsilon$ is strictly decreasing and $A_c < H$, the chain ladder overstates the horizon rate of faster cohorts ($s(c) > 1$) and understates that of slower ones.
\end{proposition}

\emph{In words:} the chain ladder reads a faster cohort's early events as a higher level. With Weibull timing (shape 1.5, median 5 years), a 7-year horizon and a cohort observed to age 2 that is 30\% faster than the reference, the chain ladder overstates the horizon rate by 19\%, so it reports an increase for any true decline smaller than 16\% (\cref{fig:schematic}b). The problem is known in reserving: \citet{berquist1977} adjust for changes in settlement speed, \citet{taylor1981} shows that the direction of a related distortion depends on the shape of the payment curve, and \citet{meyers2015} finds that chain-ladder-type models overstate paid losses and attributes this in part to faster settlement; in demography, timing shifts distort period fertility, the tempo effect \citep{bongaarts1998}. \Cref{prop:cl} gives the chain-ladder version: the error ratio is $R_c(1)/R_c(s(c))$, and its sign is set by the monotonicity of the elasticity (for increasing elasticity, as for a Gompertz cumulative hazard, it reverses).

\subsection{Calendar shocks that differ across groups are separable}\label{sec:calendar}

A shock at a calendar date hits each cohort at a different age, so it can look like a change in cohort level or speed; this is the age--period--cohort problem: because calendar time equals entry time plus age, linear trends in the three cannot be told apart \citep{kuang2008}. We let measured covariates $z_g(t)$ of group $g$ (a state, sector or technology centre), such as house-price growth, act on the sub-distribution hazard, $\tilde\lambda_g(a \mid c) = \nu_g\, \theta(c)\, s(c)\, \lambda_0(s(c)a)\, \exp\{\beta^\top z_g(c+a)\}$, with $\nu_{g_0} = 1$ for a reference group $g_0$; measured covariates are the mechanism-based route around the age--period--cohort identity \citep{heckman1985,winship2008}, as in credit-risk vintage models \citep{breeden2007}. This assumes that (C1) shocks act multiplicatively on that hazard \citep{beyersmann2008}; (C2) groups share $\theta(c)$, $s(c)$ and $\Lambda_0$ up to the multiplier $\nu_g$; (C3) the covariates are not driven by the outcomes; and (C4) $\beta$, learned from differences between groups, also applies to changes common to all groups.

\begin{theorem}[Calendar shocks that differ across groups]\label{thm:calendar}
On the hazard scale, let the contrasts $\Delta z_{gg'}(t) = z_g(t) - z_{g'}(t)$ vary over time, in that the differences $\Delta z_{gg'}(t) - \Delta z_{gg'}(t')$ span $\R^{\dim z}$, and let \cref{thm:id} apply to the shock-adjusted cumulative hazard. Then $\beta$, $\nu_g$, $\theta(c)$, $s(c)$, $\Lambda_0$ and each cohort's horizon rate along any specified calendar path (an assumed future path of the covariates) are identified.
\end{theorem}

\emph{In words:} comparing groups within the same cohort and age cancels everything except the shock. This is a standard difference-in-differences argument; we state it because it justifies the calendar step of the full method (\cref{sec:method}), a Poisson regression with cohort$\times$age and group fixed effects that separates shocks from cohort change. Unmeasured common shocks are absorbed into cohort level or speed, and without variation between groups even a common linear trend is not identified (\cref{app:proofs-id}). Theorems \ref{thm:id} and \ref{thm:calendar} say what is identified with unlimited data; the next section asks how much a young cohort's data actually reveal.

%% file: sections/limits.tex
\section{In practice, young cohorts barely reveal their speed}\label{sec:limits}

\Cref{thm:id} infers a cohort's speed from the log--log slope of its early curve. This section asks how precisely a young cohort's few early events determine that slope, and finds that, compared with a cohort whose speed is known, they reveal little at any sample size.

We work on the mixture-cure scale, where $\Lambda_0 = F_0$ is a distribution function with density $f_0$ and $G_c(a) = \theta_c F_0(s_c a)$, writing $\theta_c = \theta(c)$ and so on. Let cohort $c$ have $n_c$ units observed to rescaled age $u_c = s_c A_c$, with horizon $u_{H,c} = s_c H$. Its horizon rate is $p_c = \theta_c F_0(u_{H,c})$, its maturity $m_c = F_0(u_c)/F_0(u_{H,c})$, and its expected number of events so far $D_c = n_c p_c m_c$. Let $\eta(u) = 1 + d\log f_0/d\log u$, the sensitivity of the event density to the speed, let $V_c$ be the variance of $\eta(u)$ across the rescaled ages $u$ at which the cohort's events so far occur, and let $I_c = 1 + (\epsilon(u_c) - \epsilon(u_{H,c}))^2/V_c$. We call $I_c$ the \emph{speed inflation factor}: the factor by which not knowing the speed inflates the asymptotic variance. It weighs how strongly the extrapolation to the horizon depends on the speed against how much the observed event ages reveal about the speed.

\begin{theorem}[The price of an unknown speed]\label{thm:precision}
Let event counts in age bins be Poisson and $F_0$ known.
\begin{enumerate}[label=(\roman*),itemsep=1pt,topsep=2pt]
\item The asymptotic variance of $\log\hat p_c$ is $1/D_c$ if $s_c$ is known and $I_c/D_c$ if it is not. No regular estimator (one whose limit is stable under small changes of the parameters) does better, and the bound remains valid when $F_0$ is unknown.
\item If $f_0(u) \propto u^{\gamma-1}\{1 + bu^\rho + o(u^\rho)\}$ near zero with $b \neq 0$, then $I_c \asymp m_c^{-2\rho/\gamma}$ (same order) as $m_c \to 0$. For Weibull timing and a long horizon, $I_c \sim 12/m_c^2$ whatever the shape.
\item For log-normal timing, $I_c \sim 4\log^2(1/m_c)$ as $m_c \to 0$.
\item Let $g_s(a) = sf_0(sa)/F_0(sA_c)$ be the density of event ages in the window at speed $s$. For any two speeds $s, s'$ with valid levels that give the same $D_c$, every estimator has, under one of the two,
$\P\{|\log\hat p_c - \log p_c| \ge \tfrac12|\log(m_c(s)/m_c(s'))|\} \ge \tfrac14\exp\{-D_c\,\mathrm{KL}(g_s\,\|\,g_{s'})\}$, with $\mathrm{KL}$ the Kullback--Leibler divergence.
\end{enumerate}
\end{theorem}

\emph{In words:} when the speed is unknown, a young cohort's horizon rate is estimated as if the cohort had $I_c$ times fewer events, and $I_c$ depends on maturity and on how events begin, not on the sample size. With Weibull timing (shape 1.5, median 5 years), a 7-year horizon and two years of follow-up ($m \approx 0.24$), the standard error is 7.6 times larger, so matching the known-speed precision takes 58 times more data. Part (i) is asymptotic, and for such young Weibull cohorts the asymptotics need thousands of events (\cref{app:thm6check}). Part (iv) holds at every sample size: with 100 events so far, under one of two speeds, every estimator is off by a factor of at least 1.45 with probability at least $1/(4e)$, whereas the known-speed standard error is 10\%. When events begin very slowly, as under a log-normal curve, the early shape is more informative, and part (i) is accurate from 30 events.

\emph{What was known and what is new.} That incidence and timing are hard to separate under short follow-up is a known difficulty of cure models \citep{farewell1986,yu2004}. \citet{meeker1987} tabulated the asymptotic variance of a defective fraction against the share of defectives already failed, found a likelihood ridge when few had failed, and warned that inference is imprecise when fewer than about 80\% of the defectives have failed; without parametric assumptions the eventual fraction is identified only under sufficient follow-up \citep{maller1992,maller1996}, and identifiability needs conditions on the timing distribution \citep{li2001,hanin2014}. \citet{rubio2023} relate such problems to near-redundancy, which arises when the baseline is close to a Weibull and which they detect numerically. Epidemic nowcasting meets the same problem and lets the reporting-delay distribution change over time \citep{hohle2014,mcgough2020}. Part (i) is a direct Fisher-information calculation. \Cref{thm:precision} adds three things: the closed-form factor $I_c$ for a fixed-horizon target, which separates how much the extrapolation depends on the speed from how much the observed event ages reveal about it; the rate at which $I_c$ grows as maturity falls ($m^{-2}$ under a Weibull onset, $\log^2(1/m)$ under a log-normal one); and a lower bound at finite sample sizes that makes rigorous the two-point indistinguishability \citet{rubio2023} use as a diagnostic.

In a Gaussian approximation, estimating a cohort's own speed beats borrowing a neighbour's exactly when $\omega^2 + \tau^2 > 1/(D_cV_c)$, where $\omega^2$ is the variance of speed changes between adjacent cohorts and $\tau^2$ the uncertainty of the borrowed speed (\cref{prop:shrink}); shrinking towards the neighbour does at least as well as either. We call a cohort's speed \emph{noise-limited} when this condition fails: its own data then pin the speed down less well than a neighbour's speed predicts it.

\paragraph{Calibrated intervals.} Any continuation of a cohort's curves beyond $A_c$ fits the data equally well: the identified set of $p_H(c)$ is $[G_c(A_c), 1 - G^{\mathrm{comp}}_c(A_c)]$, and every point estimate extrapolates. We therefore calibrate split-conformal intervals \citep{vovk2005,lei2018} on the method's own past errors on now-mature cohorts at the target's age, widened by a factor $\Gamma \ge 1$ that allows future errors to exceed past ones \citep{rambachan2023}; they cover $p_H(c)$ if the target's error is exchangeable with those errors (\cref{prop:intervals}), which drift can break, so we report empirical coverage. \Cref{sec:experiments} tests these results on data.

%% file: sections/experiments.tex
\section{Experiments}\label{sec:experiments}

We ask four questions. Does correcting for maturity fix windows, classifiers and drift detectors on real data (\cref{sec:e1})? Which parts of a correction beat the chain ladder (\cref{sec:backtests})? Do cohort speeds help where \cref{thm:precision} says they are hard to learn (\cref{sec:speed})? And do the answers hold on new data, in a pre-registered study (\cref{sec:replication})?

\paragraph{Methods compared.}\label{sec:method} To measure what each idea adds, we build a correction in steps that can be switched on one at a time (Algorithm 1, \cref{app:estimator}), the rows of \cref{tab:main}. It starts from the chain ladder (CL), which completes each young cohort with the growth that mature cohorts showed from its age to the horizon. It then uses each unit's \emph{exact age} since entry instead of its cohort's, on the hazard scale, and adds a level per group; then adds a \emph{calendar step} that estimates covariate effects from differences between groups (\cref{thm:calendar}), giving the \emph{calendar-adjusted chain ladder}; and adds a \emph{speed step} that gives each cohort its own speed, smoothed towards those of its older neighbours (\cref{thm:id,prop:shrink}), giving the \emph{full method}. A plug-in formula turns the fitted curves into each cohort's horizon rate. We also compare baselines that extrapolate from old cohorts (dropping recent cohorts, the last mature cohort, and the reserving methods Bornhuetter--Ferguson, Cape Cod and generalised Cape Cod; \cref{app:baselines}) and a standard competing-risks hazard model. A cause's \emph{hazard} at a given age is the chance that a unit still at risk has that outcome in the next month; the model fits one hazard for the event and one for the competing exit, each with an age baseline shared by all cohorts, cohort effects and covariates, and accumulates them into each cohort's incidence at the horizon (\cref{app:baselines}). Methods with covariates are scored on the \emph{forecast} path, which holds covariates at their last-12-month mean, and on the \emph{realised} path, which uses their actual values and exists only in backtests.

\paragraph{Data.} \emph{Mortgages:} Freddie Mac, 200{,}000 loans per origination year (\emph{vintage}) 1999--2019, with the 7-year default rate as target and prepayment as competing exit. At 13 year-end snapshots (2006--2018) we score the six youngest vintages, against the 7-year rate at \emph{crisis} snapshots (2006--2012) and the incidence up to February 2020 at \emph{pre-COVID} ones (2013--2018). Groups are states, and the covariates are house-price growth and unemployment. \emph{Patents:} USPTO PatEx \citep{graham2018patex}, 100{,}000 utility applications per filing year 2001--2019, with the 72-month grant share as target and abandonment as competing exit. At 11 snapshots (2007--2017) we score the filing years one to four years before the snapshot (43 cohort--snapshot \emph{pairs}; filing years before 2004 are used only for training). Groups are technology centres, and the covariates are the pending inventory (applications filed and not yet decided) and an indicator for software applications after the 2014 \emph{Alice} decision, a US Supreme Court ruling on the patent eligibility of software. Each snapshot uses only outcomes dated up to the snapshot, taken from today's records (\cref{sec:discussion}). A snapshot's error is the mean absolute relative error over its scored cohorts, summarised by the median over snapshots. The \emph{error ratio} of two methods is the geometric mean of their absolute-error ratios, with a bootstrap interval that resamples cohorts and a $p$-value from an exact cluster sign-flip test (\cref{app:estimator,app:inference}). With 13--25 cohort clusters, and the cohorts scored at one snapshot sharing one fit, these $p$-values ignore some dependence and are approximate; intervals that also resample snapshots are wider (\cref{app:inference}).

\paragraph{What was pre-registered.} The mortgage specification was committed before the data were loaded, after development on a 50{,}000-loan sample of the same vintages, so the run is new with respect to loans but not to calendar history. Its one primary comparison, the full method against the chain ladder on the realised path, was not confirmed as run (error ratio 0.81 [0.56, 1.17], $p = 0.30$). An audit of the code behind the main tables and figures, by three AI agents given only the code, the specifications and the paper (\cref{app:experiments}), then found that the recorded loan age restarts when a loan is modified, so every mortgage number in this paper counts ages from the first payment. With the ages corrected (post hoc), the registered statistics disagree: the error ratio is 0.56 and its bootstrap interval, [0.34, 0.93], excludes 1, but the full method beats the chain ladder at only 9 of 13 snapshots (sign test $p = 0.27$; \cref{tab:mortgage-ages,tab:freddie}). We therefore treat the mortgage results as exploratory.

The patent protocol was written before any backtest but committed together with the results, so its timing cannot be verified, and we treat the patent results as exploratory too. Its two primary comparisons, on the realised path, were the full method against the chain ladder (better at 10 of 11 snapshots, Holm-adjusted sign-test $p = 0.023$), which rests on an inventory covariate we later found to be driven by the outcomes, and against the calendar-adjusted chain ladder (6 of 11, not significant; \cref{tab:patex}).

Everything else on mortgages and patents is post hoc and is listed with its effect in \cref{tab:deviations}; in the tables, $\dagger$ marks methods added after the protocols. Except for the sign tests in \cref{tab:freddie,tab:patex}, the mortgage and patent $p$-values in the tables come from the cluster sign-flip test, which we adopted after the runs because it uses every cohort--snapshot pair while treating each cohort, scored at several snapshots, as one unit; it gives 0.045 for the corrected mortgage primary and 0.002 for the patent one (Holm-adjusted 0.004). Apart from the patent primary comparisons, no mortgage or patent $p$-value is adjusted for multiplicity. The real-data runs keep the frozen version of the plug-in formula; a later correction changes the estimates little but moves the corrected mortgage primary to $p = 0.061$ (\cref{app:corrected}). The replication of \cref{sec:replication}, in contrast, was pre-registered in full and publicly time-stamped before its data were downloaded (\cref{app:replication}); it is our confirmatory evidence.

\subsection{Maturity correction cuts the error of windows and classifiers, least in the crisis}\label{sec:e1}

Do adaptive windows fail on real data as \cref{thm:A} predicts? They do (\cref{fig:teaser}b,c): as the sample grows, the error of naive Lepski windows rises, to 80\% in simulation (with similar results for ADWIN) and 90\% on mortgages, while the error of the maturity-weighted rule ends far lower, at 1\% and 50\% (28\% with the calendar step). On real data the gain comes from correcting maturity, not from choosing the window: the rule does not beat the best window chosen in hindsight, and with the conservative constants of \cref{thm:B} it is the worst rule at small samples (\cref{tab:e2}). When maturities are misstated, its error grows with the sample, which \cref{cor:misspec} bounds but does not remove (\cref{tab:e1}).

Correcting labels has the same effect on learned models (\cref{tab:classifiers}). Training on maturity-corrected labels, which treat each unit without an event so far as unlabelled and take its cohort's maturity as the probability that its event would already be seen \citep{bekker2019}, cuts the error of the predicted default rate of recent vintages at 12 of 13 snapshots: by more than half after 2012, but only by 18--42\% in the 2006--2012 crisis, where predictions stay too low. Maturities from the full method, the chain ladder and a delayed-feedback model all help. A loan-level competing-risks hazard classifier without calendar covariates does not help: its single age baseline is set by the crisis vintages. The correction raises the level of each vintage's rate, although in the crisis it stays 1.3--1.6 times too low (\cref{app:e5}); within a vintage the ranking barely changes, and the ranking distortion across loan features that appears in the crisis period remains (\cref{prop:ranking}; \cref{app:e5}).

For drift detectors the effect depends on the data. On mortgages, corrected labels cut ADWIN's false ``rate fell'' alarms (alarms with no matching alarm on the true outcomes) on the monthly default rate from 15 to 2--4 but miss more true changes, so with the full method false and missed alarms together do not fall (17 against 15; \cref{tab:alarms}); in the replication they fall for ADWIN (\cref{sec:replication}). On mortgages the false alarms of the Page--Hinkley detector \citep{page1954} do not change, but with chain-ladder labels (not with full-method labels) it detects the 2005--2008 rise a year sooner (\cref{app:e5}).

\begin{table}[t]
\centering\small
\caption{\textbf{Maturity-corrected labels cut the error of predicted default rates, most after 2012.} Logistic default classifiers trained at each snapshot and scored on 50{,}000 held-out loans of each of the six youngest vintages: median over snapshots of the absolute relative error of the predicted 84-month default rate, and the area under the ROC curve (AUC) pooled over the six vintages. ``Events so far'' trains on provisional labels; ES-DFM is the elapsed-time-sampling delayed-feedback model; the hazard row is a multinomial logistic model on loan-months with monthly age and vintage effects. All rows are post hoc ($^\dagger$: method added after the protocol); boosted trees, ratios and tests: \cref{app:e5}.}
\label{tab:classifiers}
\input{tables/classifiers.tex}
\end{table}

\begin{figure}[t]
\centering
\includegraphics[width=1.0\linewidth]{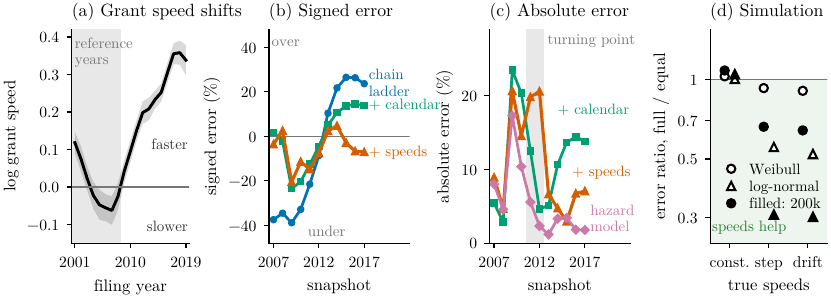}
\caption{\textbf{Patent grant speed shifts, the chain ladder errs in the predicted direction, and cohort speeds fail at the turning point.} (a) Calendar-adjusted log grant speed by filing year, relative to 2001--2008 ($\pm 2$ s.e.). (b) Mean signed relative error of the scored filing years' 72-month grant share (above 0: over-prediction), forecast path: chain ladder, calendar-adjusted chain ladder (+ calendar) and full method (+ speeds). The calendar step uses the pending inventory, which includes the scored cohorts' own applications and is therefore driven by the outcomes. (c) Mean absolute relative error of the calendar-adjusted chain ladder, the full method and the competing-risks hazard model (forecast path); shaded: the 2011--2012 turning point. (d) Simulation: error ratio of the full method to equal speeds (the calendar-adjusted chain ladder) when true speeds are constant, change in a step or drift; open markers 20{,}000 and filled 200{,}000 units per cohort year; below 1, speeds help (\cref{app:sims}).}
\label{fig:speed}
\end{figure}

\subsection{Calendar covariates (mortgages) and exact ages (patents) beat the chain ladder; standard hazard models do as well or better}\label{sec:backtests}

Which parts of a correction beat the chain ladder? \Cref{tab:main} adds one step at a time. On mortgages the calendar step gives the gain (the calendar-adjusted chain ladder against the chain ladder: 0.54, $p = 0.013$, forecast path), and neither exact ages nor cohort speeds help significantly (by a secondary measure, speeds do on the forecast path; \cref{app:inference}). On patents exact ages give the gain, and speeds again give no significant gain. The patent calendar step's apparent gain comes from the scored cohorts' own share of the pending inventory, which is driven by the outcomes and violates (C3); when the inventory excludes the scored filing years, the step is not significant (\cref{app:patents}). The full method thus beats the chain ladder through exact ages and calendar covariates, never through cohort speeds.

How do established models compare? Baselines that extrapolate from old cohorts fail on mortgages after default rates moved tenfold, and on patents the full method is significantly better only than the generalised Cape Cod among them (\cref{tab:baselines}). A standard competing-risks hazard model is about as accurate as the full method on the mortgage forecast path (error ratio of the hazard model to the full method 1.10 [0.77, 1.69] with the corrected censoring weights, 0.91 with the pre-registered ones; worse in 2009 and 2011--2012) and more accurate on the realised path, but on mortgages it also uses loan features and more covariates than the full method (\cref{app:baselines}), so that comparison mixes model and information. On patents, given the same information as the full method, it is about twice as accurate (median error 3.4\% against 7.1\%, $p = 0.026$), with or without filing-year effects and with the inventory that excludes the scored filing years (\cref{fig:speed}c; \cref{app:baselines,app:patents}). On mortgages its cohort effects matter: without them it beats the chain ladder but not the full method.

\begin{table}[t]
\centering\footnotesize
\caption{\textbf{Calendar covariates (mortgages) and exact ages (patents), not cohort speeds, beat the chain ladder; competing-risks hazard models do as well or better.} Median over snapshots of the mean absolute relative error; error ratio to the chain ladder (CL) [95\% cohort-cluster bootstrap interval] and exact cluster sign-flip $p$. Rows from CL to the full method each add a step; rows without covariates are the same on both paths; old-cohort baselines are in \cref{tab:baselines}. Mortgage rows use censoring weights corrected, post hoc, for refinances under the federal relief programme HARP (Home Affordable Refinance Program; \cref{app:freddie}), except the pre-registered-weights row, which keeps the registered weights and computes the incidence observed by the snapshot as the raw share of units with the event (\cref{app:estimator}). Mortgage loan ages count from the first payment (post hoc correction; as-run values in \cref{tab:mortgage-ages}). $^{\ddagger}$Pre-registered weights, with loan ages corrected post hoc (\cref{sec:experiments}). Patent calendar rows in the upper blocks use the pending inventory, which violates (C3); the bottom block uses the inventory without the scored filing years. $^\dagger$Method added after the protocols. The hazard models have cohort (vintage or filing-year) effects.}
\label{tab:main}
\setlength{\tabcolsep}{4pt}
\input{tables/main_results.tex}
\end{table}

\subsection{A young cohort's own speed is not learned; borrowed speeds help only when a change persists}\label{sec:speed}

Do cohort speeds help where timing really changes? Patent grant timing does shift (\cref{fig:speed}a): applications are examined in filing order \citep{mpep708}, so those filed in 2004--2008 waited longer. The chain ladder errs as \cref{prop:cl} predicts, under-predicting while the scored filing years are slower than the reference and over-predicting once they are faster (\cref{fig:speed}b). Speeds smoothed from older neighbouring cohorts roughly halve the remaining error once the speed-up persists (2014--2017) but enlarge it at the 2011--2012 turning point (\cref{fig:speed}c; \cref{tab:patex-errors}). So the step meant to fix the chain ladder's wrong-direction errors makes more of them: of the 41 patent pairs whose rate changed, on the realised path the full method gets the direction wrong in 8, the chain ladder in 5, the calendar-adjusted chain ladder in 3 and the hazard model, which estimates no cohort speeds, in 2 (\cref{tab:wrongdir}). The youngest cohorts' own speeds are noise-limited, with asymptotic speed inflation factors from 35 to more than $10^5$, and the condition of \cref{prop:shrink} does not predict where speeds help (\cref{app:speedcrit}).

Simulation shows why (\cref{fig:speed}d; \cref{tab:sim-e2e}). After a change in speed, speeds help most when the onset of events is informative, as with log-normal timing, and much less when it is a power of age, as with Weibull timing, where the youngest cohort's own speed is not learned at all and any gain comes from smoothing towards older cohorts' speeds. A direct check confirms the variance of \cref{thm:precision} (\cref{app:thm6check}).

\subsection{A pre-registered replication confirms that maturity correction matters and finds no significant gain from cohort speeds}\label{sec:replication}

The conclusions so far are exploratory, so we tested them on two public data sets that none of us had used, in a study pre-registered in full and publicly time-stamped before the data were downloaded (\cref{app:replication}). \emph{Loans:} 1.37 million SBA 7(a) small-business loans approved 1991--2019; event charge-off, competing exit payoff, 72-month horizon, states as groups, 109 scored pairs. \emph{Civil cases:} 3.9 million federal civil cases filed 1989--2016; event settlement, competing exit any other ending, 36-month horizon, 90 district courts as groups, 69 pairs. Methods and metrics are those of \cref{sec:method}, with the corrected plug-in. Three hypotheses (H1--H3 in \cref{tab:replication}) were tested on the forecast path by one-sided tests against margins set in advance, with Holm's correction for multiple testing within each data set; a hypothesis counts as confirmed overall only if it is confirmed in both data sets.

Maturity correction (H1) is confirmed on both data sets: corrected labels cut the error of the cohort rates that classifiers predict by 73\% and 91\%. This compares them with provisional labels only, a low bar; on mortgages a delayed-feedback model did better still (\cref{tab:classifiers}). The hazard model's non-inferiority (H3) is confirmed as registered on both data sets, but it holds robustly only on civil cases, where the model is also more accurate than the full method. On loans the pre-registered test passes only because payoff dates before mid-2005 are placeholders; fitted on the months with real dates, the hazard model fails the non-inferiority test (error ratio 1.15 [1.01, 1.32]; \cref{app:replication}). Cohort speeds (H2) helped significantly in neither data set, but the test that they cut error by less than 15\% passed only on loans; on civil cases the interval reaches 0.78, below the 0.85 margin, so by our own rule H2 is not confirmed overall. Nor did the full method beat the chain ladder in the replication (1.04 on loans, and 0.88 and not significant on civil cases; \cref{tab:rep-main}). All values include six fixes from an audit of the replication code by AI agents that had not seen the results; none changes a verdict, and neither does dropping the mass-tort cases (multidistrict litigation, a fifth of the 2013--2014 civil filings; \cref{app:replication}).

The secondary analyses replicate the window theorem: as the sample grows, naive Lepski windows shrink to the youngest cohort, and the median error of the maturity-weighted rule is 7--27\% of theirs (\cref{tab:rep-windows}). Two findings differ from the earlier data. First, the calendar step hurt on the forecast path in both data sets, so the calendar gain on mortgages does not generalise; exact ages helped on loans, but only borderline on civil cases and not at all there without the mass-tort cases (\cref{tab:rep-steps}). Second, corrected labels cut ADWIN's false and missed alarms together, by 77--100\% under the any-match rule used for mortgages and by 39--73\% one-to-one; for Page--Hinkley with full-method labels they rose on loans (\cref{tab:rep-alarms}).

\paragraph{Calibrated intervals.} Counting only pairs with at least nine past errors to calibrate on (pre-specified for mortgages, post hoc for patents; \cref{tab:deviations}), the full method's calibrated 90\% intervals cover all 39 mortgage pairs with the corrected censoring weights and 19 of 22 patent pairs on the realised path (\cref{app:corrected}), and 89\% of the loan pairs and only 77\% of the civil pairs in the replication.

\begin{table}[t]
\centering\footnotesize
\caption{\textbf{Pre-registered replication: maturity correction helps, a standard hazard model passes the registered non-inferiority test, and cohort speeds help significantly in neither data set.} Error ratio on the forecast path [95\% cohort-cluster bootstrap interval], Holm-adjusted one-sided cluster sign-flip $p$, and verdict. H1: classifier on maturity-corrected labels (chain-ladder maturity) over classifier on events so far; confirmed if the test rejects a ratio of 1 or above. H2: full method over calendar-adjusted chain ladder; confirmed if the test rejects a ratio of 0.85 or below, i.e.\ that speeds cut error by 15\% or more. H3: hazard model over full method; confirmed if the test rejects a ratio of 1.15 or above. Loans: 109 pairs, 24 cohort clusters (H1: 84 and 19); civil cases: 69 pairs, 25 clusters. Overall, under the pre-registered rule that a hypothesis must be confirmed in both data sets, H1 and H3 are confirmed and H2 is not. Values after correcting six implementation errors found in an audit by AI agents; the run as first done gives the same verdicts (\cref{tab:rep-asrun}). $^{\ast}$Rests on placeholder payoff dates before mid-2005; fitted only on later months (post hoc), the ratio is 1.15 [1.01, 1.32] on the 79 pairs from 2005, against 0.91 as specified on the same pairs (\cref{app:replication}).}
\label{tab:replication}
\setlength{\tabcolsep}{4pt}
\begin{tabular}{@{}lcccccc@{}}
\toprule
 & \multicolumn{3}{c}{SBA loans} & \multicolumn{3}{c}{Federal civil cases} \\
\cmidrule(lr){2-4}\cmidrule(l){5-7}
Hypothesis & ratio [95\% CI] & $p$ & verdict & ratio [95\% CI] & $p$ & verdict \\
\midrule
H1: maturity correction helps & 0.27 [0.20, 0.38] & $<$0.001 & confirmed & 0.09 [0.07, 0.12] & $<$0.001 & confirmed \\
H2: speeds cut error by $<$15\% & 1.06 [0.85, 1.31] & 0.034 & confirmed & 1.08 [0.78, 1.52] & 0.088 & inconclusive \\
H3: hazard model at least as accurate & 0.86 [0.71, 1.05] & 0.016 & confirmed$^{\ast}$ & 0.62 [0.39, 0.95] & 0.011 & confirmed \\
\bottomrule
\end{tabular}
\end{table}

%% file: tables/classifiers.tex
\begin{tabular}{@{}lcccc@{}}
\toprule
 & \multicolumn{2}{c}{error in default rate} & \multicolumn{2}{c}{AUC} \\
\cmidrule(lr){2-3}\cmidrule(l){4-5}
Training labels & 2006--12 & 2013--18 & 2006--12 & 2013--18 \\
\midrule
Realised outcomes (oracle) & 3\% & 3\% & 0.854 & 0.790 \\
\addlinespace
Events so far & 66\% & 59\% & 0.810 & 0.739 \\
Mature cohorts only & 78\% & 116\% & 0.735 & 0.777 \\
Delayed-feedback model \citep{chapelle2014} & 38\% & 23\% & 0.822 & 0.795 \\
Elapsed-time weights (ES-DFM) \citep{yang2021esdfm} & 56\% & 53\% & 0.816 & 0.798 \\
Positive--unlabelled, chain-ladder maturity & 49\% & 26\% & 0.823 & 0.786 \\
Positive--unlabelled, full-method maturity & 54\% & 25\% & 0.817 & 0.755 \\
Loan-month competing-risks hazard$^\dagger$ & 71\% & 144\% & 0.809 & 0.782 \\
\bottomrule
\end{tabular}

%% file: tables/main_results.tex
\begin{tabular}{@{}lccccccc@{}}
\toprule
 & \multicolumn{4}{c}{Mortgages (78 pairs, 18 vintages)} & \multicolumn{3}{c}{Patents (43 pairs, 13 filing years)} \\
\cmidrule(lr){2-5}\cmidrule(l){6-8}
 & \multicolumn{2}{c}{median error} & ratio to CL & $p$ & median & ratio to CL & $p$ \\
 & crisis & pre-COVID & [95\% CI] & & error & [95\% CI] & \\
\midrule
\multicolumn{8}{@{}l}{\emph{Forecast path: covariates held at their last-12-month mean}} \\
Events so far & 68\% & 55\% & 1.71 [1.26, 2.46] & 0.004 & 51\% & 2.26 [1.73, 3.07] & $<$0.001 \\
\addlinespace
Chain ladder (CL) & 59\% & 35\% & 1 &  & 26\% & 1 &  \\
CL + exact ages + groups$^\dagger$ & 59\% & 37\% & 1.02 [0.95, 1.10] & 0.576 & 13\% & 0.63 [0.54, 0.75] & 0.001 \\
Calendar-adjusted CL & 60\% & 24\% & 0.54 [0.39, 0.80] & 0.013 & 13\% & 0.38 [0.27, 0.50] & $<$0.001 \\
Full method & 54\% & 26\% & 0.44 [0.28, 0.72] & 0.012 & 7\% & 0.35 [0.20, 0.59] & 0.004 \\
Competing-risks hazard model$^\dagger$ & 51\% & 17\% & 0.49 [0.34, 0.73] & 0.004 & 3.4\% & 0.18 [0.13, 0.25] & $<$0.001 \\
\addlinespace
\multicolumn{8}{@{}l}{\emph{Realised path: actual future covariates (not available at the snapshot)}} \\
Full method & 39\% & 24\% & 0.51 [0.32, 0.83] & 0.020 & 7\% & 0.34 [0.21, 0.56] & 0.002 \\
\quad pre-registered weights$^{\ddagger}$ & 43\% & 23\% & 0.56 [0.34, 0.93] & 0.045 &  &  &  \\
Competing-risks hazard model$^\dagger$ & 15\% & 14\% & 0.33 [0.19, 0.56] & 0.001 & 3.9\% & 0.17 [0.11, 0.24] & $<$0.001 \\
\addlinespace
\multicolumn{8}{@{}l}{\emph{Patents, inventory without the scored filing years, forecast path}$^\dagger$} \\
Calendar-adjusted CL & & & & & 13.4\% & 0.59 [0.52, 0.69] & $<$0.001 \\
Full method & & & & & 17.1\% & 0.50 [0.31, 0.77] & 0.008 \\
Competing-risks hazard model & & & & & 7.5\% & 0.23 [0.11, 0.47] & $<$0.001 \\
\bottomrule
\end{tabular}

%% file: sections/discussion.tex
\section{Discussion}\label{sec:discussion}

Immature outcomes look like drift, and methods that treat provisional labels as final fail in ways that more data make worse. The theory accounts for each failure the data show, and the main conclusion, that maturity correction matters, held in a pre-registered replication. These results complement work on drift under label delay, which assumes labels that are late but final \citep{zliobaite2010,grzenda2020}.

\paragraph{Why maturity matters so much.} Provisional labels fall short by a fraction $1 - m_c$, which is large exactly where drift is sought (classifiers trained on them under-predict the default rate of recent vintages, with a median error of 59--66\%; \cref{tab:classifiers}), and adaptive rules amplify it, because every comparison of a young with an old cohort reports a difference (\cref{thm:A}). Dividing by expected maturity removes this first-order bias, which is why maturities from the chain ladder, the full method and a delayed-feedback model all did far better than provisional labels, although they differed among themselves (\cref{app:e5,app:replication}). It also explains why standard competing-risks hazard models did well: on mortgages their vintage effects, together with calendar covariates, make them a maturity correction, and on patents a model with a monthly age baseline for each cause beat the full method even without cohort effects.

\paragraph{Why cohort speeds add little.} Speeds target the smaller errors that remain, but a young cohort's own speed is nearly unknowable: for the youngest scored cohort at every snapshot, the asymptotic speed inflation factor of \cref{thm:precision} is in the tens to more than $10^5$ (\cref{app:speedcrit}). The full method therefore smooths each young cohort's speed towards those of its older neighbours, which helps when a timing change persists and hurts at a turning point. On patents and SBA loans it got the direction of change wrong more often than the chain ladder it was meant to fix. The hazard models, which share one age baseline per cause across cohorts and so estimate no cohort speeds, made the fewest such errors on mortgages, and on patents on the realised path (\cref{tab:wrongdir}). On mortgages, the vintages whose timing departs from a single speed are those whose later years overlap the housing crisis or COVID-19, and one more calendar covariate (mark-to-market loan-to-value) removes most departures before 2013 (\cref{app:freddie}): timing moves with the calendar, which group contrasts capture. In none of the four data sets did cohort speeds improve significantly on equal speeds, by the sign-flip test or by the pre-registered sign tests; only on the mortgage forecast path does a secondary measure favour them (\cref{app:inference}). This mirrors demography, where complex methods rarely improve the completion of young fertility cohorts \citep{bohkewald2018}.

\paragraph{Which method to use.} The evidence suggests three rules.
\begin{enumerate}[leftmargin=*,itemsep=1pt,topsep=2pt]
\item \emph{Always} correct provisional labels for maturity, estimated on mature cohorts, before they reach a window rule, drift detector or classifier.
\item \emph{Prefer a standard competing-risks hazard model}: an age baseline for each cause, cohort effects, and calendar covariates or unit-level covariates only where a backtest shows they help and they are not driven by the outcomes (C3). On the forecast path that is available in practice, it was about as accurate as the full method on mortgages (no significant difference, with more inputs) and more accurate on patents and civil cases; on SBA loans the result depends on how placeholder payoff dates are treated. On mortgages its cohort effects mattered: without them it beat the chain ladder but not the full method. Calendar covariates need the same check, since in the replication the calendar step hurt. In the replication the full method did not beat the chain ladder (error ratios 1.04 and 0.88, not significant).
\item \emph{Add cohort-specific speeds} only if a rolling-origin backtest (re-estimating at each past snapshot with only the data then available) shows a gain. \Cref{prop:shrink} flags cohorts whose own speed is noise-limited, which here was every youngest cohort, but it did not predict where speeds helped, and on mortgages marginal likelihood selected cohort speeds at all 13 snapshots, rolling-origin validation at only 4 of the 12 where it could be run (\cref{app:estimator,app:experiments}).
\end{enumerate}
In every case, intervals calibrated on the method's past errors (\cref{prop:intervals}) indicate how far the extrapolation can be trusted, though they under-covered on patents and civil cases (below), and should be widened after a suspected change in timing ($\Gamma = 1.25$--1.5 with Weibull timing in simulation).

\paragraph{Limitations.} Four limits matter most. First, the maturities rest on an extrapolation beyond the observed ages that the data cannot check, the sufficient-follow-up problem of cure models \citep{maller1996}. The window rule is only as good as they are (\cref{cor:misspec}), and our intervals, which assume that cohorts are exchangeable, under-covered on patents (83--86\%) and civil cases (77\%) against a nominal 90\%; weighted conformal methods could relax the assumption \citep{tibshirani2019,barber2023}. Second, the calendar step rests on (C1)--(C4), as any age--period--cohort separation must rest on outside assumptions \citep{heckman1985}; its patent gain came from a covariate that violates (C3), and it hurt in the replication. Third, maturities per cohort do not fix the ranking across features (\cref{prop:ranking}), which matters when units within a cohort are ranked, as in credit or venture screening; a pre-specified test of one feature-dependent maturity failed (\cref{app:featmat}). Fourth, the mortgage and patent comparisons are exploratory and rest on 18 and 13 cohort clusters. The replication is confirmatory, but its hazard-model result on loans depends on how placeholder payoff dates are treated, and it did not confirm that cohort speeds add less than 15\%. Smaller limits: our chain ladder uses annual cohorts at their median age, so a chain ladder on monthly cohorts may absorb part of the exact-age gain; on mortgages the hazard model used more inputs than the full method; all four backtests use today's records and revised covariates rather than those available at each snapshot; the mortgage data do not allow for loans that enter the records late; and patent applications abandoned in favour of a follow-on (continuation) application count as not granted, while PatEx omits applications abandoned before publication.

\paragraph{Outlook.} Monitoring systems should report each cohort's maturity next to every label-based metric, and drift benchmarks should include provisional labels. Since a young cohort's speed is so hard to learn from the cohort itself, better estimates need information from outside it: calendar covariates, intermediate events such as a funding round before an exit, and models that share timing across related series. Broader impact: \cref{app:impact}.

%% file: sections/ai_use.tex
\section*{Use of AI tools}
We used large language models (Claude, Anthropic) throughout this work: to help derive and check proofs, to write and run experiment code, to draft and edit the text, and, as three independent agents given only the code, the specifications and the paper, to audit the code (\cref{app:experiments}). The authors directed the work, made the design decisions, reviewed the results and take full responsibility for the paper.

%% file: sections/app_map.tex
\section*{Guide to the appendices}

\Cref{tab:map} links each claim of the main text to its evidence in the appendices: notation (\cref{app:scales}), proofs (\cref{app:proofs}), further theory (\cref{app:theory}), the ablation estimator (\cref{app:estimator}), experiments (\cref{app:experiments}), data (\cref{app:data}), related work (\cref{app:related}) and broader impact (\cref{app:impact}). The status of each analysis is in \cref{tab:deviations}.

\begin{table}[!htbp]
\centering\footnotesize
\caption{Where to find the evidence for each claim.}
\label{tab:map}
\setlength{\tabcolsep}{4pt}
\renewcommand{\arraystretch}{1.15}
\begin{tabular}{@{}p{5.6cm}p{2.6cm}p{7.2cm}@{}}
\toprule
Claim & Main text & Evidence in the appendices \\
\midrule
Provisional labels make adaptive windows and ADWIN collapse onto the youngest cohort & \cref{thm:A}; \cref{fig:teaser} & proof \cref{app:proofA}; simulations \cref{tab:e1}; mortgages \cref{tab:e2} \\
Maturity-weighted windows are minimax up to a log factor (typical case); misstated maturities leave an error & \cref{thm:B}, \cref{cor:misspec} & proof \cref{app:proofB}; \cref{tab:e1} (speeds differ) \\
One snapshot identifies a cohort's speed; equal-speed corrections err in a predictable direction & \cref{thm:id}, \cref{prop:cl}; \cref{fig:schematic} & proofs \cref{app:proofs-id}; simulation \cref{fig:speed-sim}; \cref{tab:wrongdir} \\
Calendar shocks that differ across groups are separable & \cref{thm:calendar} & proof \cref{app:proofs-id}; simulation ``Calendar shocks'' (\cref{app:sims}) \\
A young cohort's own speed is nearly unknowable & \cref{thm:precision} & proof \cref{app:proofs-limits}; \cref{tab:precision}; direct check \cref{tab:thm6check}; real data \cref{tab:speed-criterion} \\
Maturity-corrected labels cut the error of predicted rates; the crisis ranking bias remains & \cref{tab:classifiers}; \cref{prop:ranking} & \cref{tab:classifiers-full}, \cref{app:e5}; feature-dependent maturity \cref{app:featmat} \\
Corrected labels cut ADWIN's false alarms; on mortgages they miss more true changes, in the replication the total falls (not for Page--Hinkley with full-method labels) & \cref{sec:e1}, \cref{sec:replication} & \cref{tab:alarms,tab:rep-alarms} \\
Calendar covariates (mortgages) and exact ages (patents), not speeds, beat the chain ladder (exploratory) & \cref{tab:main} & tests \cref{tab:inference,tab:inference-sens}; mortgages \cref{app:freddie}; patents \cref{app:patents} \\
Competing-risks hazard models do as well or better & \cref{tab:main}; \cref{fig:speed}c & \cref{tab:baselines}, \cref{app:baselines}; patent hazard model \cref{app:patents}; \cref{tab:corrected} \\
Borrowed speeds help only when a change persists & \cref{fig:speed}; \cref{sec:speed} & \cref{tab:patex-errors,tab:wrongdir,tab:speed-criterion,tab:sim-e2e} \\
Calibrated intervals indicate how far the extrapolation can be trusted (they under-covered on patents and civil cases) & \cref{prop:intervals} & \cref{app:intervals}; coverage in \cref{app:experiments}, \cref{app:corrected}, \cref{app:freddie} \\
The conclusions tested on new data, in a pre-registered study (H2 not confirmed overall) & \cref{sec:replication}; \cref{tab:replication} & \cref{app:replication}: \cref{tab:rep-main,tab:rep-steps,tab:rep-wrongdir,tab:rep-windows,tab:rep-alarms} \\
\bottomrule
\end{tabular}
\end{table}

%% file: sections/app_scales.tex
\section{Notation and the two scales}\label{app:scales}

\begin{table}[h]
\centering\small
\caption{Notation.}
\label{tab:notation}
\resizebox{\linewidth}{!}{\begin{tabular}{@{}ll@{}}
\toprule
$c$, $T$, $A_c = T - c$ & cohort, snapshot, age of cohort $c$ at the snapshot \\
$G_c$, $G^{\mathrm{comp}}_c$ & cumulative incidence of the event and of the competing exit \\
$H$, $p_H(c) = G_c(H)$ & horizon and horizon rate (the target) \\
$m_c = G_c(A_c)/p_H(c)$ & maturity: share of horizon events already observed \\
$\psi$, $K_c = \psi \circ G_c$ & scale (identity or $-\log(1-\cdot)$) and transformed incidence \\
$\theta(c)$, $s(c)$, $\Lambda_0$ & level, speed and shared timing shape, $K_c(a) = \theta(c)\Lambda_0(s(c)a)$ \\
$\epsilon(u) = u\lambda_0(u)/\Lambda_0(u)$ & elasticity of the timing shape (log--log slope) \\
$R_c(\sigma)$ & development from age $A_c$ to $H$ at speed $\sigma$ \\
$j = 1, \dots, N$; $M_k = \sum_{j \le k} m_j$ & cohorts of a window from the youngest; effective size per unit \\
$\delta$; $\bar a_k$ & drift bound per cohort; maturity-weighted age gap of window $k$ \\
$g$, $\nu_g$, $z_g(t)$, $\beta$ & group, group effect, calendar covariates and their coefficients \\
$n_c$, $D_c$, $V_c$, $I_c$ & cohort size, expected events so far, variance of $\eta$ over them, speed inflation factor (\cref{sec:limits}) \\
$J$, $\hat q_c$, $\Gamma$, $\Gamma^\ast$ & calibration errors, their conformal quantile, sensitivity multiplier, breakdown value \\
\bottomrule
\end{tabular}}
\end{table}

On the \emph{mixture-cure scale} ($\psi$ the identity), $\Lambda_0 = F_0$ is a distribution function with $F_0(\infty) = 1$ and $\theta(c)$ is the cohort's eventual rate; this is the observation model of mixture cure and delayed-feedback models. On the \emph{hazard scale} ($\psi(y) = -\log(1-y)$), $K_c$ is the cumulative sub-distribution hazard, on which calendar shocks act multiplicatively and our estimator works. The two coincide to first order when $G_c$ is small. The results differ across scales in four places.
\begin{itemize}[leftmargin=*,itemsep=1pt]
\item \Cref{thm:id} holds on either scale, because $\psi$ is increasing and known: $K_c(H)$ identifies $p_H(c) = \psi^{-1}(K_c(H))$. \Cref{thm:calendar} is a hazard-scale result: a multiplicative shock to the incidence density is a different model.
\item Without a long reference (\cref{cor:relative}), level ratios $\theta(c)/\theta(c')$ are identified on either scale. On the mixture-cure scale they are ratios of eventual rates. On the hazard scale they are ratios $K_c(H)/K_{c'}(H)$ when speeds are equal, but ratios of $p_H$ are not identified: with $\theta(c') = 2\theta(c)$, $p_H(c)/p_H(c') = (1 - e^{-x})/(1 - e^{-2x})$ depends on the unidentified $x = \theta(c)\Lambda_0(s H)$.
\item The direction in \cref{prop:cl} depends on the monotonicity of the elasticity of $\Lambda_0$ on the scale used: decreasing for the distribution functions of \cref{lem:elasticity}, constant for a Weibull cumulative hazard (no bias, but no identification), increasing for a Gompertz cumulative hazard (bias reversed). Its magnitude also depends on the scale: in the example of \cref{fig:schematic}b the overstatement is 19\% on the mixture-cure scale and smaller on the hazard scale when $p_H$ is large.
\item \Cref{thm:precision} is stated for the mixture-cure scale with Poisson counts. On the hazard scale, the same formula gives the asymptotic variance of $\log\hat K_c(H)$; that of $\log\hat p_c$ is multiplied by $\{K e^{-K}/(1 - e^{-K})\}^2$ with $K = K_c(H)$. The Poisson model omits the survivor cell of the multinomial, and on the hazard scale the telescoping step holds to first order in $G_c(A_c)$, so the result is exact only when the incidence so far is small, which is the case for young cohorts.
\end{itemize}
A generalised-gamma distribution function used as $\Lambda_0$ on the hazard scale bounds the cumulative hazard, so the fitted model is a promotion-time cure model \citep{chen1999} in which the levels and speeds vary by cohort.

%% file: sections/app_proofs.tex
\section{Proofs}\label{app:proofs}

\subsection{\Cref{thm:A}}\label{app:proofA}

\emph{The three rules.} Let $r_k$ be the radii of a rule, nonincreasing in $k$.
\begin{itemize}[leftmargin=*,itemsep=1pt]
\item Lepski's rule selects $\hat k = \max\{k : |\hat\mu_k - \hat\mu_{k'}| \le r_k + r_{k'}\ \forall k' < k\}$. It selects $k \ge 2$ only if $|\hat\mu_k - \hat\mu_1| \le r_k + r_1 =: t_k$.
\item The Goldenshluger--Lepski rule selects $\hat k = \arg\min_k \{\hat B_k + r_k\}$ with $\hat B_k = \max_{k' < k}(|\hat\mu_k - \hat\mu_{k'}| - r_k - r_{k'})_+$ and $\hat B_1 = 0$. If it selects $k \ge 2$, then $\hat B_k + r_k \le r_1$, hence $|\hat\mu_k - \hat\mu_1| \le 2r_1 =: t_k$.
\item ADWIN with cohorts as blocks ends with a window in which no split at a cohort boundary has sub-window means differing by more than its cut threshold $\varepsilon_k$. A final window $W_k$ with $k \ge 2$ therefore satisfies $|\hat\mu_{2:k} - \hat\mu_1| \le \varepsilon_k =: t_k$, where $\hat\mu_{2:k}$ is the mean over cohorts $2, \dots, k$.
\end{itemize}
``Thresholds below $p\Delta_m/4$'' means $t_k < p\Delta_m/4$ for all $k$. If the radii are $r_k = \kappa\sqrt{2\bar p L/(nk)}$ with $\bar p \ge p$ and $L = \log(2N/\alpha)$, the condition holds for Lepski's and the Goldenshluger--Lepski rule once $n > 128\kappa^2 \bar p L/(p\Delta_m)^2$. ADWIN's usual Hoeffding-type cut thresholds do not adapt to $p$, so for rare events its condition needs $n$ of order $L/(p\Delta_m)^2$.

\emph{Good event.} Let $x = p\Delta_m/8$ and $\mathcal E = \{|\hat\mu_k - p\bar m_k| \le x \text{ for } k = 1, \dots, N\}$. Each $\hat\mu_k$ is a mean of $nk \ge n$ independent Bernoulli variables with variances at most $p$. Bernstein's inequality \citep{boucheron2013} gives $\P(|\hat\mu_k - p\bar m_k| > x) \le 2\exp\{-nkx^2/(2(p + x/3))\}$. Since $x \le p/8$, $2(p + x/3) \le 25p/12$, so the exponent is at least $12nkx^2/(25p) = 3nkp\Delta_m^2/400 \ge np\Delta_m^2/134$. A union bound gives $\P(\mathcal E^c) \le 2N\exp(-np\Delta_m^2/134)$.

\emph{On $\mathcal E$.} Because $m_j \ge m_2$ for $j \ge 2$, $\bar m_k - m_1 \ge (k-1)\Delta_m/k \ge \Delta_m/2$ for $k \ge 2$, so $\hat\mu_k - \hat\mu_1 \ge p\Delta_m/2 - 2x = p\Delta_m/4 > t_k$. Lepski's and the Goldenshluger--Lepski rule therefore cannot select $k \ge 2$. For ADWIN, $\hat\mu_{2:k} = (k\hat\mu_k - \hat\mu_1)/(k-1)$ is within $(k+1)x/(k-1) \le 3x$ of its mean $p\bar m_{2:k} \ge p(m_1 + \Delta_m)$, so $\hat\mu_{2:k} - \hat\mu_1 \ge p\Delta_m - 4x = p\Delta_m/2 > t_k$, and the final window contains cohort 1 alone. This proves (i). Then $\hat\mu_{\hat k} = \hat\mu_1 \le pm_1 + x$, which gives (ii). For (iii), on $\mathcal E$ window $k \ge 2$ has $|\hat\mu_k - p| \le \max\{p(1 - \bar m_k) + x,\ x\} \le p(1 - m_1) - 3x$, because $\bar m_k \ge m_1 + \Delta_m/2$ and $1 - m_1 \ge \Delta_m$, whereas $|\hat\mu_1 - p| \ge p(1 - m_1) - x$. \qed

\subsection{\Cref{thm:B}}\label{app:proofB}

\emph{Notation.} $S_{(k)} = \sum_{j \le k} S_j$, $\hat p_k = S_{(k)}/(nM_k)$, $\bar p_k = \E\hat p_k = \sum_{j \le k} m_j p_j/M_k$, $L = \log(2N/\alpha)$, and $B_k := \delta\bar a_k = \delta\sum_{j \le k} m_j(j-1)/M_k$. Then $|\bar p_k - p_1| \le B_k$, $B_k$ is nondecreasing in $k$, and $B_1 = 0$.
\begin{itemize}[leftmargin=*,itemsep=1pt]
\item \emph{Data-driven radii:} $U_k = (\sqrt{L/2} + \sqrt{S_{(k)} + 5L/6})^2$, $\hat w_k = (\sqrt{2U_kL} + L/3)/(nM_k)$, and $\tilde w_k = \max_{k' \ge k}\hat w_{k'}$, which is nonincreasing.
\item \emph{Deterministic radii:} $w^\circ_k = (\sqrt{2\E S_{(k)}L} + L/3)/(nM_k)$ and $w^\ast_k = 2\sqrt2\sqrt{p_{\max}L/(nM_k)} + 4L/(nM_k)$, which is nonincreasing.
\end{itemize}
The rule selects $\hat k = \max\{k : |\hat p_k - \hat p_{k'}| \le 2(\tilde w_k + \tilde w_{k'})\ \forall k' < k\}$.

\emph{Good event.} $S_{(k)}$ is a sum of independent Bernoulli variables with total variance at most $\E S_{(k)}$. By Bernstein's inequality, with probability at least $1 - \alpha/N$ for each $k$, $|S_{(k)} - \E S_{(k)}| \le \sqrt{2\E S_{(k)}L} + L/3$. Let $\mathcal E$ be the intersection, so $\P(\mathcal E) \ge 1-\alpha$. On $\mathcal E$: (a) $|\hat p_k - \bar p_k| \le w^\circ_k$; (b) solving the quadratic inequality in $\sqrt{\E S_{(k)}}$ gives $\E S_{(k)} \le U_k$, hence $w^\circ_k \le \hat w_k \le \tilde w_k$; (c) $S_{(k)} \le 2\E S_{(k)} + 5L/6$ gives $U_k \le 4\E S_{(k)} + 13L/3 \le 4np_{\max}M_k + 13L/3$, hence $\hat w_k \le w^\ast_k$, and since $w^\ast$ is nonincreasing, $\tilde w_k \le w^\ast_k$. Combining (a) and (b), $|\hat p_k - p_1| \le \tilde w_k + B_k$ for every $k$.

\emph{Oracle inequality.} Fix any $k_0$. If $\hat k \ge k_0$, the selected window passed its check against $k_0$, so
\[
|\hat p_{\hat k} - p_1| \le 2(\tilde w_{\hat k} + \tilde w_{k_0}) + \tilde w_{k_0} + B_{k_0} \le 5\tilde w_{k_0} + B_{k_0},
\]
using that $\tilde w$ is nonincreasing. If $\hat k < k_0$, window $\hat k + 1 \le k_0$ was rejected, so some $k' \le \hat k$ has $|\hat p_{\hat k+1} - \hat p_{k'}| > 2(\tilde w_{\hat k+1} + \tilde w_{k'})$; on $\mathcal E$ the left side is at most $\tilde w_{\hat k+1} + \tilde w_{k'} + B_{\hat k+1} + B_{k'}$, so $\tilde w_{k'} < 2B_{\hat k+1} \le 2B_{k_0}$, and $\tilde w_{\hat k} \le \tilde w_{k'}$. Hence $|\hat p_{\hat k} - p_1| \le \tilde w_{\hat k} + B_{\hat k} < 3B_{k_0}$. In both cases $|\hat p_{\hat k} - p_1| \le 5(w^\ast_{k_0} + B_{k_0})$, and $w^\ast_k \le 4\{\sqrt{p_{\max}L/(nM_k)} + L/(nM_k)\}$ gives part (i). \qed

\emph{\Cref{cor:misspec}.} With $\tilde p_k = S_{(k)}/(n\tilde M_k)$, $\E\tilde p_k - p_1 = (M_k/\tilde M_k)\sum_{j \le k}m_j(p_j - p_1)/M_k + p_1(M_k/\tilde M_k - 1)$. Since $\tilde m_j/m_j \in [1 - \bar\eta_k, 1 + \bar\eta_k]$ with $\bar\eta_k = \max_{j \le k}\eta_j \le 1/2$, $M_k/\tilde M_k \in [2/3, 2]$ and $|M_k/\tilde M_k - 1| \le \bar\eta_k/(1 - \bar\eta_k) \le 2\bar\eta_k$, so $|\E\tilde p_k - p_1| \le \tilde B_k := 2\delta\bar a_k + 2p_{\max}\bar\eta_k$, which is nondecreasing in $k$. The proof of part (i) uses only this bias bound, its monotonicity, and the concentration of $S_{(k)}$; with $\tilde M_k$ in the radii, $\E S_{(k)} \le np_{\max}M_k \le 2np_{\max}\tilde M_k$, so $w^\ast_k$ grows by at most a factor $\sqrt2$. The final step gives $|\tilde p_{\hat k} - p_1| \le 5(\sqrt2 w^\ast_{k_0} + \tilde B_{k_0}) \le 30\{\delta\bar a_{k_0} + p_{\max}\bar\eta_{k_0} + \sqrt{p_{\max}L/(n\tilde M_{k_0})} + L/(n\tilde M_{k_0})\}$. \qed

\emph{Estimated maturities.} With a common timing curve, $m_j = F(A_j)$ for the within-horizon timing distribution $F$. If $\hat F$ is the empirical distribution of $N_F$ event ages from mature cohorts not used in the windows, the DKW inequality \citep{massart1990} gives $\sup|\hat F - F| \le \varepsilon_F = \sqrt{\log(2/\alpha')/(2N_F)}$ with probability at least $1 - \alpha'$, hence $|\hat M_k - M_k| \le k\varepsilon_F$. A relative bound $|\hat F(A_j)/F(A_j) - 1| \le \eta_j$ then yields \cref{cor:misspec}; because $\varepsilon_F$ is additive, $\eta_j \approx \varepsilon_F/m_j$ is largest for the youngest cohort. With cohort-specific speeds, a common curve gives $\eta_j$ that do not shrink with $N_F$, which is the error floor seen in simulation (\cref{tab:e1}).

\emph{Gaussian radii.} With radii $\kappa\,\widehat{\mathrm{se}}_k$, where $\widehat{\mathrm{se}}_k$ is the plug-in binomial standard error of $\hat p_k$, and any fixed $\kappa > 0$, windows with $\bar p_k \neq p_1$ have a fixed gap to window 1 while the radii vanish, so they are rejected with probability tending to one, and windows with $\bar p_k = p_1$ are consistent. The naive rule is inconsistent for every $\kappa$ by \cref{thm:A}.

\emph{Part (ii).} Fix $\Delta \le \min(p_{\max} - p_{\min}, p_{\max}/2)$, let $k = k(\Delta) = \min(N, \lceil \Delta/\delta \rceil)$ and $p_0 = p_{\max} - \Delta$, so that $p_{\max}/2 \le p_0$ and $p_{\min} \le p_0$. Under $P_0$, $p_j \equiv p_0$; under $P_1$, $p_j = p_0 + (\Delta - \delta(j-1))_+ \le p_{\max}$. Both lie in the drift class, and their targets differ by $\Delta$. Only cohorts $j \le k$ differ. For one unit of cohort $j$, by the $\chi^2$ bound on the Kullback--Leibler divergence and $m_jp_0(1 - m_jp_0) \ge m_jp_0(1 - p_0)$, $\mathrm{KL}(\mathrm{Bern}(m_jp_j^{(1)}) \,\|\, \mathrm{Bern}(m_jp_0)) \le m_j^2\Delta^2/\{m_jp_0(1 - m_jp_0)\} \le m_j\Delta^2/\{p_0(1-p_0)\} \le 2m_j\Delta^2/\{p_{\max}(1-p_{\max})\}$, since $x(1-x) \ge (p_{\max}/2)(1 - p_{\max})$ on $[p_{\max}/2, p_{\max}]$. Summing over units, $\mathrm{KL}(P_1\|P_0) \le 2n\Delta^2 M_k/\{p_{\max}(1-p_{\max})\}$, which is at most 1 under the condition defining $\Delta^\ast$. By the Bretagnolle--Huber inequality \citep[Lemma~2.6 and Theorem~2.2]{tsybakov2009}, every estimator has $\max_{i} P_i(|\tilde p - p^{(i)}_1| \ge \Delta/2) \ge e^{-1}/4$. \qed

\emph{Matching.} Let $\Delta^\ast < \min(p_{\max} - p_{\min}, p_{\max}/2)$, $k_0 = k(\Delta^\ast)$ and $\varphi = p_{\max}(1 - p_{\max})/2$, so that the condition defining $\Delta^\ast$ reads $n\Delta^2M_{k(\Delta)} \le \varphi$. Either the constraint binds, $n\Delta^{\ast 2}M_{k_0} = \varphi$, and window $k_0$ has bias at most $\delta(k_0 - 1) < \Delta^\ast$ and $\sqrt{p_{\max}L/(nM_{k_0})} = \Delta^\ast\sqrt{p_{\max}L/\varphi}$; or $\Delta^\ast = k_0\delta$ with $k_0 < N$ and $n\Delta^{\ast 2}M_{k_0+1} \ge \varphi$, and window $k_0 + 1$ has bias at most $\delta k_0 = \Delta^\ast$ and the same bound on the square-root term. If that window holds at least $L/p_{\max}$ units' worth of maturity, $nM \ge L/p_{\max}$, then $L/(nM) \le \sqrt{p_{\max}L/(nM)}$, and the minimum in (i) is at most $\Delta^\ast(1 + 2\sqrt{p_{\max}L/\varphi})$. Since $p_{\max}/\varphi = 2/(1 - p_{\max})$, the upper bound is within the factor $40\{1 + 2\sqrt{2L/(1 - p_{\max})}\}$ of the lower bound $\Delta^\ast/2$. Centring the two-point construction at $p_{\max} - \Delta$ rather than at $p_{\min}$ is what removes a factor $\sqrt{p_{\max}/p_{\min}}$, which is large for rare events.

\subsection{Proofs for \cref{sec:identification}}\label{app:proofs-id}

\emph{\Cref{thm:id}.} (1) For the reference, $\Lambda_0 = K_{c^\ast}/\theta(c^\ast)$ on $[0, A_{c^\ast}]$, so $\Lambda_0$ is identified up to a constant factor and its elasticity $\epsilon$ exactly. (2) For cohort $c$, $\epsilon_c(a) := aK_c'(a)/K_c(a) = \epsilon(s(c)a)$. Since $\epsilon$ is strictly monotone, $s(c) = \epsilon^{-1}(\epsilon_c(a))/a$ for every $a \in (0, A_c]$. (3) $\theta(c)/\theta(c^\ast) = K_c(A_c)/K_{c^\ast}(s(c)A_c)$, and $K_c(H) = K_c(A_c)\Lambda_0(s(c)H)/\Lambda_0(s(c)A_c)$ involves only a ratio of $\Lambda_0$ at ages at most $A_{c^\ast}$. Then $p_H(c) = \psi^{-1}(K_c(H))$. \qed

\emph{When the reference is too short.} Let $U$ be the largest age at which $\Lambda_0$ is identified, from the reference or from cohorts whose speeds are identified on it. If $s(c)H \le U$, the argument above applies. Let $s(c)H > U$. For a smooth nondecreasing $h$ with $h = 0$ on $[0, U]$ and $h > 0$ on $(U, s(c)H]$, and small $\tau > 0$, let $\tilde\Lambda_0(u) = \Lambda_0(u)\exp\{-\tau\int_0^u h(v)\,dv/v\}$ on $[0, s(c)H]$, continued beyond $s(c)H$ by any increasing function. On $[0, s(c)H]$ its elasticity $\epsilon(u) - \tau h(u)$ is positive for small $\tau$ and strictly decreasing if $\epsilon$ is (if $\epsilon$ is increasing, replace $-\tau$ by $+\tau$); $\tilde\Lambda_0$ equals $\Lambda_0$ on $[0, U]$, so it fits every observed curve; and $\tilde\Lambda_0(s(c)H) < \Lambda_0(s(c)H)$. Hence $p_H(c)$ is point-identified if and only if $s(c)H \le U$. \Cref{ass:R} restricts the elasticity only on $(0, A_{c^\ast})$; if it is also assumed decreasing beyond $U$, then $0 < \epsilon(u) < \epsilon(U)$ for $u > U$, so $1 < \Lambda_0(s(c)H)/\Lambda_0(U) < (s(c)H/U)^{\epsilon(U)}$, which bounds $K_c(H) = K_c(A_c)\Lambda_0(s(c)H)/\Lambda_0(s(c)A_c)$.

\emph{\Cref{prop:cl}.} $K_c(H) = K_c(A_c)R_c(s(c))$, which gives the ratio. Moreover
\[
\frac{d\log R_c(\sigma)}{d\log\sigma} = \epsilon(\sigma H) - \epsilon(\sigma A_c) < 0
\]
when $\epsilon$ is strictly decreasing and $H > A_c$, so $R_c(1) > R_c(s(c))$ if $s(c) > 1$, and the reverse if $s(c) < 1$. The example uses a Weibull $F_0$ with shape 1.5 and median 5, with $A_c = 2$, $H = 7$ and $s(c) = 1.3$, which give $R_c(1)/R_c(1.3) = 1.19$. \qed

\emph{\Cref{thm:calendar}.} Within a (cohort, age) cell, all cohort and age factors cancel in log-hazard contrasts between groups. The rank condition identifies $\beta$ and $\log(\nu_g/\nu_{g'})$, hence every $\nu_g$ given $\nu_{g_0} = 1$. The adjusted cumulative hazard $\int_0^a \tilde\lambda_g(v \mid c)\,\nu_g^{-1}e^{-\beta^\top z_g(c+v)}\,dv = \theta(c)\Lambda_0(s(c)a)$ has the structure of \cref{thm:id}. \emph{Linear common trend.} Let the covariate be $z(t) = \zeta t$ for all groups, with coefficient $\beta$, and let $s \equiv 1$. The parameters $(\theta(c), \lambda_0(a), \beta)$ and $(\theta(c)e^{\beta\zeta c}, \lambda_0(a)e^{\beta\zeta a}, 0)$ give the same hazard at every $(c, a)$, because $\beta\zeta(c+a) = \beta\zeta c + \beta\zeta a$, and $\lambda_0(a)e^{\beta\zeta a}$ is again a valid baseline hazard. \qed

\subsection{Proofs for \cref{sec:limits}}\label{app:proofs-limits}

\emph{\Cref{thm:precision}.} (i) With intervals $(a_{j-1}, a_j]$ and $\mu_j = n\theta[F_0(sa_j) - F_0(sa_{j-1})]$, $\partial\log\mu_j/\partial\log\theta = 1$ and $\partial\log\mu_j/\partial\log s = \eta_j$, the event-weighted average of $\eta$ over interval $j$. The Fisher information for $(\log\theta, \log s)$ is $D\begin{psmallmatrix}1 & \bar\eta\\ \bar\eta & \bar\eta^2 + V\end{psmallmatrix}$, and by telescoping $\bar\eta = \epsilon(u_c)$. The target $\log p = \log\theta + \log F_0(sH)$ has gradient $(1, \epsilon(u_{H,c}))$, so $\mathrm{AVar}(\log\hat p) = \{1 + (\epsilon(u_c) - \epsilon(u_{H,c}))^2/V\}/D$, and $1/D$ if $s$ is known. ($V$ is the event-weighted variance of the interval averages; it tends to the continuous version as the intervals shrink.) The lower bound is the H\'ajek--Le Cam local asymptotic minimax theorem for this smooth functional of a locally asymptotically normal model. A model in which $F_0$ is unknown contains the model with $F_0$ known, so its local minimax risk is at least as large, and the bound remains valid. The statement is asymptotic in $n$ at fixed maturity; the expansions in (ii) and (iii) describe how the fixed-maturity variance scales as $m_c \to 0$.

(ii) For a Weibull with shape $\gamma$ and scale $\lambda$, the inflation factor is asymptotically $12(1 - \epsilon(u_{H,c})/\gamma)^2/(m_cF_0(u_{H,c}))^2$, which is $12/m_c^2$ for a long horizon. Indeed $\eta(u) = \gamma\{1 - (u/\lambda)^\gamma\}$. As $u_c \to 0$, the event ages are $u_cW$ with $W$ of density $\gamma w^{\gamma-1}$ on $[0,1]$, so $V = \gamma^2(u_c/\lambda)^{2\gamma}\mathrm{Var}(W^\gamma)(1+o(1))$ with $\mathrm{Var}(W^\gamma) = 1/12$; $\epsilon(u_c) \to \gamma$; and $(u_c/\lambda)^\gamma \sim F_0(u_c) = m_cF_0(u_{H,c})$. More generally, if $\eta(u) = \gamma + b\rho u^\rho + o(u^\rho)$ with $b \neq 0$ (as when $f_0(u) \propto u^{\gamma-1}(1 + bu^\rho)$), then $V = b^2\rho^2u_c^{2\rho}\{\gamma/(\gamma+2\rho) - \gamma^2/(\gamma+\rho)^2\}(1+o(1))$, $\epsilon(u_c) \to \gamma \ne \epsilon(u_{H,c})$ and $m_c \asymp u_c^\gamma$, so the inflation factor is of order $m_c^{-2\rho/\gamma}$.

(iii) For a log-normal with parameters $(\mu, \varsigma)$, let $z = (\log u - \mu)/\varsigma$. Then $\eta(u) = -z/\varsigma$ and $\epsilon(u) = \phi(z)/\{\varsigma\Phi(z)\}$. As $z_c \to -\infty$: $\epsilon(u_c) \sim |z_c|/\varsigma$; the events so far have $z$ distributed as a standard normal truncated above at $z_c$, whose variance is $z_c^{-2}(1 + o(1))$, so $V_c \sim 1/(\varsigma^2 z_c^2)$; and $\log(1/m_c) \sim z_c^2/2$. Hence the inflation factor is $\sim z_c^4 \sim 4\log^2(1/m_c)$. The approach is slow, because $\log(1/m_c)$ also contains a term in $\log|z_c|$.

(ii, exact Weibull form.) For a Weibull and $H = \infty$, write $w = (u_c/\lambda)^\gamma = -\log(1 - m_c)$. Then $\epsilon(u_c) = \gamma r(w)$ with $r(w) = we^{-w}/(1 - e^{-w})$, $\epsilon(\infty) = 0$, and $\eta = \gamma(1 - W)$, where $W = (U/\lambda)^\gamma$ for an event age $U$ in the window is a standard exponential truncated at $w$. So $I_c = 1 + r(w)^2/\mathrm{Var}(W \mid W \le w)$, which does not depend on $\gamma$ and is $\sim 12/m_c^2$ as $m_c \to 0$. The limit overstates the exact value by about 40\% at $m_c = 0.16$ and by 2\% at $m_c = 0.01$; the numbers in the text use the exact value.

(iv) Under the Poisson model, the event ages form a Poisson process on $[0, A_c]$ with intensity $n_c\theta s f_0(sa) = D_c\,g_s(a)$; counts in age intervals are a function of it and carry no more information. For two Poisson processes with intensities $\mu, \mu'$ of equal total mass $D_c$, $\mathrm{KL} = \int \mu\log(\mu/\mu') - \int\mu + \int\mu' = D_c\,\mathrm{KL}(g_s\,\|\,g_{s'})$. The horizon rate is $p_c = \theta F_0(sH) = D_c/\{n_c m_c(s)\}$, so the two log rates differ by $|\log(m_c(s)/m_c(s'))|$. The Bretagnolle--Huber inequality \citep[Theorem~2.2]{tsybakov2009} gives the bound. As $s' \to s$, $\mathrm{KL}(g_s\|g_{s'}) = V_c(\log s' - \log s)^2/2 + o(\cdot)$ and $\log(m_c(s)/m_c(s')) = \{\epsilon(u_c) - \epsilon(u_{H,c})\}(\log s - \log s') + o(\cdot)$, so choosing $|\log s' - \log s| = (D_cV_c)^{-1/2}$ recovers the order $\{(I_c - 1)/D_c\}^{1/2}$ of part (i) without an asymptotic argument when $D_cV_c$ is large. If $f_0(u) \propto u^{\gamma - 1}$ on $[0, \max(s, s')A_c]$, then $g_s = g_{s'}$ and the bound is $1/4$ for every $D_c$. For the example in the text ($\gamma = 1.5$, median 5, $A_c = 2$, $H = 7$, $s = 1$), maximising the separation subject to $D_c\,\mathrm{KL} \le 1$ over $s'$ gives half-separations in $\log p_c$ of 0.63, 0.37, 0.19 and 0.10 for $D_c = 10$, 100, 1000 and 3000, against known-speed standard errors of 0.32, 0.10, 0.03 and 0.02. At $D_c = 1000$ and 3000 the alternative is a slower cohort whose eventual rate is 8 and 2 times higher, which is valid only if $p_c$ is below about 0.09 and 0.34; the case $D_c = 100$ used in the text has no such restriction. \qed

\emph{\Cref{prop:intervals}.} (i) Any continuation of $G_c$ and $G^{\mathrm{comp}}_c$ beyond $A_c$ reproduces the data. The incidence at $H$ can take any value in $[G_c(A_c), 1 - G^{\mathrm{comp}}_c(A_c)]$: no further events, all units still at risk at $A_c$ experiencing the event before $H$, or anything in between. (ii) Under exchangeability the rank of $|r_c|$ among the $J+1$ errors is uniform, so $\P(|r_c| > \hat q_c) \le \alpha$; intersecting with the set in (i) keeps $p_H(c)$. Under the sensitivity condition, conditional on the calibration errors, $\P(|r_c| > \Gamma\hat q_c) \le \P(|\tilde r| > \hat q_c)$ for an exchangeable copy $\tilde r$, and taking expectations bounds the right side by $\alpha$. The lower clamp uses the cohort's realised incidence so far; if that is estimated under random censoring, a lower confidence bound should replace it. \qed

\emph{\Cref{prop:ranking}.} Substituting $p_H(x_1, c) = r\,p_H(x_2, c)$ gives
\[
\frac{\bar y(x_1)}{\bar y(x_2)} = r\,\frac{\sum_c w_c m_c(x_1)}{\sum_c w_c m_c(x_2)}, \qquad w_c = \pi_c\,p_H(x_2, c).
\] For a cohort with $A_c < H$, $m_c(x) = F_0(\sigma A_c)/F_0(\sigma H)$ with $\sigma = s(c)e^{\xi^\top x}$, and $d\log m_c/d\log\sigma = \epsilon(\sigma A_c) - \epsilon(\sigma H) > 0$ because $\epsilon$ is strictly decreasing. So the slower type is less mature in every immature cohort, and both maturities equal one in mature cohorts. \qed

%% file: sections/app_theory.tex
\section{Additional theory}\label{app:theory}

\subsection{Elasticity of common timing families}

\begin{lemma}\label{lem:elasticity}
Let $\Lambda_0 = F$ be a distribution function on $(0, \infty)$ such that $X = \log T$, $T \sim F$, has a log-concave density. Then the elasticity $\epsilon(u) = uF'(u)/F(u)$ is nonincreasing. For the Weibull, gamma, log-normal, log-logistic and generalised-gamma families it is strictly decreasing.
\end{lemma}

\emph{Proof.} With $F_X(x) = F(e^x)$, $\epsilon(e^x) = d\log F_X(x)/dx$. A log-concave density has a log-concave distribution function \citep{bagnoli2005}, so this derivative is nonincreasing in $x$. For the listed families the density of $X$ is strictly log-concave (Gumbel, log-gamma, normal, logistic and generalised log-gamma, respectively), and the derivative is strictly decreasing. \qed

The lemma shows that the decreasing elasticity behind \cref{prop:cl} holds for all common timing families, and also that the generalised-gamma shape used by our estimator imposes it: a timing curve with increasing elasticity, such as a Gompertz cumulative hazard, lies outside that family.

\subsection{Weak identification near zero and eventual-level ratios}

\begin{proposition}[Power-law limit]\label{prop:powerlaw}
Suppose $\Lambda_0(u) = c_{\mathrm{pl}}u^\gamma$ exactly on $[0, U_{\mathrm{pl}}]$ and on no larger interval, with $s(c)A_c \le U_{\mathrm{pl}}$, and let $\Lambda_0$ be known on $[0, U_{\mathrm{pl}}H/A_c]$. Then the cohort's own data identify only $\theta(c)s(c)^\gamma$, and the identified set for $K_c(H)$ is $\{K_c(A_c)\Lambda_0(\sigma H)/\Lambda_0(\sigma A_c) : 0 < \sigma \le U_{\mathrm{pl}}/A_c\}$, intersected with the set of \cref{prop:intervals}(i).
\end{proposition}

\emph{Proof.} On the window, $K_c(a) = \theta(c)s(c)^\gamma c_{\mathrm{pl}}a^\gamma$. A pair $(\theta', \sigma)$ reproduces it if and only if $\theta'\sigma^\gamma = \theta(c)s(c)^\gamma$ and $\sigma A_c \le U_{\mathrm{pl}}$; a larger $\sigma$ would carry the window beyond the power-law range and change its shape. Each such pair gives $K_c(H) = K_c(A_c)\Lambda_0(\sigma H)/\Lambda_0(\sigma A_c)$, which is known because $\sigma H \le U_{\mathrm{pl}}H/A_c$. As $\sigma \to 0$, the value tends to the pure power-law extrapolation $K_c(A_c)(H/A_c)^\gamma$. \qed

\begin{corollary}[Eventual-level ratios without a long reference]\label{cor:relative}
Drop the condition $A_{c^\ast} \ge s(c)H$ of \cref{thm:id}, and suppose every cohort's rescaled window lies within the reference cohort's. Under \cref{ass:R}, every speed and every level ratio $\theta(c)/\theta(c')$ is identified. On the mixture-cure scale, where $\theta(c)$ is the eventual rate, the levels are partially identified with sharp bounds
\[
\theta(c) \in \Bigl[q(c),\; q(c)\big/\max_{c'} \tfrac{q(c')}{1 - G^{\mathrm{comp}}_{c'}(A_{c'})}\Bigr],
\]
where $q(c) = \theta(c)F_0(U^\ast)$ is identified and $U^\ast$ is the largest observed rescaled age.
\end{corollary}

\emph{Proof.} The reference cohort's curve identifies the elasticity $\epsilon$ on its window, so speeds are identified as in \cref{thm:id}. In rescaled time, $K_c(u/s(c)) = \theta(c)\Lambda_0(u)$ on overlapping windows, so level ratios are identified and $\Lambda_0$ is identified up to scale. On the mixture-cure scale, write $F_0(U^\ast) \le 1$ for the unknown scale, so $\theta(c) = q(c)/F_0(U^\ast)$. An eventual rate cannot exceed one minus the competing incidence already observed, so $F_0(U^\ast) \in [\max_c q(c)/(1 - G^{\mathrm{comp}}_c(A_c)), 1]$. Both endpoints are attainable: $F_0$ may saturate at $U^\ast$, or the remaining units of the binding cohort may all experience the event eventually. \qed

The upper bound is informative only when some cohort has a large share of events relative to its remaining units; for rare events it is loose. The ratio statement concerns eventual levels and therefore relies on \cref{ass:R} holding beyond the observed window, which \cref{prop:intervals} shows is untestable. On the hazard scale, ratios of $p_H$ are not identified (\cref{app:scales}).

\emph{The chain ladder for very young cohorts.} Let $\Lambda_0(u) \sim c_{\mathrm{pl}}u^\gamma$ as $u \to 0$, let cohort $c'$ have the reference speed and cohort $c$ speed $\sigma > 1$. As $A_c \to 0$, the chain ladder's bias factor for $c$, $R_c(1)/R_c(\sigma) = [\Lambda_0(H)/\Lambda_0(\sigma H)][\Lambda_0(\sigma A_c)/\Lambda_0(A_c)]$, tends to $\sigma^\gamma\Lambda_0(H)/\Lambda_0(\sigma H)$. For a bounded $\Lambda_0$ and a long horizon this is $\sigma^\gamma$: with $\sigma = 1.3$ and $\gamma = 1.5$, any true decline smaller than 33\% is reported as an increase. In this limit the cohort's own data no longer identify its speed (\cref{prop:powerlaw}), so the limit describes the chain ladder's bias, not a regime in which a speed-aware estimator recovers the truth.

\subsection{Precision in numbers}

\begin{table}[h]
\centering\small
\caption{Standard error of $\log\hat p_c$ for one cohort (Weibull timing, shape 1.5, median 5 years; long horizon, so $m = F_0(u_c)$; $p = 3\%$; monthly ages; 10{,}000 replications). ``Robust'' is $1.4826\times$ the median absolute deviation; the plain standard deviation is much larger at low maturity because of a likelihood ridge along $ps^\gamma = \text{const}$, on which 31\% of the fits at $n = 20{,}000$ and two years stop at a parameter bound.}
\label{tab:precision}
\resizebox{\linewidth}{!}{\begin{tabular}{rrrrrr}
\toprule
$n$ & follow-up (years) & maturity $m$ & speed known & speed unknown (\cref{thm:precision}) & simulated (robust) \\
\midrule
20{,}000 & 2 & 0.16 & 0.10 & 1.85 & 1.52 \\
20{,}000 & 4 & 0.39 & 0.07 & 0.36 & 0.36 \\
20{,}000 & 10 & 0.86 & 0.04 & 0.05 & 0.05 \\
200{,}000 & 2 & 0.16 & 0.03 & 0.58 & 0.56 \\
200{,}000 & 4 & 0.39 & 0.02 & 0.11 & 0.11 \\
\bottomrule
\end{tabular}}
\end{table}

The table uses an eventual rate (long horizon). At the 7-year horizon of our mortgage backtests, \cref{thm:precision}(i) gives, for the same timing and $p_H = 3\%$, maturities 0.24 and 0.57 at two and four years, standard errors 0.084 (known speed) and 0.64 (unknown) at $n = 20{,}000$ and two years, and 0.05 and 0.11 at four years: inflation factors of 58 and 3.8. The small-maturity approximation of \cref{thm:precision}(ii) is rough at these maturities; the table uses (i). The heavy tails at low maturity also affect any error metric based on logarithms of estimates.

\begin{proposition}[When is a cohort's own speed worth estimating?]\label{prop:shrink}
Work with the Gaussian approximation to the likelihood of $\log s_c$, with variance $v^2 = 1/(D_cV_c)$. Let log-speeds follow a random walk across cohorts with known increment variance $\omega^2$, and let a neighbouring cohort's log-speed be known up to variance $\tau^2$. Under this prior, the posterior-mean log-speed has Bayes risk for $\log\hat p_c$ no larger than both the cohort's own estimate and the neighbour's speed. Own-speed estimation beats borrowing the neighbour's speed if and only if $\omega^2 + \tau^2 > v^2$; for Weibull timing and a long horizon, roughly when $m_c > \{12/(n_cp_c\gamma^2(\omega^2 + \tau^2))\}^{1/3}$. If the borrowed speed is that of a cohort $k$ steps away, $\omega^2$ is replaced by $k\omega^2$; the chain ladder borrows the reference speed.
\end{proposition}

\emph{Proof.} With a speed $\tilde s$ taken from elsewhere, $\log\hat p = \log K_c(A_c) + \log F_0(\tilde sH) - \log F_0(\tilde sA_c)$, so $\mathrm{MSE}(\log\hat p) \approx D^{-1} + (\epsilon(u_c) - \epsilon(u_{H,c}))^2\,\mathrm{MSE}(\log\tilde s)$. The own estimate has $\mathrm{MSE}(\log\tilde s) = v^2$, the neighbour's has $\omega^2 + \tau^2$, and the posterior mean, combining independent sources, has $v^2(\omega^2+\tau^2)/(v^2+\omega^2+\tau^2) \le \min(v^2, \omega^2+\tau^2)$. The Weibull form uses \cref{thm:precision}(ii). \qed

\subsection{Sensitivity bounds}

\begin{proposition}[Sensitivity bounds]\label{prop:sharp}
Under the restriction $|\log\hat p_c - \log p_H(c)| \le \Gamma q$, the set of values of $p_H(c)$ consistent with the data and the restriction is $\hat C_c(\Gamma)$ with $\hat q_c$ replaced by $q$. For a drift ratio between two cohorts it is the corresponding ratio interval. These sets are nested and increase with $\Gamma$, so the breakdown value $\Gamma^\ast = \inf\{\Gamma : \text{the drift-ratio interval contains } 1\}$ is well defined.
\end{proposition}

\emph{Proof.} The set is the intersection of the identified set of \cref{prop:intervals}(i) with the restriction interval, which is nested and increasing in $\Gamma$; for two cohorts the joint set is a product of intervals. \qed

$\Gamma^\ast$ measures how much larger than the calibrated errors the true error would have to be for a drift statement to fail. It is a robustness measure, not a probability, and narrower intervals mechanically give larger values.

\subsection{A test of the rescaling assumption}\label{app:test}

For cohort $c$, fit one speed, and separately early and late speeds split at an age $a^\ast$ fixed in advance, to the shock-adjusted curve with $\Lambda_0$ fixed from other cohorts. Let $\mathrm{LR}_c$ be twice the log-likelihood gain. Under \cref{ass:R} on $[0, A_c]$, an over-dispersed Poisson model and a consistent dispersion estimate $\hat\phi$, $\mathrm{LR}_c/\hat\phi \to \chi^2_1$ \citep{mccullagh1983}, ignoring the estimation of $\Lambda_0$, which uses many more cohorts. The test has power against shape changes inside the window and none against changes after $A_c$ (\cref{prop:intervals}).

\subsection{Features and composition}\label{app:features}

\emph{Identification.} With speeds $s(c)e^{\xi^\top x}$, for fixed $x$, $G(\cdot \mid x, c)$ has the structure of \cref{thm:id}. So $\log s(c) + \xi^\top x$ is identified for every $x$ in the support; if the vectors $(1, x)$ span $\R^{\dim x + 1}$ within a cohort, $\xi$ and $s(c)$ follow. Under the long-reference condition for the speeds $s(c)e^{\xi^\top x}$, the horizon rate $p_H(x, c)$ is identified; eventual levels are identified only up to a common factor, as in \cref{cor:relative}.

\emph{Ranking.} With entry features $x$, let speeds be $s(c)e^{\xi^\top x}$, and let a classifier trained on ``event so far'' labels target $\bar y(x) = \sum_c \pi_c(x)\, p_H(x,c)\, m_c(x)$, the rate of ``event so far'' among units with features $x$, where $\pi_c(x)$ is the share of such units in cohort $c$.

\begin{proposition}[Ranking distortion]\label{prop:ranking}
Let two types have $p_H(x_1, c) = r\,p_H(x_2, c)$ for all $c$, the same cohort shares, and speeds $e^{\xi^\top x_1} < e^{\xi^\top x_2}$. If $\epsilon$ is strictly decreasing and some cohort is younger than $H$, then $\bar y(x_1)/\bar y(x_2) < r$: the slower type is ranked below the faster one unless $r$ exceeds the inverse of a maturity ratio.
\end{proposition}

With Weibull timing (shape 1.5), a 10-year horizon, medians of 5 and 10 years, and equal cohorts founded in mid-2019 to mid-2023 and observed in mid-2026, the slower type must have a 30\% higher rate to be ranked first; with cohorts since 2010 the threshold is 7\%. More data, better classifiers or recalibration towards $\bar y$ do not remove the distortion, because it is in the target itself. The loan-level evidence is mixed: the distortion appears when the training data are immature and not otherwise (\cref{app:e5}).

\begin{proposition}[Composition shifts change timing shape]\label{prop:composition}
Let types $\ell = 1, \dots, Q$ have eventual rates $p_\ell$ and Weibull timing curves $F_\ell$ with a common shape and distinct scales, and let every cohort contain every type in shares $\pi_c(\ell) > 0$. The cohort's timing curve is $F_c = \sum_\ell w_c(\ell)F_\ell$ with event weights $w_c(\ell) = \pi_c(\ell)p_\ell/\sum_{\ell'}\pi_c(\ell')p_{\ell'}$. The curves $F_c$ and $F_{c'}$ are time rescalings of each other if and only if $w_c = w_{c'}$.
\end{proposition}

\emph{Proof.} Expanding near $0$, the coefficients of $a^{k\gamma}$ are proportional to the $k$-th moments of $\sum_\ell w_c(\ell)\delta_{\vartheta_\ell}$, with $\vartheta_\ell = \lambda_\ell^{-\gamma}$ for scales $\lambda_\ell$. A rescaling by $s$ multiplies the $k$-th moment by $s^{k\gamma}$. Equal moments of measures on finitely many positive points force equal measures, and full support then gives $s = 1$ and $w_c = w_{c'}$. \qed

A shift in composition therefore violates \cref{ass:R} at the cohort level even when no type changes; the cohort-level speed estimator then fits the best rescaling of a curve that is not one, and the intervals absorb the remainder.

\subsection{Calibrated intervals}\label{app:intervals}

Any continuation of a cohort's curves beyond $A_c$ fits the data equally well, so every estimate for a young cohort rests on an extrapolation, which we calibrate on the method's own past errors. Let $r_{c'}(a) = \log \hat p_{c'}(a) - \log p_H(c')$ be the out-of-sample error of the procedure for a now-mature cohort $c'$, run at the earlier snapshot $c' + a$. Let $\hat q_c$ be the $\lceil (J+1)(1-\alpha)\rceil$-th smallest of $J$ such absolute errors $|r_{c'}(A_c)|$, or $\infty$ if $J < 9$ at $\alpha = 0.1$, and report
\[
\hat C_c(\Gamma) = \Big[\max\{G_c(A_c),\, \hat p_c e^{-\Gamma\hat q_c}\},\ \min\{1 - G^{\mathrm{comp}}_c(A_c),\, \hat p_c e^{\Gamma\hat q_c}\}\Big].
\]

\begin{proposition}[Identified set and calibrated intervals]\label{prop:intervals}
\begin{enumerate}[label=(\roman*),itemsep=1pt,topsep=2pt]
\item With independent censoring and no further assumptions, the sharp identified set for $p_H(c)$, $A_c < H$, is $[G_c(A_c),\, 1 - G^{\mathrm{comp}}_c(A_c)]$.
\item If the target cohort's error and the $J$ calibration errors are exchangeable, $\P\{p_H(c) \in \hat C_c(1)\} \ge 1 - \alpha$. If, conditionally on the calibration errors, the target's absolute error is stochastically at most $\Gamma$ times that of an exchangeable copy, $\P\{p_H(c) \in \hat C_c(\Gamma)\} \ge 1 - \alpha$.
\end{enumerate}
\end{proposition}

Part (i) is a worst-case bound \citep{manski2003}; part (ii) is split conformal prediction \citep{vovk2005,lei2018} over cohorts, with a sensitivity parameter $\Gamma$ as in \citet{rambachan2023}. Errors taken at the target's age already contain estimation noise and the weak identification of speed, so no model-based variance is needed. Exchangeability is an assumption: calibration errors come from earlier snapshots, some share a fit, and drift can make them unrepresentative, so we report empirical coverage.

%% file: sections/app_estimator.tex
\section{The ablation estimator}\label{app:estimator}

\paragraph{Algorithm 1.} Given entry dates, groups, event and competing-exit times observed by $T$, covariates $z_g(t)$, a horizon $H$ and a \emph{calendar path} (assumed $z$ after $T$): (1)~\emph{exact ages:} estimate incidence so far by Aalen--Johansen at each unit's own age rather than its cohort's; (2)~\emph{calendar step} (\cref{thm:calendar}): Poisson regression of event counts in (cohort, age, group) cells on cohort$\times$age and group fixed effects and $\beta^\top z_g(t)$, used only if the covariates vary enough between groups; (3)~\emph{speed step} (\cref{thm:id,thm:precision}): score equal speeds, a random walk on $\log s(c)$ and a recent change point against a generalised-gamma $\Lambda_0$ fitted to the mature cohorts, choose by marginal likelihood (log-likelihood divided by the estimated over-dispersion), and refit the chosen model with the pooled shape of the mature cohorts; (4)~\emph{plug-in:} complete each unit from its age to $H$, with its cohort's observed (shock-adjusted) increments up to the cohort's median age and the fitted hazards along the calendar path after it; (5)~\emph{intervals}, calibrated as in \cref{sec:limits}. The speed step resembles nowcasting with time-varying delays \citep{hohle2014,mcgough2020}. All choices were fixed before the mortgage backtest, except the post hoc changes in \cref{tab:deviations}.

\paragraph{Calendar step.} The Poisson regression uses cells of (cohort, age month, group), cohort$\times$age and group fixed effects, the covariates $z_g(t)$ at the cell's calendar month, and log exposure in sub-distribution risk sets as offset. Under random censoring, competing exits keep a weight equal to the ratio of censoring survival functions \citep{fine1999}. The coefficients are unpenalised and their standard errors are clustered by group. The step is used only if the number of events times the smallest identifying-variation share is at least 100, a threshold calibrated in simulation before the backtests; a covariate's share is its event-weighted variance after the cohort$\times$age and group fixed effects, divided by its event-weighted total variance, over the cells with events. A second flag, which does not switch anything, reports the share of expected future events at covariate values outside the observed range.

\paragraph{Speed step.} A generalised-gamma distribution function (which contains the Weibull and gamma shapes, and the log-normal as a limit), fitted once per snapshot to the shock-adjusted cumulative hazards of the mature cohorts under equal speeds, is the common shape $\Lambda_0$ against which all candidate speed models are scored. The candidates are equal speeds; a Gaussian random walk on $\log s(c)$ with increment variance $\omega^2 \in \{0.001, 0.003, 0.01, 0.03, 0.1\}$, each value a separate candidate; and a change point, with jump $\delta_{\mathrm{cp}} \sim N(0, 0.5^2)$ shared by the cohorts from its location on, the location uniform over the immature cohorts among the last eight (on real data, vintages $T-6$ to $T$ and the six youngest filing years; the simulation of \cref{app:sims} uses all of the last eight). Each candidate's marginal likelihood is computed by a Laplace approximation under an over-dispersed Poisson model, with the log-likelihood divided by the Pearson dispersion $\hat\phi$ of the free-speed fit; the candidate with the largest value is refitted with the pooled empirical shape of the mature cohorts (aligned by their fitted speeds), and the plug-in uses this refitted shape. Without the division, likelihood-based selection at large $n$ prefers flexible models under mild misspecification \citep{grunwald2017}; the division corrects the variance, not a misspecified mean, and on mortgages the random walk was still selected at all 13 snapshots.

\paragraph{Plug-in.} The observed part of each cell's incidence comes from the Aalen--Johansen estimator (in the pre-registered run, from the share of units with the event). The remainder up to $H$ uses, from each unit's age to its cohort's median age, the cohort's observed shock-adjusted increments pooled over its units, and after that age the fitted sub-distribution hazards along the calendar path; cells are averaged with their entry weights. The \emph{forecast} path holds each covariate at its mean over the scored cohort's own cells in the 12 months before the snapshot (weighted by the units present each month; a group with no cell takes the mean over groups; the hazard models use the plain monthly mean); for patents, the \emph{Alice} indicator stays at its value at the snapshot. The frozen code holds $F_0$ at its horizon value for rescaled ages beyond $H$; for a cohort with fitted speed $s > 1$ the formula needs $F_0(sH)$. \Cref{app:corrected} reports the corrected version, which changes the real-data estimates of such cohorts by 0.2--1.2\% in the median and no conclusion.

\paragraph{Calibration errors.} An error is the log error of one quarterly cohort, estimated at an earlier year-end snapshot at an age within six months of the target's, whose horizon truth is complete at the current snapshot with two months to spare. An interval pools the errors from several earlier snapshots (for mortgages, a median of 23 errors from 6 snapshots). Errors from one snapshot share a fit, so the effective number of independent errors is smaller than $J$.

\paragraph{Baselines.} \emph{Events so far} is the cohort's incidence of the event observed by the snapshot: the Aalen--Johansen estimate, with the competing exit as a competing risk, at the median age of its units at the snapshot (in whole months). A cohort is mature at a snapshot when its median unit is $H$ months old. \emph{Drop recent cohorts} is the mean of the rates of the three most recent mature cohorts (twelve when cohorts are quarters), each taken at the scored cohort's horizon $h$; \emph{last mature cohort} is the same with the most recent mature cohort (four quarters). Here $h = H$, except in the pre-COVID mortgage set, where $h = \min(84, \text{age in February 2020})$ months, so both baselines target the scored estimand. In the window simulation, drop recent is the naive rate of the older half of the cohorts. The \emph{chain ladder} scales each cohort's cumulative incidence at its median age by the pooled development of the mature cohorts. \emph{+ exact ages} applies the same idea to cumulative hazards at each unit's own age, and \emph{+ groups} adds a level per group. The calendar-adjusted chain ladder adds the calendar step with equal speeds. The full method adds the speed step.

\paragraph{Error ratio.} For two methods, the error ratio is $\exp\{\overline{\log(|\mathrm{err}_1|/|\mathrm{err}_2|)}\}$ over cohort--snapshot pairs, with $10^{-6}$ added to each absolute relative error, as pre-specified (the smallest error is $7\times 10^{-5}$, so a floor would give the same values). Its interval resamples cohorts (vintages or filing years) with replacement, and the sign-flip test flips the signs of the cohort sums of the pair log ratios, exactly over all $2^C$ patterns (or $10^6$ random ones when $C > 20$).

\paragraph{Settings.} The calendar-step guard requires the number of events times the smallest share of identifying covariate variation to be at least 100; the threshold was set from simulation before the registered mortgage run (at a median statistic of 90, 85\% of coefficient estimates are within 25\% of the truth). Random-walk increment variances are chosen from $\{0.001, 0.003, 0.01, 0.03, 0.1\}$; the change point has prior standard deviation 0.5, at a location among the immature cohorts of the last eight; intervals require $J \ge 9$ calibration errors at the 90\% level. Seeds are fixed in every script; all runs used one laptop CPU.

%% file: sections/app_experiments.tex
\section{Additional experiments}\label{app:experiments}

All experiments run on CPUs: one 11-core laptop, with the largest runs taking 46 to 95 minutes and the end-to-end simulation about 4.5 CPU hours. An independent audit of the code behind \cref{tab:classifiers,tab:main}, \cref{fig:teaser,fig:speed} and \cref{tab:e1,tab:e2,tab:baselines,tab:inference,tab:alarms,tab:wrongdir,tab:classifiers-full,tab:patex,tab:patex-errors,tab:sim-e2e,tab:thm6check} (three independent reviews by AI agents, each given only the code, the written specifications and the paper) found that the recorded mortgage loan age restarts at a loan modification; all mortgage results are reported with ages counted from the first payment, and the mortgage analyses outside the audit (\cref{tab:harp-check,tab:speed-criterion,tab:inference-sens,tab:overid}, \cref{app:featmat}) were rerun the same way. The replication's code was audited in the same way (\cref{app:replication}).  The code that reproduces every table and figure is at \url{https://github.com/Vela-Research/drift-or-immaturity}.

\paragraph{Status of each analysis.} \Cref{sec:experiments} states what was pre-registered, and \cref{tab:deviations} lists every deviation with its effect. Three details add to that. The mortgage specification (estimand, evaluation sets, method, baselines, guard thresholds and primary comparison) was developed on the official 50{,}000-loan sample of the same vintages, so the registered run is new with respect to loans but not to calendar history. The patent protocol was written after a check of timing alone and before any backtest, with the frozen code, but entered version control together with the results, so its timing cannot be verified independently. The borrower-followed mortgage estimand, the end-to-end simulation and the speed criterion are post hoc.

\begin{table}[!htbp]
\centering\footnotesize
\caption{Deviations from the protocols and post hoc analyses.}
\label{tab:deviations}
\begin{tabular}{@{}p{1.4cm}p{3.0cm}p{3.6cm}p{1.9cm}p{4.2cm}@{}}
\toprule
Data & Protocol & Change & Found & Effect \\
\midrule
Mortgages & Truth censors HARP refinances & Censoring weights of the calendar step corrected for HARP & after the run & both reported; forecast path vs CL $p = 0.032$ (pre-registered weights), 0.012 (corrected) \\
Mortgages & Registered truth & Borrower-followed estimand added & after the run & 0.60 ($p = 0.026$); forecast 0.42 ($p = 0.007$) \\
Mortgages & Mark-to-market LTV listed as a sensitivity & Used in the equal-shape test & after the run & rejections 14--36\% to 0--29\% \\
Mortgages & Tests on pair files rounded to 6 decimals & All tests from full-precision values & post hoc & forecast path vs CL $p$ 0.032 from both \\
Both & Sign test and bootstrap & Exact cluster sign-flip test (primary); wild, two-way, leave-one-out, MAE ratio, floors & after the run & \cref{app:inference} \\
Both & Full method only & Step-by-step decomposition, exact ages & after the run & \cref{tab:main} \\
Both & Specified baselines & BF, Cape Cod, GCC, CL + exact ages + groups, hazard model, classifiers & post hoc & \cref{app:baselines,app:e5} \\
Both & Coverage with calibration rule ``median unit $H$ months old''; width upper/lower & Strict rule ($H$ + 2 months); log-width & post hoc & mortgages 39 of 40 to 38 of 39; patents 27 to 25 of 30; width ratio 0.45 to log-width ratio 0.54 \\
All & Upper interval limit $1 - G^{\mathrm{comp}}_c(A_c)$ & Code caps at 1; for $J < 9$ uses the largest error & from the start & none on coverage counts \\
All & Plug-in $F_0(sH)$ & Code holds $F_0$ at its horizon value beyond $H$; corrected rerun & post hoc & real data: estimates of cohorts with $s > 1$ change by 0.2--1.2\% (median); the mortgage primary comparison moves from $p = 0.045$ to 0.061 (\cref{app:corrected}) \\
Patents & Sign test on the realised path (primary) & Sign-flip $p$ in the text; forecast path reported alongside & after the run & Holm $p$ 0.023 (sign test) and 0.004 (sign-flip) \\
Patents & Inventory of all applications & Inventory without the scored cohorts' own applications & post hoc & 0.34 to 0.51 \\
Patents & Coverage over all pairs ($J \ge 1$) & Main text reports $J \ge 9$ & post hoc & 25 of 30 against 19 of 22 \\
Mortgages & Feature-maturity spec (pre-specified) & Timing by credit score computed for 2006--2010 before the spec; one snapshot run first as a code check; Step 0 extras & before the results & \cref{app:featmat} \\
SBA, civil & Replication, time-stamped & Six decisions on data structure (\cref{app:replication}); pending civil cases censored & before any method run & define the civil sample; H1 on loans uses 84 of 109 pairs \\
SBA & Hazard model on payoff dates as given & Fit only on months from 2005-08 & after the run & H3 on loans 0.91 to 1.15 on 79 pairs (\cref{app:replication}) \\
SBA, civil & Code as run & Six fixes from the audit of the replication code & after the run (audit) & no verdict changes (\cref{tab:rep-asrun}) \\
Civil & Interval evidence as run & Consistent random-walk evidence & after the run & $\Gamma$ for 90\% coverage 3 to 2 (\cref{app:replication}) \\
Civil & All case types kept & Without mass-tort cases & post hoc & same verdicts; exact-age step no longer helps (\cref{tab:rep-asrun}) \\
Mortgages & Loan age as recorded (restarts at a modification) & Age from the first-payment month; consistent random-walk evidence for quarterly calibration cohorts & after the run (audit) & every mortgage number (\cref{tab:mortgage-ages}); primary 0.81 ($p = 0.30$) to 0.56 (sign-flip $p = 0.045$; sign test $p = 0.27$) \\
Simulation & Change point at immature cohorts only (code) & Uniform over the last eight cohorts, as stated; consistent random-walk evidence; coverage only with $J \ge 9$ & after the run (audit) & cost of speeds when speeds are equal 2--10\% to 1--8\%; gain when they change 0.30--0.90 to 0.30--0.93; coverage after a speed change (200k) 75--88\% to 83--96\% \\
\bottomrule
\end{tabular}
\end{table}

\paragraph{Pre-specified outputs not shown elsewhere.} On the COVID-affected mortgage set, the full method beat the chain ladder at 4 of 7 snapshots (median difference 3.2 points). With $\omega^2$ chosen by rolling-origin validation instead of marginal likelihood, the full method's ratio to the chain ladder is 0.54 [0.28, 1.03], and equal speeds are chosen at 8 of 12 snapshots (and at 2006, where no rolling origin exists, by default). The pre-specified interval is the symmetric $\hat p\,e^{\pm \Gamma q}$ with $\Gamma = 1$, with $q$ the $\lceil (J+1)(1-\alpha) \rceil$-th smallest absolute calibration error; under this rule, coverage was 97.5\% (mortgages; width ratio 0.45) and 90.0\% (patents; 0.80). A signed version, $[\hat p\,e^{-q_{0.95}},\ \hat p\,e^{-q_{0.05}}]$ with $q_{0.05}$ and $q_{0.95}$ the empirical 5\% and 95\% quantiles of the signed calibration errors $\log(\text{estimate}/\text{truth})$ (no finite-sample correction, no widening), clipped to [events so far, 1], covers 47.5\% of the 40 mortgage pairs with $J \ge 9$ (47.1\% of the 51 with $J \ge 1$) at the nominal level of 90\% ($\alpha = 0.10$): the direction of past errors did not carry over to the scored pairs. The calendar-step guard counts events, not independent groups, which overstates the information when there are few groups, as for the eight technology centres.

\subsection{Simulations}\label{app:sims}

\paragraph{Windows.} The 972 settings cross Weibull shape (0.8, 1.5, 3), median time to event (3, 7, 12 years), the target's eventual rate (1\%, 4.5\%, 20\%), drift (none; $\pm 2\%$ of the target per year of cohort age; a step of $+30\%$ for all but the 5 youngest cohorts), the number of cohorts (10, 15, 25) and the age of the youngest cohort (1, 3, 8 years), each at four sample sizes with 500 replications; the target is the youngest cohort's eventual rate (no horizon). \Cref{tab:e1} gives all rules. The rule with the proven constants is conservative; the threshold multipliers $\kappa_N = \sqrt{2\log(N/\alpha)}$ (Bonferroni), $\kappa = 1$ and $\kappa$ chosen by rolling-origin validation behave alike. With no drift, the maturity-weighted rule keeps at least 90\% of the cohorts in 74\% of settings at $n = 10^3$ and in all settings at $10^6$. Wherever the youngest cohort's maturity is below 0.75 (864 of 972 settings), the naive Lepski window is shorter at $n = 10^6$ than at $10^3$. ADWIN in the window comparisons uses $\delta = 0.05$, split over the cohort comparisons ($\delta' = \delta/K$); the drift detector of \cref{app:e5} uses $\delta = 0.002$ with $\delta' = \delta/\ln n$ and a range term.

\begin{table}[!htbp]
\centering\small
\caption{Median relative error across 972 simulation settings, by units per cohort $n$. $\kappa_N = \sqrt{2\log(N/\alpha)}$, $\alpha = 0.05$, with $N$ the number of cohorts.}
\label{tab:e1}
\begin{tabular}{lcccc}
\toprule
Rule & $10^3$ & $10^4$ & $10^5$ & $10^6$ \\
\midrule
Naive Lepski ($\kappa = \kappa_N$) & 54\% & 65\% & 76\% & 80\% \\
Naive ADWIN & 52\% & 71\% & 79\% & 80\% \\
Naive full window & \multicolumn{4}{c}{30\% (pooled over $n$)} \\
Drop recent (older half) & \multicolumn{4}{c}{27\% (pooled over $n$)} \\
Maturity-weighted full window & \multicolumn{4}{c}{16\% (pooled over $n$)} \\
Maturity-weighted Lepski, proven constants & 16.5\% & 15.3\% & 10.1\% & 4.1\% \\
Maturity-weighted Lepski, $\kappa = \kappa_N$ & 18.2\% & 12.0\% & 5.9\% & 1.0\% \\
Maturity-weighted Lepski, $\kappa = 1$ & 15.1\% & 6.1\% & 2.2\% & 0.7\% \\
Maturity-weighted Lepski, rolling-origin $\kappa$ & 17.4\% & 6.9\% & 2.3\% & 0.7\% \\
Maturity-weighted Lepski, estimated timing ($\kappa = \kappa_N$) & 19.2\% & 12.1\% & 5.9\% & 1.1\% \\
Oracle fixed window & 7.5\% & 3.1\% & 1.2\% & 0.4\% \\
\addlinespace
\multicolumn{5}{@{}l}{\emph{Speeds differ across cohorts (maturity-weighted Lepski, $\kappa = \kappa_N$)}} \\
Faster young cohorts, own maturities & 16.7\% & 10.8\% & 5.1\% & 1.0\% \\
Faster young cohorts, one timing curve & 25--27\% & 29\% & 37\% & 50\% \\
Faster young cohorts, chain-ladder maturities & 23.3\% & 21.8\% & 25.5\% & 28.2\% \\
Slower young cohorts, own maturities & 20.3\% & 13.6\% & 7.2\% & 1.2\% \\
Slower young cohorts, one timing curve & 20--21\% & 23\% & 30\% & 36\% \\
Slower young cohorts, chain-ladder maturities & 20.0\% & 18.6\% & 24.2\% & 29.6\% \\
\multicolumn{5}{@{}l}{\emph{Maturities from the frozen speed model (reduced grid, 200 replications)}} \\
Equal speeds & 23.0\% & 18.3\% & 9.3\% & 3.5\% \\
Faster young cohorts & 24.1\% & 26.4\% & 34.1\% & 15.1\% \\
Slower young cohorts & 27.1\% & 19.7\% & 32.0\% & 30.6\% \\
\bottomrule
\end{tabular}
\end{table}

\begin{figure}[!htbp]
\centering
\includegraphics[width=0.6\linewidth]{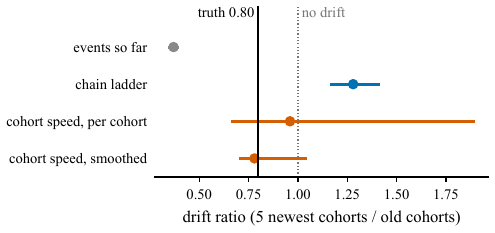}
\caption{Estimated drift ratio of eventual rates (5 newest cohorts over the 9 cohorts with at least 12 years of follow-up), median and 5--95\% range over 100 replications; truth 0.80.}
\label{fig:speed-sim}
\end{figure}

\paragraph{Cohort speed and the chain ladder.} With 20 cohorts of 20{,}000 units, recent cohorts that speed up while their eventual rates fall by 20\% lead the chain ladder to report an increase (\cref{fig:speed-sim}). Timing is Weibull (shape 1.5, median 4.6 years); speeds are 1 for cohorts followed at least 12 years and rise linearly to 1.6 for the youngest, whose eventual rate falls linearly from 4.0\% to 3.0\%. This simulation uses a speed-aware estimator with log-linearly smoothed speeds, a simpler smoother than that of Algorithm 1 (\cref{app:estimator}); the per-cohort speed estimator implied directly by \cref{thm:id} is nearly uninformative at these maturities, as \cref{thm:precision} predicts.

\paragraph{Calendar shocks.} With 15 cohorts and 10 groups hit by a crisis with group-specific intensity (100 replications), the shock coefficient is recovered (median 0.499, 5--95\%: 0.47--0.53, truth 0.5). The calendar-adjusted estimator gives a drift ratio of 0.80 against a truth of 0.755; the gap comes from the log-linear speed smoothing of this simulation, and with the true speeds the ratio is 0.76. Calendar-free estimators see spurious drift (0.42), and the chain ladder reverses its direction (1.38). With a single national shock series the coefficient is not identified (median $-3.7$), as \cref{thm:calendar} states.

\paragraph{Model choice.} With truly equal speeds and mild over-dispersion, untempered marginal likelihood selected equal speeds in only 1 of 8 replications, while the tempered version used by the estimator selected them in all 8 and still detected a real speed jump.

\paragraph{End-to-end simulation.} Monthly cohorts: 8 cohort years of 12 entry months, 6 groups, a random-walk level (incidence at $H = 36$ months about 6\%) and two calendar covariates; the truth is the realised incidence at $H$. Factors: speeds constant, a step (log 1.4, random sign, at one of the three youngest cohorts) or a drift (log 1.5 over the five youngest); calendar effect on or off; 20{,}160 or 201{,}600 units per cohort year; Weibull (shape 1.5) or log-normal ($\sigma = 0.8$) timing. The scored ages are 6, 18 and 30 months (Weibull maturities 0.09, 0.43, 0.82), with four earlier snapshots for calibration; 7{,}200 replications in all. The speed model's change point may start at any of the last eight cohorts (re-run, post hoc; \cref{tab:deviations}). \Cref{tab:sim-e2e} reports the frozen and the corrected plug-in on the same seeds; the frozen defect is large here because $F_0(H) = 0.5$. With the corrected plug-in:
\begin{itemize}[leftmargin=*,itemsep=1pt,topsep=2pt]
\item \emph{Speeds} help in every setting where they change and cost 1--8\% where they do not. With Weibull timing the youngest cohort's log speed is not learned (median ratio of fitted to true log speed 0 at age 6 months; 0.53 at 18 and 0.55 at 30 months for a step at 200{,}000 units), so its gain comes from smoothing towards older cohorts' speeds; with log-normal timing it is learned (0.54--0.81 for the youngest cohort after a step; 0.33--0.52 under drift).
\item \emph{Direction.} The full method recovers the sign of year-to-year changes of at least 5\% in 94--100\% of cases with Weibull timing and 89--98\% with log-normal timing, against 77--81\% for the chain ladder and 59--63\% for events so far (Weibull, calendar effect on).
\item \emph{Criterion.} With the true information and the true squared log speed in place of $\omega^2 + \tau^2$ (an oracle version of the condition, $(\log s_c)^2 > 1/(D_c V_c)$), \cref{prop:shrink} predicts both timing shapes: where it holds, a free speed has error ratio 0.77 (Weibull) and 0.36 (log-normal) against equal speeds, and 1.12 and 1.30 where it fails.
\item \emph{Intervals.} After a speed change in large cohorts, coverage at $\Gamma = 1$ ranges from 83\% to 96\%, below 90\% in some settings (at the youngest age every pair has 6 calibration errors and so gets no interval under the rule); a larger $\Gamma$ restores at least 90\% ($\Gamma = 1.25$--1.5 with Weibull timing), and with constant speeds $\Gamma = 1$ suffices.
\end{itemize}

\begin{table}[!htbp]
\centering\small
\caption{End-to-end simulation: error ratio of the full method to equal speeds (calendar-adjusted chain ladder) on the realised path, frozen and corrected plug-in, and coverage of the corrected nominal 90\% intervals at $\Gamma = 1$. Replications: 500 (Weibull; both calendar settings) and 200 (log-normal) for the error ratios; 500, 80 and 40 with intervals. Coverage: pairs with at least 9 calibration errors (ages 18 and 30 months). Re-run with the change point uniform over the last eight cohorts (\cref{tab:deviations}).}
\label{tab:sim-e2e}
\begin{tabular}{@{}llcccccc@{}}
\toprule
 & & \multicolumn{2}{c}{frozen} & \multicolumn{2}{c}{corrected} & \multicolumn{2}{c}{coverage} \\
\cmidrule(lr){3-4}\cmidrule(lr){5-6}\cmidrule(l){7-8}
Timing, calendar effect & speeds & 20k & 200k & 20k & 200k & 20k & 200k \\
\midrule
Weibull, on & constant & 1.09 & 1.20 & 1.03 & 1.08 & 98\% & 98\% \\
 & step & 1.03 & 0.92 & 0.93 & 0.66 & 95\% & 86\% \\
 & drift & 0.99 & 0.85 & 0.91 & 0.64 & 96\% & 85\% \\
Weibull, off & constant & 1.09 & 1.11 & 1.04 & 1.05 & 100\% & 99\% \\
 & step & 1.00 & 0.84 & 0.90 & 0.65 & 96\% & 83\% \\
 & drift & 0.96 & 0.84 & 0.90 & 0.60 & 97\% & 84\% \\
Log-normal, on & constant & 1.02 & 1.07 & 1.01 & 1.05 & 100\% & 99\% \\
 & step & 0.68 & 0.40 & 0.56 & 0.31 & 96\% & 96\% \\
 & drift & 0.73 & 0.54 & 0.52 & 0.30 & 95\% & 85\% \\
\bottomrule
\end{tabular}
\end{table}

\paragraph{Direct check of \cref{thm:precision}.}\label{app:thm6check} A single cohort with $F_0$ known ($H = 84$ months, median 60 months), Poisson event counts in monthly bins, and maximum likelihood for $(\log\theta, \log s)$ over speeds in $[1/4, 4]$; 4{,}000 replications per cell (\cref{tab:thm6check}). With the speed known, $D_c\mathrm{Var}(\log\hat p_c)$ is 0.94--1.12, as theory says. With log-normal timing, $D_c\mathrm{Var}(\log\hat p_c)/I_c$ is 0.96--1.15 in all 25 cells. With Weibull timing the asymptotic variance is reached at maturity 0.69 from about 100--300 events, at 0.4 from about 1{,}000 and at 0.2 only with about 3{,}000; at maturity 0.1 or less the speed estimate sits at the boundary of its range in 17--84\% of replications, so the variance is capped by the range rather than by $I_c$, and the median absolute error of $\log\hat p_c$ is 0.27--0.67 for every $D_c$ up to 3{,}000. This is the weak-identification regime that part (iv) describes.

\begin{table}[!htbp]
\centering\small
\caption{Direct check of \cref{thm:precision}: $D_c\mathrm{Var}(\log\hat p_c)/I_c$ and, in brackets, the share of speed estimates at the boundary of $[1/4, 4]$, by timing shape, maturity and expected events so far $D_c$.}
\label{tab:thm6check}
\resizebox{\linewidth}{!}{\begin{tabular}{@{}llcccccc@{}}
\toprule
Timing & maturity & $I_c$ & $D_c = 30$ & 100 & 300 & 1{,}000 & 3{,}000 \\
\midrule
Weibull 1.5 & 0.05 & 2179 & 0.01 (0.84) & 0.04 (0.71) & 0.09 (0.56) & 0.20 (0.41) & 0.35 (0.32) \\
 & 0.10 & 463 & 0.05 (0.69) & 0.12 (0.51) & 0.23 (0.40) & 0.41 (0.29) & 0.66 (0.17) \\
 & 0.20 & 93 & 0.14 (0.48) & 0.27 (0.35) & 0.45 (0.27) & 0.73 (0.13) & 0.96 (0.02) \\
 & 0.40 & 12.5 & 0.36 (0.33) & 0.60 (0.21) & 0.84 (0.08) & 1.01 (0.00) & 1.01 (0.00) \\
 & 0.69 & 2.0 & 0.81 (0.19) & 0.95 (0.05) & 0.96 (0.00) & 1.01 (0.00) & 1.00 (0.00) \\
Log-normal 0.8 & 0.05 & 23.3 & 1.15 (0.01) & 1.03 & 1.01 & 1.01 & 0.96 \\
 & 0.10 & 14.1 & 1.09 & 1.07 & 1.02 & 0.98 & 0.98 \\
 & 0.21 & 6.9 & 1.13 & 1.03 & 1.03 & 1.00 & 1.01 \\
 & 0.39 & 2.9 & 1.11 & 1.05 & 1.03 & 1.04 & 1.02 \\
 & 0.70 & 1.3 & 1.04 & 1.00 & 1.04 & 0.99 & 0.97 \\
\bottomrule
\end{tabular}}
\end{table}

\subsection{Baselines}\label{app:baselines}

\Cref{tab:baselines} adds every baseline on the same pairs, truth and scoring.

\emph{Old-cohort baselines.} Bornhuetter--Ferguson uses the last mature cohort's horizon rate as prior; Cape Cod pools all cohorts weighted by their development; Generalised Cape Cod uses decay 0.75 per cohort year; all use the chain ladder's timing curve on the incidence scale. They fail on mortgages, because an old cohort's rate is a poor prior after 2008, and do well on patents: the full method against Bornhuetter--Ferguson gives 0.91 ($p = 0.78$; forecast path 0.95, $p = 0.87$), against Cape Cod 0.79 ($p = 0.38$; forecast path 0.82, $p = 0.48$; median error 6.9\% against 8.5\%), and against Generalized Cape Cod 0.51 ($p = 0.015$).

\emph{Mortgage hazard model.} One proportional hazard per cause (default, prepayment) at each snapshot on the loan-months the snapshot shows: Poisson regression on cells (first-payment month $\times$ state $\times$ credit-score band $\times$ loan-to-value band $\times$ age) with a free monthly age baseline, state and vintage effects, loan features (credit score, loan-to-value, debt-to-income, rate spread, purpose, occupancy), and by state and month house-price growth, unemployment, log mark-to-market loan-to-value and the refinance incentive (note rate minus the market rate) with its positive part. Its estimate is the incidence seen at the snapshot plus the survival times the product integral of the two hazards to $H$. Its vintage effects matter: without them it still beats the chain ladder (realised 0.40 [0.27, 0.65], $p = 0.005$; forecast 0.70, $p = 0.029$) but not the full method (0.79, $p = 0.19$; 1.57, $p = 0.058$). With realised covariates it beats the full method at 12 of 13 snapshots; on the forecast path it is worse in 2009, 2011 and 2012, when it carries the house-price fall forward, and better in the other years.

\begin{table}[!htbp]
\centering\small
\caption{All baselines on the same pairs and truth (post hoc): median over snapshots of the mean absolute relative error; error ratio to the chain ladder with a 95\% cohort-cluster bootstrap interval; exact cluster sign-flip $p$. Mortgages with corrected censoring weights.}
\label{tab:baselines}
\resizebox{\linewidth}{!}{\input{tables/baselines.tex}}
\end{table}

\subsection{Inference}\label{app:inference}

The test procedure (\cref{tab:inference}), adopted after the runs (\cref{tab:deviations}), computes for each comparison the mean over pairs of the log ratio of absolute errors ($10^{-6}$ added); its primary test is the exact cluster sign-flip test over all $2^C$ sign vectors of the cohort sums ($C$ = 18 mortgage and 13 patent cohorts). As robustness it reports the cohort-cluster bootstrap ($B = 2{,}000$), a wild cluster bootstrap \citep{cameron2008}, a two-way bootstrap over cohorts and snapshots, and leave-one-snapshot-out ranges. Wild and sign-flip $p$-values agree. Two-way intervals are wider. For mortgages they include 1 for the full method against the chain ladder on the realised path in every version, and on the forecast path exclude it with the corrected censoring weights and with borrowers followed; on the forecast path they exclude 1 for the calendar step and for the full method against the chain ladder with exact ages and groups only with the corrected weights. For patents they exclude 1 for the full method against the chain ladder on both paths, for the exact-age step, and for the calendar step on the forecast path. \Cref{tab:inference-sens} gives the ratio of mean absolute errors, which favours the full method more on mortgages (0.34, $p = 0.002$, corrected weights, realised path), and the log ratio at larger error floors and with 5\% trimming. The calendar step helps in every version. The speed step is not significant by the log ratio in any version, but on the mortgage forecast path it is by the ratio of mean absolute errors (0.69--0.79, $p \le 0.013$); on mortgages exact ages also look better by that ratio (0.84--0.85) than by the log ratio (0.97--1.01), and $p$-values near 0.05 move to either side.

\begin{table}[!htbp]
\centering\small
\caption{The test procedure adopted after the runs: ratio with cohort-cluster interval, exact cluster sign-flip $p$ (primary), wild cluster bootstrap $p$, two-way (cohort$\times$snapshot) interval, range of the ratio and largest $p$ when one snapshot is left out.}
\label{tab:inference}
\resizebox{\linewidth}{!}{\input{tables/inference.tex}}
\end{table}

\begin{table}[!htbp]
\centering\small
\caption{Secondary metric and sensitivity: ratio of mean absolute errors with its $p$; error floors $10^{-4}$ and $10^{-3}$; 5\% trimming ($p$ in brackets).}
\label{tab:inference-sens}
\resizebox{\linewidth}{!}{\input{tables/inference_sensitivity.tex}}
\end{table}

\subsection{The criterion for cohort speeds}\label{app:speedcrit}

For each scored pair, $D$ is the cohort's events so far and $V$ the variance of $\eta$ over them, from the generalised gamma fitted to the reference; the own-speed variance is $\hat\phi/(DV)$ with the over-dispersion $\hat\phi$ of the selection step; $\omega^2$ is the increment variance of the selected random walk, and $\tau^2$ the posterior variance of the older neighbour's log speed. The condition of \cref{prop:shrink} holds for 56\% of patent and 44\% of mortgage pairs, and its share rises with maturity: 0\% below maturity 0.15 (median speed inflation factors 88 and $2.6\times10^6$), 100\% and 90\% above 0.7. The youngest scored cohort fails it at every snapshot (inflation factors 35--275 on patents; on mortgages 63--286 at 2006--2011 and above $2\times10^5$ at 2012--2018, median $2.0\times10^5$, because the fitted onset is nearly a power of age and a 4--6-month vintage has almost no defaults). At the patent snapshots 2011 and 2012 it is the only scored cohort to fail, and the method moved its speed to 1.31, against 0.93 and 1.02 for its neighbour; its estimate was 34\% and 40\% against truths of 66\% and 67\% (with the corrected plug-in, 35\% and 42\%). At the same time the fitted reference shape moved from near its log-normal limit (2007--2009) towards a Weibull-like onset, so early grants came sooner without a general speed-up.

The condition does not, however, separate the pairs where speeds help from those where they hurt (patents: error ratio 0.83 where it holds and where it fails; with the corrected plug-in 0.99 and 0.77; rank correlation of the gain with the margin 0.13--0.19): the youngest cohorts at 2013--2017 fail it, yet their speeds of 1.12--1.37 were right and helped. It says where a speed is noise-limited, not whether moving it helps, and it depends on $\hat\phi$: without it, the condition holds for every patent pair. \Cref{tab:speed-criterion} gives the snapshot-level results.

\begin{table}[!htbp]
\centering\small
\caption{Per snapshot: selected speed model, over-dispersion, share of scored cohorts that meet the condition of \cref{prop:shrink}, error ratio of the full method to the calendar-adjusted chain ladder on both paths, and the fitted generalized-gamma shape ($a$ and the early exponent $ac$).}
\label{tab:speed-criterion}
\resizebox{\linewidth}{!}{\input{tables/speed_criterion.tex}}
\end{table}

\subsection{Corrected plug-in}\label{app:corrected}

For a cohort with fitted speed $s > 1$ the plug-in needs $F_0(sH)$; the frozen code holds $F_0$ at its horizon value beyond $H$. The corrected version continues the pooled $F_0$ beyond $H$ by the fitted generalised gamma and changes nothing else. On the real data the estimates of cohorts with $s > 1$ rise by 1.2\% (patents) and 0.2--0.3\% (mortgages) in the median (at most 9\% on patents and 6--9\% on mortgages), and the error ratios move by at most 0.06, but the mortgage primary comparison (pre-registered weights) moves from $p = 0.045$ to 0.061 (\cref{tab:corrected}); all these $p$-values are post hoc. Real-data interval coverage is the same for both versions: at $\Gamma = 1, 1.25, 1.5, 2, 3$ the mortgage intervals (realised path) cover 38, 39, 39, 39 and 39 of 39 pairs with the pre-registered weights and all 39 with the corrected ones, and the patent intervals 19, 20, 21, 22 and 22 of 22, so $\Gamma = 1$ (1.25 for patents on the realised path) already gives 90\%; since $\Gamma$ is read off the evaluated pairs, this is descriptive.

\begin{table}[!htbp]
\centering\small
\caption{Frozen and corrected plug-in on the real data: error ratio (sign-flip $p$).}
\label{tab:corrected}
\resizebox{\linewidth}{!}{\begin{tabular}{@{}lcccc@{}}
\toprule
Comparison & Patents & Mortgages, pre-registered weights & corrected weights & borrower-followed \\
\midrule
Full method vs CL & 0.34 (0.002) $\to$ 0.36 (0.004) & 0.56 (0.045) $\to$ 0.58 (0.061) & 0.51 (0.020) $\to$ 0.53 (0.031) & 0.60 (0.026) $\to$ 0.63 (0.039) \\
Forecast path vs CL & 0.35 (0.004) $\to$ 0.36 (0.007) & 0.53 (0.032) $\to$ 0.52 (0.024) & 0.44 (0.012) $\to$ 0.42 (0.008) & 0.42 (0.007) $\to$ 0.41 (0.004) \\
Speed step, realised path & 0.83 (0.49) $\to$ 0.88 (0.62) & 1.00 (0.99) $\to$ 1.04 (0.78) & 1.04 (0.73) $\to$ 1.09 (0.48) & 1.13 (0.43) $\to$ 1.17 (0.27) \\
Full vs Cape Cod & 0.79 (0.38) $\to$ 0.83 (0.45) &  & 0.18 $\to$ 0.19 & 0.21 $\to$ 0.22 \\
Full vs hazard model, realised &  & 1.72 (0.005) $\to$ 1.78 (0.002) & 1.57 (0.005) $\to$ 1.63 (0.001) & 1.53 (0.013) $\to$ 1.59 (0.003) \\
Full vs hazard model, forecast &  & 1.10 (0.45) $\to$ 1.06 (0.63) & 0.91 (0.82) $\to$ 0.87 (0.67) & 0.69 (0.14) $\to$ 0.67 (0.099) \\
\bottomrule
\end{tabular}}
\end{table}

\subsection{Mortgages}\label{app:freddie}

\begin{figure}[!htbp]
\centering
\includegraphics[width=0.9\linewidth]{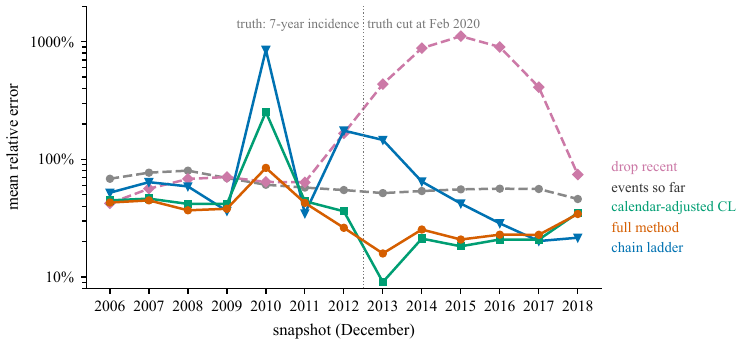}
\caption{Mortgages (pre-registered weights, corrected loan ages): mean absolute relative error of the six youngest vintages at each snapshot (log scale). The vertical line separates the crisis set (7-year truth) from the pre-COVID set (truth cut at February 2020).}
\label{fig:freddie}
\end{figure}

\Cref{tab:freddie} gives the pre-specified comparisons in four versions: as pre-registered; with corrected censoring weights (post hoc); with HARP refinances counted as prepayments (robustness; this needs no censoring assumption and ignores the borrowers' later defaults); and following borrowers into their relief loans (the exact estimand; post hoc). \Cref{fig:freddie} gives the errors by snapshot. On the forecast path the full method beats the chain ladder in every version, at 9 to 11 of the 13 snapshots; with HARP refinances counted as prepayments, the sign-flip $p$ is 0.022 (forecast path 0.014). Exact ages do not help on mortgages, so the plain chain ladder is the stronger reference there (\cref{tab:inference} gives the steps and further comparisons).

\emph{Loan ages.} The recorded loan age restarts when a loan is modified, which put later defaults inside the snapshot and raised the 7-year truth by 1--4\%. Every mortgage result in the paper counts ages from the first-payment month; \cref{tab:mortgage-ages} sets the key results beside the run as first done.

\begin{table}[!htbp]
\centering\footnotesize
\caption{Mortgages: key results with the loan ages as recorded (as run) and with ages counted from the first payment (post hoc). The recorded loan age restarts when a loan is modified; this put later defaults inside the snapshot (a leak) and raised the 7-year truth by 1--4\%. The as-run entry of the pre-registered primary is its registered verdict (not confirmed); every other entry is post hoc, and all $p$-values come from the sign-flip test adopted after the run.}
\label{tab:mortgage-ages}
\resizebox{\linewidth}{!}{\begin{tabular}{@{}lll@{}}
\toprule
 & as run (loan ages as recorded) & ages since the first payment \\
\midrule
Chain ladder, median error (crisis / pre-COVID) & 44\% / 28\% & 59\% / 35\% \\
Pre-registered primary: full method vs CL, realised path & 0.81 [0.56, 1.17], $p = 0.303$ & 0.56 [0.34, 0.93], $p = 0.045$ \\
Full method vs CL, realised path & 0.74 [0.54, 1.05], $p = 0.130$ & 0.51 [0.32, 0.83], $p = 0.020$ \\
Full method vs CL, forecast path & 0.73 [0.53, 0.98], $p = 0.065$ & 0.44 [0.28, 0.72], $p = 0.012$ \\
Calendar step (calendar-adjusted CL vs CL), forecast & 0.68 [0.53, 0.87], $p = 0.010$ & 0.54 [0.39, 0.80], $p = 0.013$ \\
Speed step (full method vs calendar-adjusted CL), forecast & 1.06 [0.86, 1.27], $p = 0.565$ & 0.83 [0.62, 1.10], $p = 0.251$ \\
Hazard model vs full method, realised & 0.45 [0.36, 0.61], $p < 0.001$ & 0.64 [0.51, 0.83], $p = 0.005$ \\
Hazard model vs full method, forecast & 0.80 [0.48, 1.26], $p = 0.386$ & 1.10 [0.77, 1.69], $p = 0.815$ \\
Classifier level error 2006--12: events so far / PU, full-method maturity & 68\% / 35\% & 66\% / 54\% \\
Interval coverage, $J \ge 9$, $\Gamma = 1$ (realised path) & 36 of 39 & 39 of 39 \\
Single speed rejected, vintages 1999--2012 / 2013--2019 & 29\% / 29\% & 36\% / 14\% \\
HARP relief loans' default ratio, 60 months & 0.32 & 0.29 \\
\bottomrule
\end{tabular}}
\end{table}

\emph{HARP censoring.} Borrowers who refinanced into HARP default at 0.29 times the rate of comparable loans still at risk (\cref{tab:harp-check}), so censoring them is informative and the registered truth overstates defaults for the 2005--2009 vintages. The difference mixes selection (a HARP refinance requires a current loan) and the effect of the programme.

\emph{Timing.} After accounting for over-dispersion, the test of \cref{app:test}, split at 36 months, rejects a single speed for 36\% of the 1999--2012 vintages and 14\% of the 2013--2019 vintages (\cref{tab:overid}); every rejected vintage (2005--2009, 2018) has faster late defaults, and these vintages' later years overlap the housing crisis or COVID-19. Adding mark-to-market LTV as a third covariate (post hoc) changes the rejections to 0--7\% (1999--2012) and 14--29\% (2013--2019). The front-loading of the 2004--2008 vintages that appears in cause-specific risk sets is largely an artefact of prepayment waves; this check used the development sample.

\emph{Windows on real data.} \Cref{tab:e2} gives the windows and errors of all rules on mortgages.

\emph{Intervals.} With the corrected ages and the corrected censoring weights, no mortgage interval misses the truth (with the pre-registered weights, one does at $\Gamma = 1$; \cref{app:corrected}). As first run, the three misses were pre-COVID pairs whose target is cut before 84 months, where the 84-month calibration errors do not apply. The full method's intervals have a median log-width 0.53--0.56 times that of chain-ladder intervals built by the same rule. The guards: the calendar step was on at every snapshot of every data set (events times identifying-variation share 2{,}346--12{,}930 for mortgages, against the threshold 100); the share of expected future events at covariate values outside the observed range was 49\%, 23\% and 9\% at 2006--2008 and 10--31\% at 2016--2018 on the realised path, and 0 on the forecast path.

\begin{table}[!htbp]
\centering\small
\caption{Mortgages: pre-specified comparisons (13 snapshots; 78 vintage--snapshot pairs). The calendar-adjusted chain ladder is the full method with equal speeds. Error ratios with 95\% vintage-cluster bootstrap intervals of the log ratio. Sign tests over snapshots are optimistic because a vintage is scored at several snapshots.}
\label{tab:freddie}
\resizebox{\linewidth}{!}{\input{tables/freddie_tests.tex}}
\end{table}

\begin{table}[!htbp]
\centering\small
\caption{HARP relief loans: default incidence after the refinance against the incidence, from the parent's age at refinance, of the loans of the parent's vintage, state and LTV band still at risk. Independent censoring implies a ratio of 1.}
\label{tab:harp-check}
\input{tables/harp_check.tex}
\end{table}

\begin{table}[!htbp]
\centering\small
\caption{Test of a single speed per vintage (split at 36 months; over-dispersion-adjusted quasi-likelihood ratio, $\chi^2_1$). $^{*}$: rejected at 5\%.}
\label{tab:overid}
\resizebox{0.85\linewidth}{!}{\input{tables/overid.tex}}
\end{table}

\begin{table}[!htbp]
\centering\small
\caption{Adaptive windows on mortgages (13 snapshots; target: the 7-year default rate of the vintage of the preceding year). Windows are means and errors medians over the snapshot--subsample runs (10 subsamples at 10k and 50k loans per vintage, one sample at 200k). $\kappa_N = \sqrt{2\log(N/\alpha)}$, $\alpha = 0.05$.}
\label{tab:e2}
\resizebox{0.9\linewidth}{!}{\input{tables/e2.tex}}
\end{table}

\subsection{Loan-level prediction}\label{app:e5}

\paragraph{Classifiers.} At each of the 13 December snapshots we train logistic regression and gradient-boosted trees on 50{,}000 loans per vintage as the snapshot shows them (origination features and a vintage term), and score 50{,}000 other loans of each of the six youngest vintages against default within 84 months (corrected censoring weights). The training labels are: events so far; mature cohorts only; drop recent; the delayed-feedback model of \citet{chapelle2014} (logistic only); importance weights in the style of ES-DFM \citep{yang2021esdfm}; and positive--unlabelled correction in EM form \citep{bekker2019} (two EM iterations from the provisional-label fit, without a convergence test; the largest mean change of a predicted probability in the last iteration is 0.006) with the labelling probability of each origination month taken from the chain ladder, the calendar-adjusted chain ladder or the full method, all on the forecast path. Final labels, not available at the snapshot, are the reference. \Cref{tab:classifiers-full} gives both models. Three chain-ladder variants appear here: the positive--unlabelled labels use a yearly chain ladder; the alarm streams use a monthly link-ratio chain ladder, which the calendar-adjusted and full-method streams also use where those methods give no estimate; and the backtests use the cohort chain ladder of \cref{app:estimator}.

\paragraph{Classifier results.} Against events so far, the positive--unlabelled correction with the full method's maturity cuts the level error to 0.32 (logistic) and 0.38 (boosted trees) times as a geometric mean over the 78 pairs ($p < 0.001$, 12 of 13 snapshots; ratio of mean absolute errors 0.64 and 0.70, of medians over snapshots 0.43); against mature cohorts only, to 0.26 and 0.26; against drop recent, to 0.29 and 0.37; and against ES-DFM-style weights, to 0.48 and 0.51. By the geometric mean over pairs, the full method's maturity is better than the chain ladder's (0.65, $p = 0.021$, logistic; 0.71, $p = 0.097$, boosted trees), although the chain ladder's has the lower median error in the crisis (\cref{tab:classifiers}) and not significantly different from the delayed-feedback model (0.85, $p = 0.57$), and all corrections remain 1.3--1.6 times too low at the crisis snapshots. Following borrowers into their relief loans gives the same picture (0.30 against events so far, $p < 0.001$; against the delayed-feedback model 0.90, $p = 0.64$). Over the six youngest vintages pooled, the AUC rises in the median, by 0.017 with the full method's maturity (range $-0.019$ to 0.20) and 0.051 with the chain ladder's (range 0.006--0.26); within one vintage it changes by at most 0.001, so the gain is in ranking vintages. The ranking distortion of \cref{prop:ranking} stays, because a maturity per origination month cannot depend on the features: the ratio of the level ratios of credit scores 720 and above to those below 660 is 0.59 (events so far), 0.59--0.68 (positive--unlabelled), 0.75 (delayed-feedback model) and 0.93--0.94 (final labels), pooled over the six test vintages (per vintage--snapshot pair, events so far has median 0.82 in the crisis and 1.02 after 2012).

\paragraph{Loan-month hazard classifier (post hoc).} A multinomial logistic regression on the loan-months the snapshot shows, with outcome default, prepayment or none, monthly age effects (ages 1--84), vintage effects and the same origination features; a loan leaves the risk set at its first event or at the end of its follow-up; months without an event are subsampled (3.6--15\%) with inverse weights; and the prediction is $\sum_a S(a-1)h_{\mathrm{default}}(a)$ with $S(a) = \prod (1 - h_{\mathrm{default}} - h_{\mathrm{prepay}})$. Its level error is 71\% (2006--2012) and 144\% (2013--2018), with AUC 0.809 and 0.782. Its error ratio is 1.09 [0.68, 1.69] against events so far ($p = 0.73$), 3.39 [1.84, 5.77] against positive--unlabelled correction with the full method's maturity ($p = 0.003$) and 2.89 against the delayed-feedback model ($p = 0.008$). It is too low early in the crisis and far too high from 2013, because its single age profile is set by the crisis vintages; unlike the backtest hazard model of \cref{app:baselines}, it has no calendar covariates.

\paragraph{Drift detectors.} ADWIN ($\delta = 0.002$) on the monthly default rate of each snapshot ends with a window whose mean is 0.31 of the true rate for events so far and 0.92 for the full method. \Cref{tab:alarms} counts its ``rate fell'' alarms against the 89 raised on the stream of true outcomes, under two rules for matching alarms: corrections cut false alarms and miss more true ones. Page--Hinkley \citep{page1954}, two-sided on the monthly rate weighted by the month's loans, with $\delta = 0.005$ and threshold 50 and restarted after each alarm (the replication uses the same settings), does not separate the streams in false alarms; on monthly snapshots it detects the 2005--2008 rise in September 2006 on the true outcomes, December 2007 with chain-ladder labels and December 2008 with events so far and with the calendar-adjusted and full-method labels.

\begin{table}[!htbp]
\centering\small
\caption{ADWIN ``rate fell'' alarms on the monthly default rate over the December snapshots, against the 89 alarms raised on the true outcomes. Any match: an alarm is confirmed if a true alarm in the same direction and snapshot lies within six origination months, so several alarms may match one true alarm. One-to-one: each alarm and each true alarm is matched at most once. A month enters a stream with at least 24 months of age and 2{,}000 loans; only alarms at origination months younger than 84 months count, also among the 89 true alarms; the one-to-one rule uses the same six-month tolerance.}
\label{tab:alarms}
\begin{tabular}{lcccccc}
\toprule
 & \multicolumn{3}{c}{Any match} & \multicolumn{3}{c}{One-to-one} \\
\cmidrule(lr){2-4}\cmidrule(l){5-7}
Training labels & alarms & false & missed & matched & false & missed \\
\midrule
Events so far & 120 & 15 & 0 & 88 & 32 & 1 \\
Chain ladder & 78 & 3 & 6 & 69 & 9 & 20 \\
Calendar-adjusted chain ladder & 73 & 4 & 16 & 62 & 11 & 27 \\
Full method & 71 & 2 & 15 & 62 & 9 & 27 \\
\bottomrule
\end{tabular}
\end{table}

\begin{table}[!htbp]
\centering\small
\caption{Loan-level classifiers, logistic regression / gradient-boosted trees: median over snapshots of the predicted over realised 7-year default rate (level ratio), the median absolute relative error of that rate (level error), and AUC.}
\label{tab:classifiers-full}
\resizebox{\linewidth}{!}{\begin{tabular}{@{}lcccccc@{}}
\toprule
 & \multicolumn{2}{c}{level ratio} & \multicolumn{2}{c}{level error} & \multicolumn{2}{c}{AUC} \\
\cmidrule(lr){2-3}\cmidrule(lr){4-5}\cmidrule(l){6-7}
Training labels & 2006--12 & 2013--18 & 2006--12 & 2013--18 & 2006--12 & 2013--18 \\
\midrule
Final labels (not available) & 1.01 / 1.01 & 0.98 / 0.99 & 3\% / 3\% & 3\% / 6\% & 0.854 / 0.864 & 0.790 / 0.828 \\
Events so far & 0.32 / 0.32 & 0.46 / 0.47 & 66\% / 67\% & 59\% / 58\% & 0.810 / 0.812 & 0.739 / 0.737 \\
Mature cohorts only & 0.23 / 0.16 & 1.64 / 2.06 & 78\% / 79\% & 116\% / 184\% & 0.735 / 0.721 & 0.777 / 0.772 \\
Drop recent & 0.19 / 0.16 & 1.87 / 1.22 & 71\% / 71\% & 94\% / 42\% & 0.729 / 0.719 & 0.764 / 0.770 \\
Delayed-feedback model & 0.64 & 1.13 & 38\% & 23\% & 0.822 & 0.795 \\
ES-DFM-style weights & 0.39 / 0.38 & 1.35 / 1.35 & 56\% / 55\% & 53\% / 55\% & 0.816 / 0.819 & 0.798 / 0.815 \\
PU, CL maturity & 0.69 / 0.67 & 1.11 / 1.10 & 49\% / 49\% & 26\% / 21\% & 0.823 / 0.830 & 0.786 / 0.817 \\
PU, calendar-adjusted CL maturity & 0.74 / 0.73 & 0.90 / 0.88 & 53\% / 62\% & 25\% / 28\% & 0.805 / 0.807 & 0.761 / 0.761 \\
PU, full-method maturity & 0.77 / 0.75 & 0.78 / 0.77 & 54\% / 54\% & 25\% / 30\% & 0.817 / 0.811 & 0.755 / 0.762 \\
\bottomrule
\end{tabular}}
\end{table}

\emph{Feature-dependent maturity (pre-specified).}\label{app:featmat} Current corrections use one maturity per origination month, so they cannot repair the ranking distortion of \cref{prop:ranking}. We pre-specified a test (specification committed before any real-data run of the method; a diagnostic of default timing by credit score had been computed earlier, \cref{tab:deviations}) of a maturity that depends on loan features: an accelerated-failure-time model for the default age given default within $H$, fitted on the mature vintages at each snapshot with origination features, gives each training loan its own maturity, which enters the positive--unlabelled EM in place of the cohort-level one (\texttt{pu\_feat}); a stratified chain ladder by credit-score and LTV bands is the nonparametric variant. The primary measure was the mean $|\log D|$ over pairs, where $D$ is the level ratio of credit scores of 720 and above over that of scores below 660, compared with the chain-ladder maturity (\texttt{pu\_cl}). A simulation run first passed (two loan types with speeds 1 and 0.6: $D$ = 0.95 with \texttt{pu\_cl}, 1.00 with \texttt{pu\_feat}).

On the real data the hypothesis is not supported: $|\log D|$ is 0.315 against 0.285 ($p = 0.15$), and the level-error ratio is 1.10, at its limit. The measure was poorly chosen: with final labels the median $D$ is 1.11 (crisis) and 1.30 (after 2012), and $|\log D| = 0.288$, so it mostly measures the misfit of the additive model across score bands. Measured post hoc as the distance to the final-label $D$, \texttt{pu\_feat} moves $D$ towards it for logistic regression, though not significantly (0.224 to 0.193, $p = 0.072$), and not for boosted trees, and the LTV distortion falls from 0.434 to 0.380 ($p = 0.006$; final labels 0.279). The timing effect itself is real but unstable: the log time ratio of scores of 720 and above against below 660 ranges from 0.03 to 0.74 across the 13 snapshots, and in an earlier check fits on early and late mature vintages differed by up to a factor of 30, because the crisis vintages, in which high-score loans defaulted late, dominate the mature data. Like a cohort's own speed, a feature's effect on timing is hard to learn when timing itself drifts.

\emph{An earlier experiment.} A first loan-level experiment (on the loan ages as first recorded; snapshots 2008, 2012 and 2016; a mixture model with one speed per vintage and no calendar step) is superseded by the one above: it reduced the ranking distortion at 2008 (0.87 against 0.35 for events so far) but under-predicted the level there by about 60\%, over-predicted it about threefold at 2012, and was the most distorted model at 2016.

\subsection{Patent applications}\label{app:patents}

\Cref{tab:patex,tab:patex-errors} give the pre-specified comparisons and the errors by snapshot. The 43 pairs are four filing years at each of 11 snapshots, less filing year 2003 at the 2007 snapshot, which is a training year (2001--2003 are used only for training).

\emph{Pre-specified comparisons.} The two primary comparisons (the full method against the chain ladder and against the calendar-adjusted chain ladder) are Holm-corrected; by the pre-specified sign test the first gives Holm $p = 0.023$ (\cref{tab:patex}), and by the sign-flip test adopted later Holm $p = 0.004$ on the realised path and 0.008 on the forecast path; it holds in three sensitivity analyses (published applications only, no 2001--2003 training cohorts, 96-month horizon; ratios 0.35--0.51, all intervals excluding one). \Cref{tab:inference} gives the steps and the comparison with a chain ladder with exact ages and groups (post hoc). Against drop recent and the last mature cohort the full method gives 0.54 ($p = 0.12$) and 0.61 ($p = 0.10$), winning at 7 of 11 snapshots against each on the realised path. With at least one calibration error, the strict interval rule covers 25 of 30 pairs (83\%); where fewer than nine calibration errors exist, the limit uses the largest one and carries no 90\% guarantee.

\emph{The inventory violates (C3).} The realised path uses the realised inventory, which counts the scored cohorts' own pending applications; the scored filing years hold 58--68\% of the pending inventory at a snapshot. When the inventory excludes them, the covariate's coefficient moves from $-1.29$ to $-0.14$ to between $-0.03$ and $+0.10$, the calendar step is no longer significant (0.86, $p = 0.30$; forecast path 0.93, $p = 0.13$), and the full method against the chain ladder gives 0.51 ($p = 0.011$). The calendar gain on patents is therefore the backlog of the scored cohorts themselves, known at the snapshot but driven by their outcomes, not an office-wide calendar effect. The inventory is counted from today's PatEx file, which holds published applications only.

\emph{Cause-specific hazard model (post hoc).} One Poisson regression per cause (grant, abandonment) at each snapshot on the application-months it shows, in cells of filing month $\times$ technology centre $\times$ age month, with a free monthly age baseline per cause, centre and filing-year effects, log pending inventory and the \emph{Alice} indicator, as in models of examination duration \citep{harhoff2009,tong2018}; the estimate is the share granted by the snapshot plus the product integral of the two hazards to 72 months, on the same pairs and scoring. Median error 3.9\% (realised) and 3.4\% (forecast); full method against it 1.99 [1.30, 3.13] ($p = 0.008$) and 1.90 [1.16, 3.07] ($p = 0.026$); calendar-adjusted chain ladder against it 2.39 ($p = 0.002$) and 2.07 ($p = 0.017$). Without filing-year effects: 4.0\% and 4.4\%. With class group and the first-inventor-to-file indicator: 3.2\% and 2.7\%. With the inventory without the scored filing years: 12.8\% and 7.5\%; against the full method with the same inventory 1.90 ($p = 0.021$) and 2.14 ($p = 0.020$). Wrong direction in 2 and 3 of 41 pairs. The inventory coefficient is $-1.3$ to $-0.3$ per standard deviation on grants and smaller on abandonments; without the scored years' applications it is about 0. The youngest filing years' grant effects are large and unstable (for 2011: $+0.97$ at the 2011 snapshot, $+0.49$ at 2012).

\begin{table}[!htbp]
\centering\small
\caption{Wrong direction of change (post hoc): number of scored pairs whose estimate lies on the wrong side of the latest mature cohort's rate, among pairs whose true rate differs from it by at least 2 points (patents, 41 of 43 pairs) or 10\% (mortgages, 74 of 78 pairs; reference vintage $T-7$ at the scored pair's horizon). Methods without covariates have one path.}
\label{tab:wrongdir}
\begin{tabular}{lcccc}
\toprule
 & \multicolumn{2}{c}{Patents (41)} & \multicolumn{2}{c}{Mortgages (74)} \\
\cmidrule(lr){2-3}\cmidrule(l){4-5}
Method & realised & forecast & realised & forecast \\
\midrule
Events so far & 28 & 28 & 15 & 15 \\
Chain ladder & 5 & 5 & 10 & 10 \\
Cape Cod, all cohorts & 3 & 3 & & \\
Calendar-adjusted chain ladder & 3 & 3 & 6 & 9 \\
Full method & 8 & 9 & 5 & 9 \\
Competing-risks hazard model & 2 & 3 & 1 & 3 \\
\bottomrule
\end{tabular}
\end{table}

\begin{table}[!htbp]
\centering\small
\caption{Patent applications: pre-specified comparisons (11 snapshots, 43 filing-year--snapshot pairs; 13 filing-year clusters). Sign-test $p$ without correction; with Holm's correction over the two primary comparisons, 0.023 and 1.0.}
\label{tab:patex}
\resizebox{0.95\linewidth}{!}{\input{tables/patex.tex}}
\end{table}

\begin{table}[!htbp]
\centering\small
\caption{Patent applications: mean absolute relative error of the 72-month grant share of the scored filing years, by snapshot. CL: chain ladder; Cal.-adj.: calendar-adjusted; Full: full method; (F): forecast path, otherwise realised path.}
\label{tab:patex-errors}
\resizebox{\linewidth}{!}{\input{tables/patex_errors.tex}}
\end{table}

\subsection{The pre-registered replication}\label{app:replication}

\paragraph{Design.} The study was written down in full before any of its data were downloaded: data, units, events, horizons, groups, covariates, snapshots, methods, hypotheses, margins, tests and the reporting rule. It was committed and time-stamped with OpenTimestamps (attested in Bitcoin blocks 973110 and 973112).
\begin{itemize}[leftmargin=*,itemsep=1pt,topsep=2pt]
\item \emph{SBA loans.} 7(a) loans approved 1991--2019 (public FOIA file as of 30 June 2026), excluding cancelled and committed loans, borrowers outside the 50 states and DC, loans whose outcome date is missing, and 4 loans with an event before approval: 1{,}369{,}660 loans and 193{,}667 charge-offs. Event charge-off, competing exit payoff, horizon 72 months. Groups are states; covariates are state unemployment and house-price growth. At year-end snapshots 2000--2018 the six youngest approval years are scored. No outcome after February 2020 is used, because COVID relief payments broke the link between covariates and losses, so pairs at 2014--2018 are scored against the incidence up to then, with the 12-month rule of the mortgage pre-COVID set: 109 pairs, 24 cohort clusters.
\item \emph{Federal civil cases} (FJC Integrated Database). Cases filed 1989--2016 in 90 district courts under federal-question or diversity jurisdiction, filed originally or removed from state court, excluding prisoner and social-security cases: 3{,}909{,}947 cases and 1{,}360{,}589 settlements. Event settlement, competing exit any other ending, horizon 36 months. Groups are districts. Covariates are the log number of active district judges, from the Biographical Directory of Article III judges (vacancies are set by the confirmation process, not by case outcomes), and state unemployment. At snapshots 1994--2016 the three youngest filing years are scored: 69 pairs, 25 clusters.
\item \emph{Methods and tests.} The methods are those of \cref{sec:backtests}, with the corrected plug-in, which is pre-specified here. The classifiers for H1 are logistic regressions trained on the units of the last ten years and scored on a random half of the scored cohorts. Each hypothesis is tested on the forecast path by a one-sided cluster sign-flip test of the mean log error ratio, shifted by the margin (all $2^{19}$ sign vectors for H1 on loans, otherwise $10^6$ random ones), with Holm's correction over the three hypotheses within each data set.
\end{itemize}

\needspace{5\baselineskip}\paragraph{Changes.}
\begin{itemize}[leftmargin=*,itemsep=1pt,topsep=2pt]
\item \emph{Before any method was run}, feasibility checks on the data's structure led to six decisions, logged with their reasons:
  \begin{itemize}[leftmargin=*,itemsep=0pt,topsep=1pt]
  \item the four territorial districts were dropped;
  \item months with no active judge use $\log 1$;
  \item only original and removed cases are kept, by the codebook labels;
  \item H1 on loans uses only the snapshots 2000--2013, whose truth covers the full 72 months that the classifiers predict;
  \item windows and detectors use only snapshots with that truth;
  \item wrong-direction counts on truncated pairs compare with the latest mature cohort at the same horizon.
  \end{itemize}
  Pending civil cases, which at first looked as if they ended before they were filed because of a placeholder end date, are censored. Implementation choices logged before the run: the strict interval rule; truth for H1 = the Aalen--Johansen incidence of the test units; and the hazard model's product integral runs to each pair's own horizon. The truncation rule of the mortgage pre-COVID set and the FRED copy of the unemployment series were applied as written.
\item \emph{After the run} we found that every SBA payoff before mid-2005 carries one of two placeholder dates (277{,}277 loans in the file, 269{,}693 of them in the sample).
  \begin{itemize}[leftmargin=*,itemsep=0pt,topsep=1pt]
  \item The frozen methods hardly use payoff dates. Dating these payoffs at random and rerunning at three snapshots (2002, 2006, 2016) leaves every cell-based method unchanged; at 2002, events so far and the chain ladder move by at most 0.24\%, Bornhuetter--Ferguson and Cape Cod by at most 0.08\%. A variant without the calendar step moves by up to 13\%, but it was never used, because the calendar guard was on at every snapshot. Windows and detectors were not checked.
  \item The hazard model's competing exit does use them, so its run as specified, on the dates as given, sees no exits before 2005.
  \item Fitted only on months from August 2005 (post hoc), it is not non-inferior: 1.15 [1.01, 1.32] against the full method, $p = 0.50$, on 79 pairs. As specified, on the same pairs, it is 0.91 [0.76, 1.08] ($p = 0.009$).
  \end{itemize}
\item \emph{After the run, an audit of the code} by three AI-agent reviewers that had not seen the results (one rebuilt the key numbers with its own code, and they matched) found six implementation errors, all fixed:
  \begin{itemize}[leftmargin=*,itemsep=0pt,topsep=1pt]
  \item civil cases: the original filing was chosen after, not before, the filters (106 cases; 3{,}909{,}947 remain);
  \item the classifiers' training set left out more units than the specification says;
  \item the window rules used $\kappa = \sqrt{2\log(N/\alpha)}$ instead of the specified $z_{\alpha/N}$;
  \item loan pairs whose truth is cut at February 2020 were counted in the interval coverage, although the intervals are calibrated for 72 months;
  \item two-sided $p$-values used other random sign vectors than specified, and Holm's correction used rounded $p$-values (both move $p$ by at most 0.0024).
  \end{itemize}
  The paper reports the corrected values. \Cref{tab:rep-asrun} sets them beside the run as first done; no verdict changes.
\item \emph{Mass-tort cases (post hoc).} Cases filed directly into a multidistrict-litigation docket are kept, as the decisions before the run say. They are up to 22.7\% of a filing year (2013), and from July 2016 the same kind of case is coded as an origin that was dropped. Without them, every verdict is the same, but the exact-age step no longer helps on civil cases (\cref{tab:rep-asrun}).
\end{itemize}

\paragraph{Results.} \Cref{tab:rep-main} gives every method and \cref{tab:rep-steps} each step. The calendar guard was on at every snapshot. The random walk was chosen at every loan snapshot and at all but the first civil one.
\begin{itemize}[leftmargin=*,itemsep=1pt,topsep=2pt]
\item \emph{Hazard model.} Its error ratio to the full method is in \cref{tab:replication}; a two-sided test of the full method against it gives $p = 0.19$ on loans and $p = 0.046$ on civil cases.
\item \emph{Wrong direction} (\cref{tab:rep-wrongdir}). On loans, the chain ladder and its exact-age variants make the fewest errors on the forecast path (8), the full method 14 and the hazard model 13; on the realised path the hazard model makes 5. On civil cases such errors are rare: 2 for the chain ladder, 1 for the full method and none for the hazard model.
\item \emph{Intervals.} With $J \ge 9$, the full method's 90\% intervals cover 54 of 61 loan pairs (89\%; only pairs whose target is the full 72 months, the calibration target) and 44 of 57 civil pairs (77\%), against 48 of 56 and 47 of 57 for chain-ladder intervals; reaching 90\% takes $\Gamma = 1.25$ on loans and $\Gamma = 2$ on civil cases (with the consistent random-walk evidence, a post hoc fix listed in \cref{tab:deviations}; 3 before).
\item \emph{Broader civil event} (settled or voluntarily dismissed). Speeds again do not help: 1.23 [0.94, 1.67], non-superiority $p = 0.012$. The hazard model is again more accurate: 0.64 [0.45, 0.87] against the full method, superiority $p = 0.006$.
\item \emph{Classifier maturity.} With the full method's maturity, classifier error is 0.36 and 0.09 times that of events so far: no better than with the chain ladder's maturity (1.31 [0.90, 1.88], $p = 0.20$; 1.03 [0.68, 1.57], $p = 0.90$).
\end{itemize}

\needspace{6\baselineskip}\paragraph{Limitations of the replication.}
\begin{itemize}[leftmargin=*,itemsep=1pt,topsep=2pt]
\item Units are defined by their records as of 2026: a loan cancelled later, or a case's attributes as last recorded, are known to the backtest, so it is cleaner than a real-time forecast. The covariates are today's revised series, not those published at each snapshot.
\item The SBA charge-off date is a processing date, with bursts in 2006, 2008 and 2010.
\item Civil cases include bulk filings and bulk endings on single days. Clustering by filing year allows for dependence within a cohort, not across cohorts.
\item Transfers and statistical closings count as competing exits, although such a case can still settle in another court.
\end{itemize}

\begin{table}[!htbp]
\centering\footnotesize
\caption{Replication: the run as first done, after the audit's corrections (reported in the paper), and, for civil cases, without mass-tort cases (post hoc). Ratios [95\% interval] on the forecast path; verdicts as in \cref{tab:replication}. Coverage: full method, $J \ge 9$; for loans after correction, only pairs with the full 72-month target. Windows: maturity-weighted rule, all units, median error. Civil coverage uses the consistent random-walk evidence (post hoc, \cref{tab:deviations}).}
\label{tab:rep-asrun}
\setlength{\tabcolsep}{4pt}
\resizebox{\linewidth}{!}{\begin{tabular}{@{}llccc@{}}
\toprule
 & & as first run & corrected (reported) & without mass-tort cases \\
\midrule
SBA loans & H1 & 0.27 [0.19, 0.38], confirmed & 0.27 [0.20, 0.38], confirmed & \\
 & H2 & 1.06 [0.85, 1.31], confirmed & same & \\
 & H3 & 0.86 [0.71, 1.05], confirmed & same & \\
 & interval coverage & 68 of 75 & 54 of 61 & \\
 & classifier maturity, full / CL & 1.33 [0.91, 1.88] & 1.31 [0.90, 1.88] & \\
 & windows & 20\% & 26\% & \\
\addlinespace
Civil cases & H1 & 0.08 [0.06, 0.11], confirmed & 0.09 [0.07, 0.12], confirmed & 0.08 [0.05, 0.11], confirmed \\
 & H2 & 1.10 [0.81, 1.54], inconclusive & 1.08 [0.78, 1.52], inconclusive & 0.92 [0.69, 1.21], inconclusive \\
 & H3 & 0.61 [0.38, 0.93], confirmed & 0.62 [0.39, 0.95], confirmed & 0.62 [0.42, 0.91], confirmed \\
 & exact-age step & 0.80 [0.64, 0.99] & 0.80 [0.64, 0.99] & 0.90 [0.69, 1.16] \\
 & calendar step & 1.17 [1.01, 1.48] & 1.17 [1.01, 1.48] & 1.16 [1.02, 1.45] \\
 & interval coverage & 44 of 57 & 44 of 57 & \\
 & classifier maturity, full / CL & 1.38 [1.02, 1.89] & 1.03 [0.68, 1.57] & \\
 & windows & 1.8\% & 1.4\% & \\
\bottomrule
\end{tabular}}
\end{table}

\begin{table}[!htbp]
\centering\footnotesize
\caption{Replication: every method on the same pairs (forecast path unless stated). Median over snapshots of the mean absolute relative error; error ratio to the chain ladder (CL) [95\% cohort-cluster bootstrap interval]; two-sided cluster sign-flip $p$ ($10^6$ random sign vectors).}
\label{tab:rep-main}
\setlength{\tabcolsep}{4pt}
\begin{tabular}{@{}lcccccc@{}}
\toprule
 & \multicolumn{3}{c}{SBA loans (109 pairs, 24 cohorts)} & \multicolumn{3}{c}{Civil cases (69 pairs, 25 cohorts)} \\
\cmidrule(lr){2-4}\cmidrule(l){5-7}
Method & error & ratio to CL & $p$ & error & ratio to CL & $p$ \\
\midrule
Events so far & 61\% & 3.35 [2.54, 4.38] & $<$0.001 & 31.5\% & 8.57 [6.47, 11.33] & $<$0.001 \\
Drop recent & 42\% & 2.50 [1.43, 4.37] & 0.006 & 7.3\% & 2.42 [1.43, 3.88] & 0.003 \\
Last mature cohort & 30\% & 2.01 [1.22, 3.32] & 0.015 & 7.5\% & 2.67 [1.68, 4.14] & $<$0.001 \\
Bornhuetter--Ferguson & 28\% & 1.23 [0.80, 1.97] & 0.397 & 3.5\% & 1.10 [0.87, 1.43] & 0.532 \\
Cape Cod, all cohorts & 36\% & 1.34 [0.77, 2.46] & 0.349 & 2.8\% & 0.91 [0.72, 1.21] & 0.546 \\
\addlinespace
Chain ladder (CL) & 34\% & 1 & & 3.6\% & 1 & \\
CL + exact ages & 28\% & 0.87 [0.81, 0.94] & 0.002 & 3.1\% & 0.80 [0.64, 0.99] & 0.050 \\
CL + exact ages + groups & 28\% & 0.87 [0.80, 0.94] & 0.002 & 2.9\% & 0.70 [0.51, 0.93] & 0.015 \\
Calendar-adjusted CL & 29\% & 0.98 [0.88, 1.08] & 0.659 & 2.9\% & 0.81 [0.68, 1.00] & 0.042 \\
Full method & 29\% & 1.04 [0.79, 1.34] & 0.810 & 3.3\% & 0.88 [0.65, 1.20] & 0.423 \\
Competing-risks hazard model & 19\% & 0.89 [0.65, 1.20] & 0.505 & 2.0\% & 0.54 [0.36, 0.80] & 0.007 \\
\addlinespace
\multicolumn{7}{@{}l}{\emph{Realised path}} \\
Calendar-adjusted CL & 30\% & 0.94 [0.80, 1.09] & 0.510 & 3.2\% & 0.86 [0.70, 1.08] & 0.227 \\
Full method & 34\% & 1.09 [0.80, 1.44] & 0.572 & 2.9\% & 0.91 [0.66, 1.27] & 0.607 \\
Competing-risks hazard model & 22\% & 0.97 [0.70, 1.29] & 0.865 & 2.3\% & 0.62 [0.44, 0.90] & 0.016 \\
\bottomrule
\end{tabular}
\end{table}

\begin{table}[!htbp]
\centering\footnotesize
\caption{Replication: the effect of each step, as the ratio to the previous row of \cref{tab:rep-main} (forecast path) [95\% interval] and two-sided $p$. Exact ages help on loans and, borderline, on civil cases (not without the mass-tort cases, \cref{tab:rep-asrun}); calendar covariates hurt in both; cohort speeds help in neither.}
\label{tab:rep-steps}
\begin{tabular}{@{}lcccc@{}}
\toprule
 & \multicolumn{2}{c}{SBA loans} & \multicolumn{2}{c}{Civil cases} \\
\cmidrule(lr){2-3}\cmidrule(l){4-5}
Step & ratio [95\% CI] & $p$ & ratio [95\% CI] & $p$ \\
\midrule
Exact ages & 0.87 [0.81, 0.94] & 0.002 & 0.80 [0.64, 0.99] & 0.050 \\
Groups & 0.99 [0.97, 1.01] & 0.442 & 0.87 [0.75, 0.98] & 0.042 \\
Calendar covariates & 1.13 [1.03, 1.24] & 0.021 & 1.17 [1.01, 1.48] & 0.046 \\
Cohort speeds & 1.06 [0.85, 1.31] & 0.606 & 1.08 [0.78, 1.52] & 0.672 \\
\bottomrule
\end{tabular}
\end{table}

\begin{table}[!htbp]
\centering\footnotesize
\caption{Replication: wrong direction of change. Number of scored pairs whose estimate lies on the wrong side of the latest mature cohort's rate at the pair's horizon, among pairs whose truth differs from it by at least 10\%.}
\label{tab:rep-wrongdir}
\begin{tabular}{@{}lcccc@{}}
\toprule
 & \multicolumn{2}{c}{SBA loans (83)} & \multicolumn{2}{c}{Civil cases (19)} \\
\cmidrule(lr){2-3}\cmidrule(l){4-5}
Method & forecast & realised & forecast & realised \\
\midrule
Events so far & 31 & 31 & 9 & 9 \\
Chain ladder & 8 & 8 & 2 & 2 \\
CL + exact ages & 8 & 8 & 1 & 1 \\
CL + exact ages + groups & 8 & 8 & 0 & 0 \\
Calendar-adjusted CL & 9 & 8 & 0 & 1 \\
Full method & 14 & 11 & 1 & 1 \\
Competing-risks hazard model & 13 & 5 & 0 & 0 \\
\bottomrule
\end{tabular}
\end{table}

\begin{table}[!htbp]
\centering\footnotesize
\caption{Replication: adaptive windows as the sample grows. Target: the horizon rate of the cohort of the year before the snapshot (loans 2000--2014, civil cases 1994--2016). Both rules use $\kappa = z_{\alpha/N}$, as pre-specified. Median relative error and, in brackets, the mean window length in cohorts; below 100\%, ten random subsamples.}
\label{tab:rep-windows}
\begin{tabular}{@{}lcccc@{}}
\toprule
 & \multicolumn{2}{c}{SBA loans} & \multicolumn{2}{c}{Civil cases} \\
\cmidrule(lr){2-3}\cmidrule(l){4-5}
Share of units & naive Lepski & maturity-weighted & naive Lepski & maturity-weighted \\
\midrule
10\% & 95\% (1.2) & 24\% (4.4) & 18\% (1.4) & 3.9\% (8.1) \\
30\% & 96\% (1.0) & 19\% (3.3) & 19\% (1.1) & 2.1\% (3.2) \\
100\% & 96\% (1.0) & 26\% (2.1) & 19\% (1.0) & 1.4\% (1.8) \\
\bottomrule
\end{tabular}
\end{table}

\begin{table}[!htbp]
\centering\footnotesize
\caption{Replication: ``rate fell'' alarms at immature entry months over the December snapshots (loans 2000--2013, civil cases 1994--2016), against the alarms raised on the true outcomes (in brackets). Any match: false alarms and missed true alarms under the rule of \cref{tab:alarms}. One-to-one: false and missed alarms when each alarm and each true alarm is matched at most once.}
\label{tab:rep-alarms}
\setlength{\tabcolsep}{4pt}
\begin{tabular}{@{}llcccccc@{}}
\toprule
 & & \multicolumn{3}{c}{SBA loans} & \multicolumn{3}{c}{Civil cases} \\
\cmidrule(lr){3-5}\cmidrule(l){6-8}
Detector & Labels & alarms (true) & any match & one-to-one & alarms (true) & any match & one-to-one \\
\midrule
ADWIN & events so far & 246 (98) & 100 / 0 & 149 / 1 & 67 (44) & 11 / 0 & 23 / 0 \\
 & CL-corrected & 90 (98) & 3 / 8 & 16 / 24 & 45 (44) & 0 / 0 & 4 / 3 \\
 & full-method-corrected & 85 (98) & 5 / 18 & 18 / 31 & 52 (44) & 1 / 0 & 11 / 3 \\
\addlinespace
Page--Hinkley & events so far & 60 (43) & 15 / 0 & 19 / 2 & 50 (43) & 3 / 0 & 8 / 1 \\
 & CL-corrected & 44 (43) & 7 / 7 & 9 / 8 & 44 (43) & 1 / 0 & 3 / 2 \\
 & full-method-corrected & 28 (43) & 4 / 13 & 4 / 19 & 44 (43) & 1 / 0 & 3 / 2 \\
\bottomrule
\end{tabular}
\end{table}

%% file: tables/baselines.tex
\begin{tabular}{@{}lccccccc@{}}
\toprule
 & \multicolumn{4}{c}{Mortgages (78 pairs, 18 vintages)} & \multicolumn{3}{c}{Patent applications (43 pairs, 13 filing years)} \\
\cmidrule(lr){2-5}\cmidrule(l){6-8}
 & \multicolumn{2}{c}{median error} & ratio to CL & $p$ & median & ratio to CL & $p$ \\
 & crisis & pre-COVID & [95\% CI] & & error & [95\% CI] & \\
\midrule
Events so far & 68\% & 55\% & 1.71 [1.26, 2.46] & 0.004 & 51\% & 2.26 [1.73, 3.07] & $<$0.001 \\
Drop recent & 64\% & 658\% & 4.49 [2.31, 8.39] & 0.001 & 13\% & 0.62 [0.42, 0.94] & 0.049 \\
Last mature cohort & 72\% & 417\% & 3.53 [2.24, 5.33] & 0.001 & 11\% & 0.55 [0.37, 0.82] & 0.021 \\
Bornhuetter--Ferguson & 62\% & 314\% & 2.10 [1.38, 3.09] & 0.007 & 10\% & 0.37 [0.23, 0.56] & 0.001 \\
Cape Cod & 69\% & 233\% & 2.83 [1.72, 4.78] & 0.001 & 9\% & 0.43 [0.30, 0.61] & 0.001 \\
Generalised Cape Cod (decay 0.75) & 69\% & 193\% & 2.76 [1.68, 4.47] & 0.002 & 16\% & 0.65 [0.53, 0.81] & 0.005 \\
\addlinespace
Chain ladder (CL) & 59\% & 35\% & 1 &  & 26\% & 1 &  \\
CL + exact ages & 59\% & 37\% & 1.01 [0.94, 1.08] & 0.723 & 14\% & 0.65 [0.56, 0.77] & 0.001 \\
CL + exact ages + groups & 59\% & 37\% & 1.02 [0.95, 1.10] & 0.576 & 13\% & 0.63 [0.54, 0.75] & 0.001 \\
\addlinespace
Hazard model, realised path & 15\% & 14\% & 0.33 [0.19, 0.56] & 0.001 &  &  &  \\
Hazard model, forecast path & 51\% & 17\% & 0.49 [0.34, 0.73] & 0.004 &  &  &  \\
Hazard model w/o vintage effects, realised path & 26\% & 20\% & 0.40 [0.27, 0.65] & 0.005 &  &  &  \\
Hazard model w/o vintage effects, forecast path & 54\% & 35\% & 0.70 [0.55, 0.94] & 0.029 &  &  &  \\
\addlinespace
Calendar-adjusted CL, realised path & 42\% & 20\% & 0.49 [0.27, 0.86] & 0.040 & 12\% & 0.40 [0.31, 0.51] & $<$0.001 \\
Calendar-adjusted CL, forecast path & 60\% & 24\% & 0.54 [0.39, 0.80] & 0.013 & 13\% & 0.38 [0.27, 0.50] & $<$0.001 \\
Full method, realised path & 39\% & 24\% & 0.51 [0.32, 0.83] & 0.020 & 7\% & 0.34 [0.21, 0.56] & 0.002 \\
Full method, forecast path & 54\% & 26\% & 0.44 [0.28, 0.72] & 0.012 & 7\% & 0.35 [0.20, 0.59] & 0.004 \\
\bottomrule
\end{tabular}

%% file: tables/inference.tex
\begin{tabular}{@{}lcccccc@{}}
\toprule
 & ratio [95\% CI] & sign-flip $p$ & wild $p$ & two-way 95\% CI & leave-one-out & worst $p$ \\
\midrule
\multicolumn{7}{@{}l}{\emph{Mortgages, pre-registered weights} (78 pairs, 18 cohort clusters, 13 snapshots)} \\
Full method vs CL & 0.56 [0.34, 0.93] & 0.045 & 0.045 & [0.26, 1.14] & 0.50--0.67 & 0.073 \\
Full method, forecast path, vs CL & 0.53 [0.34, 0.86] & 0.032 & 0.026 & [0.25, 1.09] & 0.47--0.62 & 0.059 \\
Step: speeds, realised path & 1.00 [0.78, 1.29] & 0.985 & 0.985 & [0.70, 1.48] & 0.94--1.06 & 0.973 \\
\midrule
\multicolumn{7}{@{}l}{\emph{Mortgages, corrected weights} (78 pairs, 18 cohort clusters, 13 snapshots)} \\
Full method vs CL & 0.51 [0.32, 0.83] & 0.020 & 0.021 & [0.18, 1.19] & 0.44--0.68 & 0.056 \\
Full method, forecast path, vs CL & 0.44 [0.28, 0.72] & 0.012 & 0.007 & [0.20, 0.93] & 0.39--0.53 & 0.027 \\
Full method vs CL + exact ages + groups & 0.50 [0.31, 0.81] & 0.016 & 0.014 & [0.18, 1.14] & 0.43--0.66 & 0.040 \\
Full method, forecast path, vs CL + exact ages + groups & 0.44 [0.29, 0.71] & 0.011 & 0.005 & [0.20, 0.88] & 0.39--0.51 & 0.021 \\
Step: exact ages & 1.01 [0.94, 1.08] & 0.723 & 0.726 & [0.90, 1.12] & 1.00--1.03 & 1.000 \\
Step: groups & 1.01 [1.00, 1.02] & 0.061 & 0.052 & [1.00, 1.03] & 1.01--1.01 & 0.143 \\
Step: calendar, realised path & 0.48 [0.27, 0.85] & 0.035 & 0.029 & [0.17, 1.16] & 0.40--0.59 & 0.087 \\
Step: calendar, forecast path & 0.53 [0.38, 0.79] & 0.012 & 0.005 & [0.29, 0.93] & 0.49--0.62 & 0.044 \\
Step: speeds, realised path & 1.04 [0.82, 1.31] & 0.729 & 0.722 & [0.69, 1.63] & 0.96--1.11 & 0.996 \\
Step: speeds, forecast path & 0.83 [0.62, 1.10] & 0.251 & 0.241 & [0.49, 1.31] & 0.75--0.91 & 0.575 \\
\midrule
\multicolumn{7}{@{}l}{\emph{Mortgages, borrower-followed} (78 pairs, 18 cohort clusters, 13 snapshots)} \\
Full method vs CL & 0.60 [0.42, 0.88] & 0.026 & 0.019 & [0.29, 1.30] & 0.51--0.72 & 0.093 \\
Full method, forecast path, vs CL & 0.42 [0.27, 0.67] & 0.007 & 0.002 & [0.18, 0.98] & 0.35--0.50 & 0.016 \\
Full method vs CL + exact ages + groups & 0.62 [0.43, 0.91] & 0.035 & 0.024 & [0.29, 1.34] & 0.51--0.73 & 0.117 \\
Full method, forecast path, vs CL + exact ages + groups & 0.43 [0.28, 0.70] & 0.010 & 0.003 & [0.18, 1.03] & 0.36--0.51 & 0.021 \\
Step: exact ages & 0.97 [0.91, 1.01] & 0.251 & 0.252 & [0.84, 1.07] & 0.96--1.00 & 0.940 \\
Step: groups & 1.01 [1.00, 1.03] & 0.294 & 0.313 & [0.99, 1.05] & 1.00--1.01 & 0.706 \\
Step: calendar, realised path & 0.54 [0.34, 0.89] & 0.039 & 0.027 & [0.21, 1.33] & 0.43--0.68 & 0.109 \\
Step: calendar, forecast path & 0.53 [0.39, 0.80] & 0.014 & 0.006 & [0.28, 1.02] & 0.47--0.59 & 0.025 \\
Step: speeds, realised path & 1.13 [0.84, 1.48] & 0.433 & 0.417 & [0.76, 1.78] & 1.02--1.21 & 0.910 \\
Step: speeds, forecast path & 0.81 [0.64, 1.02] & 0.122 & 0.117 & [0.52, 1.23] & 0.74--0.86 & 0.289 \\
\midrule
\multicolumn{7}{@{}l}{\emph{Patent applications} (43 pairs, 13 cohort clusters, 11 snapshots)} \\
Full method vs CL & 0.34 [0.21, 0.56] & 0.002 & 0.004 & [0.14, 0.93] & 0.26--0.43 & 0.011 \\
Full method, forecast path, vs CL & 0.35 [0.20, 0.59] & 0.004 & 0.008 & [0.14, 0.98] & 0.27--0.44 & 0.018 \\
Full method vs CL + exact ages + groups & 0.53 [0.34, 0.83] & 0.021 & 0.029 & [0.22, 1.25] & 0.42--0.67 & 0.107 \\
Full method, forecast path, vs CL + exact ages + groups & 0.55 [0.34, 0.88] & 0.039 & 0.046 & [0.24, 1.36] & 0.44--0.68 & 0.161 \\
Step: exact ages & 0.65 [0.56, 0.77] & 0.001 & $<$0.001 & [0.52, 0.84] & 0.63--0.67 & 0.003 \\
Step: groups & 0.98 [0.97, 0.99] & 0.005 & $<$0.001 & [0.96, 1.00] & 0.97--0.98 & 0.011 \\
Step: calendar, realised path & 0.64 [0.46, 0.86] & 0.024 & 0.023 & [0.34, 1.02] & 0.59--0.78 & 0.070 \\
Step: calendar, forecast path & 0.60 [0.41, 0.84] & 0.021 & 0.024 & [0.29, 0.99] & 0.56--0.75 & 0.057 \\
Step: speeds, realised path & 0.83 [0.51, 1.31] & 0.490 & 0.489 & [0.32, 2.38] & 0.62--1.11 & 0.653 \\
Step: speeds, forecast path & 0.92 [0.55, 1.45] & 0.749 & 0.746 & [0.34, 2.55] & 0.70--1.21 & 0.981 \\
\bottomrule
\end{tabular}

%% file: tables/inference_sensitivity.tex
\begin{tabular}{@{}lccccc@{}}
\toprule
 & MAE ratio & $p$ & floor $10^{-4}$ & floor $10^{-3}$ & 5\% trimmed \\
\midrule
\multicolumn{6}{@{}l}{\emph{Mortgages, pre-registered weights} (78 pairs, 18 cohort clusters, 13 snapshots)} \\
Full method vs CL & 0.29 & 0.008 & 0.56 (0.045) & 0.57 (0.046) & 0.62 (0.024) \\
Full method, forecast path, vs CL & 0.35 & 0.005 & 0.53 (0.032) & 0.54 (0.030) & 0.59 (0.026) \\
Step: speeds, realised path & 0.72 & 0.355 & 1.00 (0.984) & 1.00 (0.969) & 1.09 (0.520) \\
\midrule
\multicolumn{6}{@{}l}{\emph{Mortgages, corrected weights} (78 pairs, 18 cohort clusters, 13 snapshots)} \\
Full method vs CL & 0.34 & 0.002 & 0.51 (0.020) & 0.52 (0.019) & 0.59 (0.010) \\
Full method, forecast path, vs CL & 0.45 & 0.002 & 0.45 (0.012) & 0.46 (0.012) & 0.49 (0.004) \\
Full method vs CL + exact ages + groups & 0.40 & 0.002 & 0.50 (0.016) & 0.51 (0.015) & 0.64 (0.019) \\
Full method, forecast path, vs CL + exact ages + groups & 0.54 & 0.003 & 0.44 (0.011) & 0.45 (0.010) & 0.53 (0.013) \\
Step: exact ages & 0.85 & 0.511 & 1.01 (0.724) & 1.01 (0.733) & 1.01 (0.717) \\
Step: groups & 0.99 & 0.806 & 1.01 (0.061) & 1.01 (0.061) & 1.01 (0.128) \\
Step: calendar, realised path & 0.47 & 0.005 & 0.48 (0.035) & 0.49 (0.034) & 0.60 (0.056) \\
Step: calendar, forecast path & 0.68 & 0.005 & 0.53 (0.012) & 0.55 (0.012) & 0.59 (0.015) \\
Step: speeds, realised path & 0.87 & 0.088 & 1.04 (0.726) & 1.04 (0.723) & 0.99 (0.898) \\
Step: speeds, forecast path & 0.79 & 0.013 & 0.82 (0.238) & 0.82 (0.179) & 0.92 (0.362) \\
\midrule
\multicolumn{6}{@{}l}{\emph{Mortgages, borrower-followed} (78 pairs, 18 cohort clusters, 13 snapshots)} \\
Full method vs CL & 0.31 & 0.002 & 0.60 (0.026) & 0.60 (0.024) & 0.53 (0.005) \\
Full method, forecast path, vs CL & 0.38 & 0.002 & 0.42 (0.007) & 0.43 (0.007) & 0.45 (0.009) \\
Full method vs CL + exact ages + groups & 0.37 & 0.002 & 0.61 (0.034) & 0.61 (0.029) & 0.61 (0.014) \\
Full method, forecast path, vs CL + exact ages + groups & 0.46 & 0.003 & 0.43 (0.010) & 0.43 (0.009) & 0.46 (0.012) \\
Step: exact ages & 0.84 & 0.216 & 0.97 (0.266) & 0.98 (0.352) & 0.99 (0.479) \\
Step: groups & 0.99 & 0.533 & 1.01 (0.302) & 1.01 (0.402) & 1.00 (0.956) \\
Step: calendar, realised path & 0.46 & 0.004 & 0.55 (0.038) & 0.55 (0.037) & 0.61 (0.056) \\
Step: calendar, forecast path & 0.67 & 0.005 & 0.53 (0.013) & 0.53 (0.012) & 0.58 (0.022) \\
Step: speeds, realised path & 0.80 & 0.219 & 1.13 (0.440) & 1.11 (0.478) & 1.11 (0.484) \\
Step: speeds, forecast path & 0.69 & 0.002 & 0.81 (0.120) & 0.81 (0.112) & 0.79 (0.044) \\
\midrule
\multicolumn{6}{@{}l}{\emph{Patent applications} (43 pairs, 13 cohort clusters, 11 snapshots)} \\
Full method vs CL & 0.43 & 0.002 & 0.34 (0.002) & 0.35 (0.002) & 0.38 (0.007) \\
Full method, forecast path, vs CL & 0.42 & 0.003 & 0.35 (0.004) & 0.36 (0.004) & 0.33 (0.001) \\
Full method vs CL + exact ages + groups & 0.69 & 0.048 & 0.53 (0.022) & 0.56 (0.024) & 0.60 (0.044) \\
Full method, forecast path, vs CL + exact ages + groups & 0.67 & 0.057 & 0.55 (0.040) & 0.57 (0.040) & 0.61 (0.074) \\
Step: exact ages & 0.65 & $<$0.001 & 0.65 (0.001) & 0.65 (0.001) & 0.64 ($<$0.001) \\
Step: groups & 0.97 & 0.001 & 0.98 (0.005) & 0.98 (0.005) & 0.97 ($<$0.001) \\
Step: calendar, realised path & 0.76 & 0.062 & 0.64 (0.025) & 0.65 (0.027) & 0.76 (0.050) \\
Step: calendar, forecast path & 0.73 & 0.049 & 0.60 (0.021) & 0.62 (0.023) & 0.75 (0.038) \\
Step: speeds, realised path & 0.91 & 0.592 & 0.84 (0.496) & 0.85 (0.520) & 0.94 (0.817) \\
Step: speeds, forecast path & 0.92 & 0.622 & 0.92 (0.753) & 0.92 (0.758) & 0.86 (0.403) \\
\bottomrule
\end{tabular}

%% file: tables/speed_criterion.tex
\begin{tabular}{@{}lccccccc@{}}
\toprule
Snapshot & speed model & $\hat\phi$ & cohorts with & \multicolumn{2}{c}{error ratio, full / equal speeds} & \multicolumn{2}{c}{reference shape} \\
 & & & condition & realised path & forecast path & $a$ & $a c$ \\
\midrule
\multicolumn{8}{@{}l}{\emph{Mortgages, corrected weights}} \\
2006 & random walk, $\omega^2 = 0.03$ & 4.8 & 67\% & 0.50 & 1.18 & 415.0 & 26.2 \\
2007 & random walk, $\omega^2 = 0.03$ & 3.3 & 83\% & 0.76 & 0.98 & 1256.8 & 44.4 \\
2008 & random walk, $\omega^2 = 0.03$ & 4.0 & 83\% & 0.85 & 0.95 & 1182.9 & 43.0 \\
2009 & random walk, $\omega^2 = 0.03$ & 5.3 & 83\% & 0.90 & 0.85 & 583.2 & 31.3 \\
2010 & random walk, $\omega^2 = 0.03$ & 12.4 & 67\% & 0.84 & 0.34 & 167.6 & 17.1 \\
2011 & random walk, $\omega^2 = 0.03$ & 17.3 & 50\% & 0.91 & 0.34 & 31.7 & 7.6 \\
2012 & random walk, $\omega^2 = 0.03$ & 17.7 & 33\% & 0.84 & 0.83 & 0.9 & 2.0 \\
2013 & random walk, $\omega^2 = 0.03$ & 33.0 & 17\% & 0.49 & 0.70 & 0.4 & 1.7 \\
2014 & random walk, $\omega^2 = 0.03$ & 24.5 & 0\% & 1.97 & 0.97 & 0.3 & 1.6 \\
2015 & random walk, $\omega^2 = 0.1$ & 24.4 & 17\% & 3.05 & 0.27 & 0.3 & 1.6 \\
2016 & random walk, $\omega^2 = 0.1$ & 23.0 & 33\% & 2.05 & 1.02 & 0.3 & 1.6 \\
2017 & random walk, $\omega^2 = 0.1$ & 22.3 & 17\% & 1.78 & 2.61 & 0.2 & 1.5 \\
2018 & random walk, $\omega^2 = 0.03$ & 20.8 & 17\% & 0.88 & 1.87 & 0.2 & 1.5 \\
\multicolumn{8}{@{}l}{Realised path: condition holds, 34 pairs, error ratio 1.05; fails, 44 pairs, 1.04} \\
\multicolumn{8}{@{}l}{Forecast path: condition holds, 34 pairs, error ratio 0.90; fails, 44 pairs, 0.77} \\
\midrule
\multicolumn{8}{@{}l}{\emph{Patent applications}} \\
2007 & random walk, $\omega^2 = 0.01$ & 25.0 & 67\% & 0.70 & 1.36 & 1185.7 & 46.9 \\
2008 & random walk, $\omega^2 = 0.003$ & 38.0 & 50\% & 2.12 & 3.33 & 230.5 & 22.7 \\
2009 & random walk, $\omega^2 = 0.003$ & 41.8 & 50\% & 0.53 & 0.42 & 107.0 & 17.9 \\
2010 & random walk, $\omega^2 = 0.01$ & 75.9 & 75\% & 0.65 & 0.61 & 11.4 & 6.6 \\
2011 & random walk, $\omega^2 = 0.03$ & 80.3 & 75\% & 1.58 & 1.56 & 4.3 & 4.5 \\
2012 & random walk, $\omega^2 = 0.03$ & 109.4 & 75\% & 14.05 & 12.62 & 2.4 & 3.6 \\
2013 & random walk, $\omega^2 = 0.01$ & 150.3 & 50\% & 1.98 & 2.50 & 1.9 & 3.4 \\
2014 & random walk, $\omega^2 = 0.003$ & 161.5 & 25\% & 0.48 & 0.44 & 1.7 & 3.3 \\
2015 & random walk, $\omega^2 = 0.003$ & 149.3 & 50\% & 0.05 & 0.06 & 1.6 & 3.3 \\
2016 & random walk, $\omega^2 = 0.003$ & 132.1 & 50\% & 0.46 & 0.49 & 1.4 & 3.2 \\
2017 & random walk, $\omega^2 = 0.003$ & 135.5 & 50\% & 0.53 & 0.55 & 1.3 & 3.1 \\
\multicolumn{8}{@{}l}{Realised path: condition holds, 24 pairs, error ratio 0.83; fails, 19 pairs, 0.83} \\
\multicolumn{8}{@{}l}{Forecast path: condition holds, 24 pairs, error ratio 0.92; fails, 19 pairs, 0.91} \\
\bottomrule
\end{tabular}

%% file: tables/freddie_tests.tex
\begin{tabular}{llcccc}
\toprule
 & & \multicolumn{2}{c}{Snapshots (13)} & \multicolumn{2}{c}{Vintage level (78 pairs)} \\
\cmidrule(lr){3-4}\cmidrule(lr){5-6}
Comparison & Run & wins & sign-test $p$ & error ratio & 95\% CI of log ratio \\
\midrule
Full method vs chain ladder (primary) & pre-reg.\ weights & 9/13 & 0.267 & 0.56 & $[-1.07,\,-0.08]$ \\
 & corrected & 9/13 & 0.267 & 0.51 & $[-1.15,\,-0.19]$ \\
 & no HARP rule & 10/13 & 0.092 & 0.57 & $[-0.94,\,-0.16]$ \\
 & follows borrowers & 10/13 & 0.092 & 0.60 & $[-0.86,\,-0.12]$ \\
\addlinespace
Calendar-adjusted CL vs chain ladder & pre-reg.\ weights & 9/13 & 0.267 & 0.56 & $[-1.24,\,+0.05]$ \\
 & corrected & 9/13 & 0.267 & 0.49 & $[-1.30,\,-0.15]$ \\
 & no HARP rule & 10/13 & 0.092 & 0.56 & $[-0.96,\,-0.14]$ \\
 & follows borrowers & 10/13 & 0.092 & 0.53 & $[-1.10,\,-0.15]$ \\
\addlinespace
Full method, forecast path, vs chain ladder & pre-reg.\ weights & 11/13 & 0.022 & 0.53 & $[-1.08,\,-0.15]$ \\
 & corrected & 11/13 & 0.022 & 0.44 & $[-1.26,\,-0.33]$ \\
 & no HARP rule & 9/13 & 0.267 & 0.51 & $[-1.04,\,-0.25]$ \\
 & follows borrowers & 10/13 & 0.092 & 0.42 & $[-1.30,\,-0.40]$ \\
\addlinespace
Full method vs calendar-adjusted CL & pre-reg.\ weights & 8/13 & 0.581 & 1.00 & $[-0.24,\,+0.26]$ \\
 & corrected & 9/13 & 0.267 & 1.04 & $[-0.20,\,+0.27]$ \\
 & no HARP rule & 6/13 & 1.000 & 1.01 & $[-0.23,\,+0.24]$ \\
 & follows borrowers & 7/13 & 1.000 & 1.13 & $[-0.18,\,+0.39]$ \\
\addlinespace
Full method vs events so far & pre-reg.\ weights & 12/13 & 0.003 & & \\
 & corrected & 12/13 & 0.003 & & \\
 & no HARP rule & 12/13 & 0.003 & & \\
 & follows borrowers & 12/13 & 0.003 & & \\
\addlinespace
Full method vs drop recent & pre-reg.\ weights & 11/13 & 0.022 & & \\
 & corrected & 11/13 & 0.022 & & \\
 & no HARP rule & 11/13 & 0.022 & & \\
 & follows borrowers & 11/13 & 0.022 & & \\
\bottomrule
\end{tabular}

%% file: tables/harp_check.tex
\begin{tabular}{lrccccc}
\toprule
 & & \multicolumn{3}{c}{Within 60 months} & \multicolumn{2}{c}{Within 36 months} \\
\cmidrule(lr){3-5}\cmidrule(lr){6-7}
Parent vintage & Relief loans in the sample & HARP path & At-risk path & Ratio & HARP path & At-risk path \\
\midrule
2003 & 38,269 & 0.75\% & 2.28\% & 0.33 & 0.48\% & 1.77\% \\
2004 & 25,717 & 1.24\% & 4.10\% & 0.30 & 0.77\% & 3.19\% \\
2005 & 38,635 & 1.95\% & 6.75\% & 0.29 & 1.29\% & 5.28\% \\
2006 & 32,340 & 3.00\% & 10.44\% & 0.29 & 1.97\% & 8.17\% \\
2007 & 39,917 & 3.08\% & 11.11\% & 0.28 & 2.00\% & 8.66\% \\
2008 & 45,754 & 2.29\% & 7.76\% & 0.29 & 1.49\% & 6.01\% \\
2009 & 27,391 & 0.49\% & 1.32\% & 0.37 & 0.32\% & 0.98\% \\
\midrule
All & 248,023 & 1.91\% & 6.56\% & 0.29 & 1.24\% & 5.10\% \\
\bottomrule
\end{tabular}

%% file: tables/overid.tex
\begin{tabular}{lrcccccc}
\toprule
 & & \multicolumn{3}{c}{Pre-registered weights} & \multicolumn{3}{c}{Corrected weights} \\
\cmidrule(lr){3-5}\cmidrule(lr){6-8}
Vintage & Events & $\log(s_{\text{late}}/s_{\text{early}})$ & $T$ & $p$ & $\log(s_{\text{late}}/s_{\text{early}})$ & $T$ & $p$ \\
\midrule
1999 & 2,146 & $-0.06$ & 0.2 & 0.649 & $-0.06$ & 0.2 & 0.651 \\
2000 & 3,084 & $-0.07$ & 0.3 & 0.581 & $-0.07$ & 0.4 & 0.512 \\
2001 & 2,076 & $+0.26$ & 0.3 & 0.599 & $+0.26$ & 0.2 & 0.641 \\
2002 & 1,902 & $+0.11$ & 0.2 & 0.682 & $+0.11$ & 0.2 & 0.685 \\
2003 & 1,819 & $+0.15$ & 0.3 & 0.606 & $+0.16$ & 0.3 & 0.596 \\
2004 & 4,837 & $+0.24$ & 1.0 & 0.311 & $+0.24$ & 1.3 & 0.254 \\
2005 & 11,324 & $+0.24$$^{*}$ & 4.7 & 0.030 & $+0.32$$^{*}$ & 6.1 & 0.014 \\
2006 & 18,611 & $+0.18$$^{*}$ & 8.4 & 0.004 & $+0.16$$^{*}$ & 10.3 & 0.001 \\
2007 & 23,455 & $+0.36$$^{*}$ & 10.9 & 0.001 & $+0.19$$^{*}$ & 12.7 & 0.000 \\
2008 & 13,701 & $+0.25$$^{*}$ & 8.4 & 0.004 & $+0.23$$^{*}$ & 6.2 & 0.013 \\
2009 & 2,751 & $+0.39$$^{*}$ & 5.3 & 0.022 & $+0.40$$^{*}$ & 5.7 & 0.017 \\
2010 & 2,510 & $+0.05$ & 0.3 & 0.592 & $+0.05$ & 0.3 & 0.583 \\
2011 & 1,972 & $+0.15$ & 0.7 & 0.409 & $+0.14$ & 0.7 & 0.415 \\
2012 & 1,904 & $+0.09$ & 0.7 & 0.392 & $+0.09$ & 0.7 & 0.398 \\
2013 & 1,805 & $+0.20$ & 1.0 & 0.319 & $+0.20$ & 1.0 & 0.310 \\
2014 & 1,624 & $+0.72$ & 2.6 & 0.105 & $+0.72$ & 2.5 & 0.115 \\
2015 & 1,341 & $+0.04$ & 0.2 & 0.670 & $+0.04$ & 0.2 & 0.664 \\
2016 & 1,544 & $-0.05$ & 0.2 & 0.670 & $-0.05$ & 0.2 & 0.666 \\
2017 & 1,907 & $+0.34$ & 2.1 & 0.149 & $+0.34$ & 2.1 & 0.148 \\
2018 & 2,069 & $+1.15$$^{*}$ & 16.5 & 0.000 & $+1.14$$^{*}$ & 16.6 & 0.000 \\
2019 & 1,754 & $-0.10$ & 0.5 & 0.481 & $-0.09$ & 0.5 & 0.475 \\
\bottomrule
\end{tabular}

%% file: tables/e2.tex
\begin{tabular}{lcccccc}
\toprule
 & \multicolumn{3}{c}{Mean chosen window (vintages)} & \multicolumn{3}{c}{Median relative error} \\
\cmidrule(lr){2-4}\cmidrule(lr){5-7}
Loans per vintage & 10k & 50k & 200k & 10k & 50k & 200k \\
\midrule
Naive Lepski & 2.91 & 1.88 & 1.00 & 79\% & 88\% & 90\% \\
Naive ADWIN & 2.50 & 1.03 & 1.00 & 84\% & 90\% & 90\% \\
\midrule
Maturity-weighted Lepski, proven constants & 9.38 & 6.00 & 5.23 & 189\% & 84\% & 64\% \\
Maturity-weighted Lepski, $\kappa = \kappa_N$ & 4.92 & 3.83 & 2.77 & 64\% & 64\% & 50\% \\
Maturity-weighted Lepski, $\kappa = 1$ & 2.51 & 1.89 & 1.54 & 58\% & 52\% & 48\% \\
Maturity-weighted Lepski, rolling-origin $\kappa$ & 3.92 & 3.02 & 2.23 & 62\% & 62\% & 50\% \\
Best maturity-weighted window in hindsight & 2.48 & 2.30 & 2.54 & 44\% & 43\% & 44\% \\
\midrule
Calendar-adjusted maturity-weighted Lepski, $\kappa = \kappa_N$ & -- & -- & 3.00 & -- & -- & 28\% \\
Calendar-adjusted CL (window 1) & -- & -- & 1.00 & -- & -- & 41\% \\
\bottomrule
\end{tabular}

%% file: tables/patex.tex
\begin{tabular}{lcccc}
\toprule
 & \multicolumn{2}{c}{Snapshots (11)} & \multicolumn{2}{c}{Filing-year level (43 pairs)} \\
\cmidrule(lr){2-3}\cmidrule(lr){4-5}
Comparison & wins & sign-test $p$ & error ratio & 95\% CI of log ratio \\
\midrule
Full method vs chain ladder (pre-specified primary) & 10/11 & 0.012 & 0.34 & $[-1.57,\,-0.57]$ \\
Full method vs calendar-adjusted CL (pre-specified primary) & 6/11 & 1.000 & 0.83 & $[-0.67,\,+0.27]$ \\
Full method vs events so far & 11/11 & 0.001 &  &  \\
Full method vs drop recent & 7/11 & 0.549 &  &  \\
Full method vs last mature cohort & 7/11 & 0.549 &  &  \\
Calendar-adjusted CL vs chain ladder & 11/11 & 0.001 & 0.40 & $[-1.19,\,-0.68]$ \\
Full method w/o calendar vs chain ladder & 8/11 & 0.227 & 0.68 & $[-0.80,\,+0.03]$ \\
Full method, forecast path, vs chain ladder & 10/11 & 0.012 & 0.35 & $[-1.61,\,-0.53]$ \\
\midrule
Wrong direction of change & \multicolumn{4}{l}{events so far 68\%, drop recent 71\%, chain ladder 12\%, calendar-adjusted CL 7\%, full method 20\%} \\
\bottomrule
\end{tabular}

%% file: tables/patex_errors.tex
\begin{tabular}{lccccccccc}
\toprule
Snapshot & Events so far & Drop recent & Last mature & CL & Full w/o calendar & Cal.-adj.\ CL & Cal.-adj.\ CL (F) & Full & Full (F) \\
\midrule
2007 & 64.8\% & 21.3\% & 21.3\% & 37.5\% & 33.9\% & 9.3\% & 5.4\% & 14.7\% & 8.9\% \\
2008 & 55.9\% & 20.2\% & 18.4\% & 34.5\% & 27.1\% & 3.3\% & 2.9\% & 4.1\% & 4.7\% \\
2009 & 59.2\% & 17.5\% & 11.8\% & 38.8\% & 32.5\% & 22.7\% & 23.5\% & 20.1\% & 20.7\% \\
2010 & 57.3\% & 8.9\% & 5.3\% & 32.8\% & 20.6\% & 20.2\% & 20.3\% & 14.5\% & 14.5\% \\
2011 & 55.7\% & 5.6\% & 7.4\% & 21.5\% & 24.8\% & 12.4\% & 12.5\% & 19.8\% & 19.8\% \\
2012 & 51.4\% & 9.8\% & 12.8\% & 8.6\% & 28.9\% & 4.4\% & 4.6\% & 20.5\% & 20.6\% \\
2013 & 46.7\% & 14.8\% & 15.0\% & 10.6\% & 12.0\% & 5.5\% & 5.1\% & 6.4\% & 6.7\% \\
2014 & 42.0\% & 15.2\% & 11.4\% & 21.7\% & 10.3\% & 11.2\% & 10.7\% & 5.2\% & 4.7\% \\
2015 & 38.7\% & 12.7\% & 6.9\% & 26.5\% & 8.8\% & 14.1\% & 13.7\% & 2.6\% & 3.0\% \\
2016 & 37.1\% & 10.3\% & 6.5\% & 26.3\% & 9.8\% & 14.8\% & 14.4\% & 6.4\% & 6.7\% \\
2017 & 35.8\% & 9.5\% & 7.9\% & 23.6\% & 11.3\% & 14.1\% & 13.8\% & 6.9\% & 7.1\% \\
\midrule
Median & 51.4\% & 12.7\% & 11.4\% & 26.3\% & 20.6\% & 12.4\% & 12.5\% & 6.9\% & 7.1\% \\
\bottomrule
\end{tabular}

%% file: sections/app_data.tex
\section{Data}\label{app:data}

\paragraph{Freddie Mac.} Single-Family Loan-Level Dataset, Release 47 (July 2026), Standard files for origination years 1999 to the first quarter of 2026, downloaded in October 2026. The last performance month is March 2026.
\begin{itemize}[leftmargin=*,itemsep=1pt]
\item Sample: Standard dataset only; simple random sample of 200,000 loans per origination year, fixed seed. Refinances into HARP relief loans are identified from all origination files and censored at refinance; 21--33\% of the sampled loans of the 2005--2008 vintages later refinanced into HARP. \Cref{tab:harp-check} counts the relief loans that are themselves in the sample, by parent vintage; that count is not a share of the parent vintage, because parents and relief loans are sampled independently. The official 50,000-loan sample files are not used for results. The user guide describes them as a simple random sample of each vintage year, but in every year the files hold 12,500 loans per origination quarter, so quarters with many originations are under-weighted.
\item Check of the censoring assumption (exploratory; \cref{tab:harp-check}, \cref{app:freddie}): the sample holds 248,023 HARP relief loans whose parents were originated in 2003--2009. Within 60 months, 1.9\% of them default. The loans of the parent's vintage, state and LTV band that were still at risk at the parent's age at refinance default at 6.6\% (ratio 0.29; 0.28--0.37 by parent vintage). So HARP censoring is informative. The difference mixes selection (a HARP refinance requires a current loan) and the effect of the programme.
\item Event definition: default is the first of (i) the first month 180 days or more past due (REO acquisition included) without a borrower-assistance, disaster or payment-deferral flag, and (ii) a zero balance with code 02 (third-party sale), 03 (short sale or charge-off) or 09 (REO disposition). Prepayment (competing) is zero balance code 01 with no earlier default. All other loans are censored at their last reported month, including repurchases (code 96) and whole-loan or reperforming-loan sales (codes 15, 16) without an earlier default, and loans active at the data end. Modified loans stay at risk. Counting flagged months would give 174{,}027 defaults instead of 142{,}664. Sensitivities: censoring at February 2020, and the raw definition.
\item Calendar covariates: FHFA state house-price index growth and BLS state unemployment; the hazard-model baseline also uses Freddie Mac's 30-year mortgage rate survey (PMMS). Licences: Freddie Mac data for research use without redistribution of loan-level data; FHFA, BLS and PMMS series and USPTO PatEx are public.
\end{itemize}

\paragraph{US patent applications (PatEx).} USPTO Patent Examination Research Dataset, release folder ``2022'' (data downloaded by the USPTO from PEDS in June 2023), file \texttt{application\_data} only.
\begin{itemize}[leftmargin=*,itemsep=1pt]
\item Sample: nonprovisional utility applications filed 2001--2019 in the technology centres 16, 17, 21, 24, 26, 28, 36 and 37; a simple random sample of 100,000 per filing year, fixed seed. The published-only sensitivity keeps applications with a pre-grant publication.
\item Event: grant (\texttt{patent\_issue\_date}). Competing exit: abandonment. The file has no abandonment date, so an application is abandoned if its current status contains ``Abandoned'' and it has no patent; the date is \texttt{appl\_status\_date}. A status that changed later (for example a revival) is not visible. All other applications are censored at the data date (June 2023).
\item Pending inventory: per technology centre and month, the number of utility applications in the file that are filed and not yet granted or abandoned. Applications filed before 2001 are in PatEx only if they were granted or published, so the inventory is too low before about 2004. Applications filed from 2022 on are mostly not public yet, so it is also too low from 2022.
\item Coverage (\cref{tab:patex-coverage}): PatEx holds only public applications. An application that was abandoned and never published is missing, so each filing year's denominator is too small, while grant timing among grants is exact. We estimate the missing share from the data, under the assumption that an abandoned application is unpublished as often as a granted one: 7.0\% for 2001, 5.6\% for 2002, 2--4\% for 2003--2014 and 1.2\% for 2019. Filing years 2001--2003 are used for training only.
\item Two implementation choices that the pre-specified protocol does not fix: (i) before July 2014 the Alice indicator is zero in everything a snapshot shows, so the snapshots 2007--2013 use the inventory covariate alone and the snapshots 2014--2017 use both; (ii) on the forecast path the inventory stays at its last-12-month mean per centre and the Alice indicator at its value at the snapshot.
\end{itemize}

\paragraph{SBA 7(a) loans (replication).} US Small Business Administration, 7(a) and 504 FOIA data, 7(a) files for fiscal years 1991 to the present, as of 30 June 2026, with the data dictionary (\url{https://data.sba.gov/dataset/7a-504-foia}); US government data, public domain. Borrower names and addresses were dropped on ingest. Calendar covariates: BLS state unemployment (LAUS, seasonally adjusted, via the FRED copies of the same series) and the FHFA all-transactions state house-price index.

\paragraph{Federal civil cases (replication).} Federal Judicial Center Integrated Database, civil cases, cumulative files to 30 June 2026, with the civil codebook, and the FJC Biographical Directory of Article III judges (export of judicial service records); US government data, public domain. Settled cases are those with disposition code 13. Calendar covariates: the monthly number of active (non-senior) Article III judges per district and BLS state unemployment.

\begin{table}[h]
\centering\small
\caption{PatEx coverage by filing year: utility applications in the eight technology centres, outcomes at the data date (June 2023), share with a pre-grant publication, and the estimated share of applications that PatEx does not hold (abandoned and never published).}
\label{tab:patex-coverage}
\input{tables/patex_coverage.tex}
\end{table}

%% file: tables/patex_coverage.tex
\begin{tabular}{lrcccc}
\toprule
Filing year & Applications & Granted & Abandoned & Pre-grant publication & Estimated missing share \\
\midrule
2001 & 268,923 & 73.7\% & 26.2\% & 80.1\% & 7.0\% \\
2002 & 274,433 & 72.5\% & 27.4\% & 84.5\% & 5.6\% \\
2003 & 278,491 & 69.8\% & 30.2\% & 91.2\% & 3.7\% \\
2004 & 294,780 & 66.1\% & 33.8\% & 93.5\% & 3.2\% \\
2005 & 310,943 & 63.1\% & 36.8\% & 94.2\% & 3.2\% \\
2006 & 324,837 & 61.6\% & 38.3\% & 94.9\% & 3.0\% \\
2007 & 334,585 & 62.7\% & 37.3\% & 95.3\% & 2.7\% \\
2008 & 328,372 & 65.0\% & 34.9\% & 95.5\% & 2.3\% \\
2009 & 303,152 & 67.8\% & 32.1\% & 95.6\% & 2.0\% \\
2010 & 323,127 & 68.7\% & 31.2\% & 95.5\% & 2.0\% \\
2011 & 341,466 & 68.8\% & 31.1\% & 95.3\% & 2.0\% \\
2012 & 364,145 & 70.4\% & 29.5\% & 94.6\% & 2.2\% \\
2013 & 384,618 & 71.0\% & 28.9\% & 94.5\% & 2.2\% \\
2014 & 397,343 & 72.1\% & 27.6\% & 94.7\% & 2.0\% \\
2015 & 397,683 & 74.7\% & 25.0\% & 94.2\% & 1.9\% \\
2016 & 398,557 & 76.7\% & 22.8\% & 93.8\% & 1.9\% \\
2017 & 408,326 & 78.0\% & 20.8\% & 93.9\% & 1.7\% \\
2018 & 412,038 & 77.2\% & 19.3\% & 94.1\% & 1.5\% \\
2019 & 433,863 & 74.9\% & 15.9\% & 94.3\% & 1.2\% \\
\bottomrule
\end{tabular}

%% file: sections/app_related.tex
\section{Extended related work}\label{app:related}

\paragraph{Delayed outcomes in learning.} Delayed-feedback models \citep{chapelle2014,yoshikawa2018,ktena2019,yang2021esdfm} use a delay distribution that does not depend on the cohort and choose training windows empirically. In bandits, learning is impossible when the environment changes on the timescale of the delay unless intermediate signals are available \citep{vernade2017,vernade2020}, and proxies can stand in for delayed outcomes \citep{mann2019}; label delay also degrades online continual learning \citep{csaba2024}. Adaptive windows \citep{mazzetto2023adaptive,han2024model,lepski1991,goldenshluger2011} and change detectors \citep{bifet2007adwin} assume final labels, work on verification latency studies drift detection under delay \citep{zliobaite2010,krempl2014,grzenda2020}, and survival modellers discard recent records to avoid censoring leakage \citep{xu2026pitfalls}.

\paragraph{Cure, hazard and demographic models.} The product of an eventual probability and a delay distribution is the mixture cure model \citep{larson1985,li2002,zhang2007}, where sufficient follow-up is a classical question \citep{maller1996,li2001}; the promotion-time cure model \citep{chen1999} bounds the cumulative hazard instead, and a Box--Cox family nests the two \citep{yin2005cjs}. Level-and-speed structures appear in extended hazard models \citep{etezadi1987,chen2001,rubio2019,rubio2023}, shape-invariant regression \citep{lawton1972,kneip1988}, and demographic models of delayed versus forgone events \citep{coale1972,bloom1984,goldstein2001,bongaarts1998}; completing incomplete fertility cohorts, complex methods rarely beat simple ones \citep{bohkewald2018}, as cohort speeds do not here.

\paragraph{Reserving, nowcasting and age--period--cohort models.} The chain ladder is the equal-speed case of our model and the age--cohort submodel of age--period--cohort analysis \citep{mack1993,renshaw1998,kuang2008,harnau2018}. Cape Cod and Bornhuetter--Ferguson pool immature periods, with a decay factor set by judgment \citep{stanard1985,gluck1997,bornhuetter1972,richman2026}; separation methods remove calendar effects \citep{taylor1977,barnett2000}; and Berquist--Sherman adjustments, growth curves and the changing-settlement-rate model let development speed change \citep{berquist1977,clark2003,guszcza2008,meyers2015}. Nowcasting corrects for reporting delay, including change points in the delay distribution \citep{lawless1994,kalbfleisch1991,hohle2014,mcgough2020}. Measured covariates are the mechanism-based route around the age--period--cohort identity \citep{heckman1985,winship2008}, as in credit-risk vintage models \citep{breeden2007} and mortgage termination models \citep{deng2000}. Our intervals build on conformal prediction, including its versions for drift and hierarchical data \citep{vovk2005,lei2018,barber2023,tibshirani2019,gibbs2021,dunn2023}.

%% file: sections/app_impact.tex
\section{Broader impact}\label{app:impact}

Naive labels penalise recent and slowly maturing groups in credit and venture investment, so correcting them can change who receives capital. Our methods estimate cohort-level rates and do not by themselves make decisions about individuals. Decisions based on corrected rates should account for the uncertainty that \cref{prop:intervals} makes explicit, and for the assumptions of \cref{sec:calendar}, which can fail when covariates respond to the outcomes.